\documentstyle[epsfig]{report}

\def\sqr#1#2{{\vcenter{\hrule height.#2pt
   \hbox{\vrule width.#2pt height#1pt \kern#1pt
    \vrule width.#2pt} 
    \hrule height.#2pt}}}
\def\dalamb {\mathchoice{\sqr{6.9}4}{\sqr{6.9}4}{\sqr{5.5}3}{\sqr{4.6}3}}
\def\meio { \frac{1}{2} }
\def\Lie { {\cal L}_{\xi} }
\def\Lies { {\cal L}_{\sigma} }
\def\LieX { {\cal L}_X }
\def\LieY { {\cal L}_Y }
\def\LieZ { {\cal L}_Z }
\def\gmn { g_{\mu \nu} }
\def\tgmn { \tilde{g}_{\mu \nu} }
\def\ymn { \gamma_{\mu\nu} }
\def\tymn { \tilde{\gamma}_{\mu \nu} }

\def\dgmn { \delta g_{\mu\nu} }
\def\tdgmn { \delta {\tilde g}_{\mu\nu} }

\def\hmn { h_{\mu \nu} }
\def\hij { h_{ij} }

\def\H { {\cal H} }
\def\tQ { \tilde{Q} }

\def\tP { \tilde{\cal P} }
\def\tq { \tilde{q} }
\def\tg { \tilde{g} }

\def\tp { \tilde{p} }
\def\tr { \tilde{r} }

\def\Gmn { G_{\mu \nu} }

\def\Tmn { T_{\mu \nu} }

\def\Pmn { \Pi_{\mu \nu} }

\def\tx { \tilde {x} }
\def\tX { \tilde {X} }
\def\tY { \tilde {Y} }
\def\tZ { \tilde {Z} }
\def\eps { \epsilon }
\def\veps { \varepsilon }
\def\taumn { \tau_{\mu \nu} }

\newcommand{\be}{\begin{equation}}
\newcommand{\ee}{\end{equation}}
\newcommand{\beq}{\begin{eqnarray}}
\newcommand{\eeq}{\end{eqnarray}}
\def\vv { \varphi }
\def\dv { \delta \varphi }
\def\Vl { V_{, \vv}}
\def\Vll { V_{, \vv \vv}}
\def\Mpl {M_{pl}}
\def\cv { {\vartheta} }

\def\CMP#1#2#3{{\it Commun.\ Math.\ Phys.}\/ {\bf#1} (19#2) #3}

\begin{document}

\begin{flushright}
	{\small BROWN-HET-1096}
\end{flushright}

\vskip 1cm

\begin{center}

{\Large The Back Reaction of Gravitational Perturbations 
and Applications in Cosmology\footnote{Work based on the Ph.D. 
thesis by the author, Brown University (1997).}}

\vskip 1.5cm

{\large
L. Raul W. Abramo$^*$}

Physics Department, Brown University, Providence, R.I. 02912 

\vskip 2.5cm

Abstract
\end{center}

\begin{quote}
\noindent We study the back reaction
of cosmological perturbations on the evolution of the
universe. The object usually employed to describe the
back reaction of perturbations is called the effective energy-momentum
tensor (EEMT) of cosmological perturbations. In this
formulation, the problem of the gauge dependence of the EEMT must
be tackled.
We advance beyond traditional results that involve only
high frequency perturbations in vacuo\cite{Isaacson}, and formulate
the back reaction problem in a gauge invariant manner for
completely generic perturbations. 
We give a quick proof that
the EEMT for high-frequency perturbations is gauge invariant
which greatly simplifies the pioneering approach by
Isaacson\cite{Isaacson}.
As applications we analyze the back reaction of gravitational waves
and scalar metric fluctuations in Friedmann-Robertson-Walker
background spacetimes.
We investigate in particular back reaction effects during
inflation in the Chaotic scenario. Fluctuations
with a wavelength much bigger than the Hubble radius
during inflation contribute a negative energy density, and
in that case back reaction counteracts any pre-existing cosmological
constant.
Finally, we set up the equations of motion for the
back reaction on the geometry and on the matter,
and show how they are perfectly consistent with
the Bianchi identities and the continuity equations.

\end{quote}

\vfill



		\chapter{Introduction}

It is a well-known fact in oceanography that waves in the sea
carry some amount of energy, which is released in the form of thermal
energy as the waves crash
on continental shorelines. This energy is responsible for a small fraction
of the total thermal energy stored in the oceans' water. 
Since waves pass through each other
without much dispersion, the problem of propagation of sea waves can be
treated linearly, and in fact oceanography analyses the several different
classes of waves (of different amplitudes and wavelengths)  independently
from each other. Waves ``average to zero" in the sense that the overall
sea level, for a constant amount of water, is set at $0$ meters. 
Nevertheless, the average energy stored in waves is not zero,
since it is not a linear but really a {\it quadratic} function of the
wave. This second order phenomenon is the simplest case of a nonlinear
feedback effect, or {\it back reaction}, in fluid dynamics.

A close analogy can be drawn between sea waves (called, ironically,
``waves of gravity" in oceanographic jargon) and the coupled matter and
metric perturbations ({\it gravitational} perturbations for short) with
which we will concern ourselves here.  The sea level and the
temperature of the sea water correspond, respectively, to the background
geometry upon which the waves (gravitational perturbations) propagate, and
their energy density; the average of a gravitational perturbation is
zero as well; the problem of propagation of perturbations can be treated
in linear theory as a first approximation, although the exact problem,
like fluid dynamics, is fundamentally nonlinear; the energy stored in the
gravitational perturbations is, similarly, a quadratic function of the
perturbations; finally, metric and density perturbations can effect the
background geometry and energy density in which they propagate.

The theory of General Relativity\cite{Landau,Weinberg,MTW}, used here to
describe gravitational systems, is much more complex than fluid dynamics in
that one of its key elements is covariance under general gauge
transformations that change the coordinate frames of observers in the
manifold spacetime. Although this constraint on Einstein's equations can be
dealt with in exact problems containing symmetries and even in the linear
theory of perturbations, to second order (when the quadratic effects
associated with the energy carried by the waves appear), a formalism
appropriate to the situation at hand has been sorely lacking in the
literature. Consequently, many difficulties arise in the treatment of
problems related to back reaction such as long calculations, lack of
well-defined physical observables and, gravest of all: gauge dependence of
the source of back reaction effects (which would correspond to the sea
temperature), called the effective energy-momentum tensor of gravitational
perturbations.

We separate gravitational perturbations in two broad classes: 
fluctuations which are of short wavelength and high frequency (HF), and
the ones which have long wavelength and low frequency (LF).  Although the
problems cited above have been partially solved in the case of HF
perturbations\cite{Isaacson}, we will see that, quite surprisingly, the
decisive contribution from back reaction in at least one case (the Chaotic
Inflationary scenario\cite{LindeSlava}) comes actually from the soft LF
perturbations, and not from the hard HF ones.

We will show how a crucial ingredient in solving these problems is the fact
that even linear gauge transformations (i.e., transformations that simply
redefine the perturbations of matter and metric) can change the
characteristics of the background geometry, as well as its dynamics.  With
that element in hand, we will show how it is possible to construct physical
observables (gauge invariant quantities) which measure the effects of the
back reaction of gravitational perturbations.

We apply this formalism to the problems of back reaction of gravitational
waves (metric perturbations which are decoupled from matter sources) and
scalar perturbations (coupled matter and metric fluctuations) in
homogeneous and isotropic Friedmann-Robertson-Walker (FRW) background
geometries.  One of the results is that, due to back reaction, the equation
of state of long wavelength gravity waves suffers corrections that have not
been included in previous treatments.

The main result of our analysis comes from an application to the case of
scalar perturbations during inflation in the chaotic model, where we have
shown that back reaction can reduce the speed and the duration of
inflation. The power of the feedback effects comes not from the amplitude
of the perturbations, which is always small, but from the fact that the
expansion of the universe redshifts the physical wavelengths of
perturbations causing an exponential growth of the population of soft
infrared (IR) modes.

As a spinoff of these ideas we suggest that back reaction might provide
a dynamical relaxation mechanism whereby a preexisting cosmological
constant is screened.

This work is organized as follows: in Chapter 2 we treat the problem of
propagation of coupled metric and matter perturbations in FRW spacetimes,
with special care given to the simple but interesting case of scalar field
matter. Chapter 3 deals with the Chaotic Inflationary scenario, and with
the process of generation of cosmological perturbations by quantum
fluctuations of the scalar field. These introductory chapters can be
skipped if the reader is familiar with modern cosmology.

We give a brief historical account of early as well as recent attempts to
tackle the back reaction of gravitational perturbations in Chapter 4, and
show how the problem in the context of HF perturbations can be formulated
in a simpler way than before. In Chapter 5 we discuss the theory of finite
gauge transformations, which we argue is fundamental to an understanding of
some conceptual as well as practical issues related to back reaction. We
also provide a covariant formulation of back reaction in generic
backgrounds.  Finally, in chapters 6 and 7 we develop applications to the
case of gravity waves and scalar perturbations.


		\chapter{The Theory of Cosmological Perturbations}


Recent years have seen a confirmation of the basic working hypotheses of
Cosmology: 1. that the universe is approximately homogeneous, with small
perturbations superimposed onto this smooth background, and 2. that the
universe is expanding, its energy density falling, and the temperature of
the primordial background radiation is decreasing.  The COBE
satellite\cite{COBE} and a series of similar experiments\cite{CMBR} have
determined this fact beyond any dispute, and now it is finally clear that
the plasma that predominated in the universe up to some $300,000$
years after the Big Bang was extremely homogeneous, in fact
its temperature did not vary more than one part in $10^5$.

The structures on large scales that we see with the help of telescopes
today (planets, clouds, galaxies, clusters of galaxies and so on) are but
the result of a few billion-years of clumping of these tiny inhomogeneities
in the plasma. The theory of the origin and evolution of these cosmological
perturbations has become a cornerstone of modern cosmology, since by virtue
of the theory of cosmological perturbations it is finally becoming possible
to falsify models of the evolution of the early universe, as well as to
test some of the underlying tenets of relativistic cosmology. The
observation of the cosmic background radiation anisotropies, together with
the theory of their origin and evolution, provides the matrix on which the
basic parameters of cosmology (expansion rate, density, ratio of baryonic
matter to total matter, deacceleration parameter and so on) have to be
fitted.

Having established the importance of the study of perturbations in
cosmology, it is to this we now turn.

\section{Gravitational instability}
\label{SecGrIn}

Soon after publishing his seminal paper on General Relativity, Einstein
used his theory to try and describe the Cosmos. It is important to notice
that by 1917, when he published his ``Kosmologische Betrachtungen zur
allgemeinen Relativit\"atstheorie"\cite{Einstein}, E. Hubble had not yet
established Slipher's suggestion that distant galaxies are receding, and
therefore as far as anyone was concerned the Universe was static. Einstein
soon realized that his theory would not lead to such a static Universe,
and in order to achieve that he introduced a ``universal constant", now
widely known as the cosmological constant $\Lambda$. In the ensuing model,
a closed universe is kept static by the repulsive gravitational force
coming from a fine-tuned cosmological constant. It just turns out that
this model is unstable: it would expand (collapse) at an accelerated rate
if the matter was slightly less (more) dense than some critical density.

The recent history of large-scale structure in the Universe
is dominated by a similar type of gravitational instability: galaxies,
for example, formed from gas clouds that collapsed by gravitational
attraction around the galactic center. By the same process, smaller (stars,
planets) and bigger (clusters and superclusters) structures also formed
from this coarsening of a quasi-homogeneous expanding fluid. An evolution
scenario that starts by forming the largest structures first is known as 
``top-down", while one in which the small structures
form first is known as ``bottom-up". The intermediate
picture, of ``scale-invariant" evolution, is the one that is currently
favored by experimental data.

In the same manner that an overdensity is unstable to gravitational
attraction and can grow to become a galaxy, for example, underdensities
are also unstable and grow with time to form voids,
regions with a much lower density of galaxies. In
that case it is useful to think of an ``effective gravitational
repulsion"\cite{Piran} acting on the matter in the
boundaries between the voids, and pushing them together into pancakes.
It is not clear which are the structures that
dominate the Universe at the present time, but it has been
argued that in the scale of clusters of galaxies, voids are in fact more
pervasive than clusters\cite{Piran}.

Perturbations in the homogeneous 
fluid of a static Universe, as often happens with instabilities in
Partial Differential Equations, grow in amplitude exponentially (their
physical size of course grows together with the expansion of the Universe).
If the expansion rate of the Universe is a power-law
in time, i.e., if the physical scales $R \sim t^m$, the expansion
counteracts the gravitational attraction and the perturbations' amplitudes 
now grow only as a power-law: $\delta q \sim t^n$. Finally, if the
universe is expanding exponentially (as it does if a cosmological constant
dominates the evolution), perturbations are asymptotically static,
that is, their amplitude tends to a constant. 
\footnote{We should note that in fact there are always two solutions for
the perturbations, one growing, the other decaying, and the
remarks above apply to the dominant (growing) modes. Neglect of the
decaying modes is usually justified, but sometimes can lead to
crucial mistakes \cite{Grischuk,SlavaDerruele}.}

Since the Universe at the present time seems to be expanding as a
power-law, we conclude that it is also slowly becoming more and more
complex, with growing voids and clusters of galaxies. But if there was a
time when the Universe inflated exponentially, then perturbations with
large enough wavelengths $\lambda > R_H$ 
would have been frozen during this period. If by
some mechanism perturbations of that scale were created with the same
amplitude during an inflationary phase, once that phase was over and a
power-law expansion subsided the perturbations could then grow and start
forming stars, galaxies, voids, etc. The spectrum of perturbations, that
is, the amplitude of each mode with physical wavelength $\lambda$, would
then be ``scale-invariant", which is consistent with the data available
now.

In mathematical language, we will solve
Einstein's theory linearized around an expanding 
homogeneous background geometry\cite{Landau,Weinberg,MTW}. 
A crucial
issue is related to the freedom of gauge,
that is, the symmetry under general coordinate transformations. 
It turns out that it is possible to build variables which
have the same value in all coordinate frames, i.e., gauge-invariant
variables, and those are of course the physically meaningful quantities.
In the next sections we consider this theory in more detail.

\section{Relativistic cosmology: background models}
\label{SecReCo}

Clearly a Newtonian approximation to the description of the Universe in
its largest scale is insufficient: since the expansion rate is now $65 \pm
15 km s^{-1} Mpc^{-1}$ (see \cite{HubbleConst} for an account of the
latest measurements), galaxies at a distance of $1000 Mpc$ from the Milky
Way, for example, are receding from us at $20\%$ of the velocity of light.
The expansion rate gets even bigger as we look back into the past, thus
only Einstein's relativistic theory is capable of providing an accurate
description.

In a perfectly homogeneous and isotropic Universe, there are no preferred
directions or axes in space, and the geometry, the metric
and all other physical parameters depends only on time. We write the
line element in this ``cosmic time" as

\be
\label{Line_El}
ds^2 = dt^2 - a^2(t)d\vec{x}^2 \quad ,
\ee
where $a(t)$ is the scale factor. The spatial coordinates $\vec{x}$
are called ``comoving coordinates", and are related to physical
distances by $\vec{r} = a(t) \vec{x}$ , hence the name ``scale factor".
If we could sprinkle dots over the universe, the physical distances between
dots would increase if the Universe was expanding and decrease if the
Universe
happened to be contracting, but the comoving distances between dots would
nevertheless remain always the same. Eq. (\ref{Line_El}) is known
as the Friedmann-Robertson-Walker (FRW) metric on flat space (we will not
discuss here the trivial but cumbersome cases of open and closed
Universes)

It will be practical, in certain applications, to use a different clock
or time-slicing, where the time variable (called ``conformal time") is
$d\eta = dt/a(t) $ . In this case we can write the line element as

\be
\label{Lin_El_Co}
ds^2 = a^2(\eta) ( d{\eta}^2 - d\vec{x}^2 ) \quad .
\ee
The time derivative of a function $f$ with respect to conformal time,
$f'$, can be related to the derivative with respect to cosmic time, 
$\dot{f}$, using the rule

\be 
\label{Conf_Cosm} 
\frac{d}{d\eta} f = a(t) \frac{d}{dt} f \quad .  
\ee
The reader can freely exchange between the cosmic and conformal time
pictures at any time. 

The equations determining the kinematics and the dynamics of matter
in General Relativity are the Einstein Field Equations\footnote{We use
units where $8 \pi G=M_{pl}^{-2}=1 $ .},

\be
\label{Eins_Eq}
\Gmn = \Tmn
\ee
where $\Gmn$ is the Einstein tensor (which expresses a curvature
of spacetime) and $\Tmn$ is the energy-momentum tensor (EMT) of matter.

We will later consider more extensively the case of scalar field
matter, where the EMT is given by

\be
\label{EMT_Sc}
T_{\mu\nu}^{S} = \varphi_{,\mu} \varphi_{,\nu} - 
\gmn \left[ \frac{1}{2} \varphi^{,\alpha} \varphi_{,\alpha} - V(\varphi)
\right] \quad,
\ee
where $V(\vv)$ is the potential of the scalar field $\vv$.

For now, let us examine the more common cases of
hydrodynamical fluids such as pressureless matter (dust) 
and radiation, whose EMTs can be expressed as

\be
\label{EMT_Hy}
\Tmn = \left( \begin{array}{cccc} 
			\rho & 0 & 0 & 0 \\
			0 & p & 0 & 0 \\
			0 & 0 & p & 0 \\
			0 & 0 & 0 & p 
		\end{array}
       \right)
\ee
where $\rho(t)$ is the energy density, and the pressure $p(t)$ is related
to the energy density by the equation of state $p=w\rho$ with $w$
constant.  For matter we have $w_d=0$, while for radiation $w_r=1/3$. It
is useful to note that when the dominant component of the Universe is a
cosmological constant, the EMT can also be written as in (\ref{EMT_Hy}),
with $w_\Lambda =-1$. 

In the homogeneous and isotropic background of a FRW Universe, with metric
(\ref{Lin_El_Co}), the field equations (\ref{Eins_Eq}) are diagonal and
the time-time and space-space parts read, respectively, 

\beq
\label{Eins_Eq_FRW_00}
 3 H^2 &=& \rho \quad , \\
\label{Eins_Eq_FRW_ij}
 -3H^2 - 2 \dot{H} &=& p
\eeq
where $H=\dot{a}/a$ is the Hubble parameter, which measures the
expansion rate of the universe. The inverse of the Hubble parameter
$H^{-1}$ has dimensions of time, and is in fact a fundamental time scale
in cosmology, since it is proportional to the age of the Universe -
around 14 billion years, give or take a couple of billion years.

The geometric fact expressed by the Bianchi identities, $G_{\mu\nu}^{;\mu}
= 0$ is reciprocated on the rhs of (\ref{Eins_Eq})  by the continuity
equations, $T_{\mu\nu}^{;\mu} =0$ (semi-colons denote covariant
derivative). The continuity equations are expressed, in a FRW Universe, by
the constraint

\be
\label{Cont_FRW}
\dot{\rho} + 3 H (\rho + p) = 0 \quad .
\ee
It is easy to see that the identity above is just the integrability
condition that must be satisfied in order that Eqs. (\ref{Eins_Eq_FRW_00})
and (\ref{Eins_Eq_FRW_ij}) have a solution.
We can find the first integral of (\ref{Cont_FRW}), which is

\be
\label{Rho_a}
\rho \sim a(t)^{-3(1+w)} \quad .
\ee

Pressureless matter will have an energy density inversely proportional to
the volume $V^{-1} \sim a^{-3}$, while radiation is not only diluted but
also suffers from redshift, $\lambda_{ph} = \lambda/a(t)$, thus its energy
density is proportional to
$a^{-4}$. The temperature in radiation is related to its energy density by
statistical mechanics, $\rho_r \sim T_r^4$, and thus $T_r \sim a^{-1}$. A
cosmological constant ($w_\Lambda = -1$) have constant energy density.  An
important consequence of these equations is that, as we run the clock
backwards (thus decreasing $a$), the energy density of radiation ( $\sim
a^{-4}$ ) increases more rapidly than the one in matter ($ \sim a^{-3}$)
while the energy density due to a cosmological constant stays constant.  We
thus conclude that no matter how small the energy density in radiation is
right now, it will necessarily be dominant in the early Universe. 

Substituting the above expression into Eq. (\ref{Eins_Eq_FRW_00})
yields the solution for the scale factor as a function of
time,

\be
\label{a_t}
a(t) \sim t^{\frac{2}{3(1+w)}}
\ee
if $w \neq -1$ and

\be
\label{a_t_CC}
a(t) \sim e^{H_0 t}
\ee
if $w=-1$. In this last case, called de Sitter spacetime, 
the scale factor grows exponentially
and the Hubble parameter is a constant, as evident in Eq.
(\ref{Eins_Eq_FRW_00}) . The fact that for non-exotic matter (such as a
cosmological constant) 
the scale factor goes to zero when $t \rightarrow 0$ is an
intrinsic feature of General Relativity, and this special case
of space-like singularity is called the Big Bang.

At the present time the Universe is dominated by pressureless matter, with
a fraction ($10^{-4}$) of the energy density contained in a background of
photons which occupy the microwave band with a temperature of
$2.73^{\circ} K$ (thus the name Cosmic Microwave Background Radiation -
CMBR). Approximately 1,000,000 years after the Big Bang, though, the
energy density is still just high enough that the temperature of the CMBR
can ionize Hydrogen atoms. After this moment (sometimes also called {\it
recombination}), when $T_{dec} \approx 3000^{\circ}K$, matter becomes
transparent to radiation and they decouple from each other.

An important consequence of recombination is that photons did not scatter
from matter after the time of decoupling $t_{dec}$ up to the present time.
In other words, they traveled in a straight line since this ``surface of
last scattering" at $t_{dec}$ (see Fig. 2.1), and if we can set
the temperature of our experiment low enough that we can detect those
photons, we will be seeing a picture of the last events that took place at
this time of decoupling. 

\begin{figure} 
\centerline{\epsfig{file=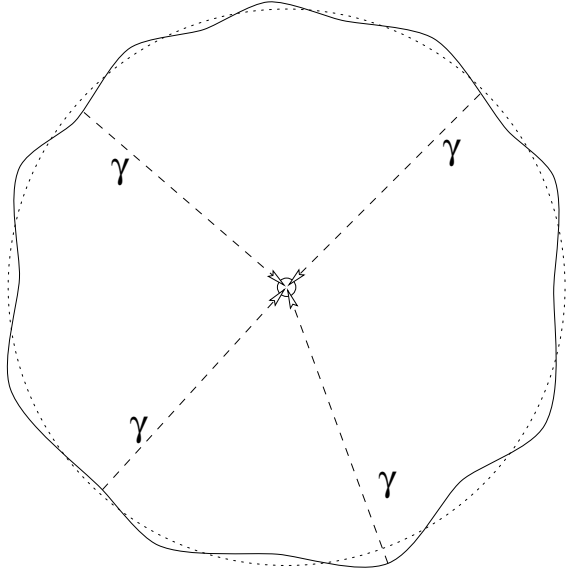,height=3.0in}}
\vspace{10pt}
\caption{\small Surface of last scattering (dotted circle) of the 
cosmic microwave background radiation. Photons that emerged
from overdense regions after the time of decoupling are more redshifted
then photons that came from underdense regions.}
\label{Fig_Su}
\end{figure}

That is the content of the map of the CMBR sky produced by the COBE
satellite. Photons coming from the surface of last
scattering have an average temperature of $2.736^{\circ}K$, with
fluctuations in the quadrupole of about one part
in $5 \times 10^{-6}$, or $10 mK$.  There is also a dipole that was
subtracted in the
CMBR map which is ascribed to a peculiar motion of our galaxy, with
respect to this ``ether"  of primordial photons, of about $370$ $km$
$s^{-1}$.

Notice that the presence of this background of radiation introduces a class
of preferred reference frames (coordinates) for observers in the Universe: 
the comoving observers\footnote{Peculiar velocities of galaxies with
respect to the comoving frame can indeed be inferred from the way in which
those galaxies feel the background radiation, and underlies the
Sunyaev-Zel'dovich method for measuring velocities of distant galaxies.}. 
It would make little sense to utilize a reference frame which is moving
with respect to the background photons, since it would introduce dipoles in
the CMBR, in the Hubble diagrams, and indeed in most all-sky observations.

It is important to stress this fact: when talking about cosmology, we are
going to consider only reference frames (or coordinate frames) which are,
within a first approximation, {\it static} with
respect to the cosmic background radiation. When we consider gauge
transformations, in this context, we will be always considering
transformations between coordinate frames inside this class of quasi-static
comoving frames. A precise mathematical formulation of this
point will be given in Section 5.1.

Before concluding this section let us examine the important case
of scalar field matter, which will be crucial in the next sections.
If the Universe is filled with a homogeneous scalar field $\vv_0 (t)$
with potential energy $V( \varphi_0 )$, the EMT is diagonal with energy
density and pressure given by

\beq
\label{Rho_Sc}
\rho_s = \frac{1}{2} \dot{\vv}_0^2 + V(\vv_0) \quad , \\
\label{Pre_Sc}
p_s = \frac{1}{2} \dot{\vv}_0^2 - V(\vv_0) \quad .
\eeq
When the potential term dominates, the EMT is identical
to the EMT of a cosmological constant $\Lambda = V(\vv_0)$.
If such a scalar field is dominant over other forms of matter, the Universe will
expand exponentially - or {\it inflate} - like de Sitter space, 
until the kinetic terms become comparable to the potential term.
If this inflationary period lasts long enough, an apparent
event horizon is created due to the exponentially fast expansion rate. 
Two observers in opposite poles of a sphere with radius
the size of the horizon $R_H$ will not be able to exchange signals,
because the expansion of the Universe is so fast that those signals cannot reach
the other poles. Of course, once inflation is over those signals can progress
towards the other observer once again, but during inflation we can define
this horizon. Clearly, the faster the expansion rate the smaller the horizon
size, and in fact, $R_H = H^{-1}$. The existence of horizons during
periods of quasi-exponential expansion will be a welcome feature of 
inflationary models that we explore later on.

In many situations (during the reheating phase after inflation, for
example) the scalar field is
oscillating coherently around the zero of the potential with amplitude 
$2 {\bar\vv}_0 $, and then on average the kinetic term equals
the potential term, implying that $\bar{p}_s=0$ and 
$\bar{\rho}_s =2 V ( \bar{\vv}_0 ) $.

The continuity equations in the case of scalar 
field matter are just the equation 
of motion for $\vv_0$ (the Klein-Gordon equation in a FRW spacetime) 
and read
 
\be
\label{EOM_vv}   
\ddot{\vv}_0 + 3 H \dot{\vv}_0 + V_{,\vv}(\vv_0) = 0 \quad .
\ee 
In general the first situation described above (potential energy dominance)
implies that the first term in (\ref{EOM_vv}) can be neglected. In that
case $\dot{\vv}_0 \cong - V'/3H$ and the scalar field is ``slowly rolling''
down towards the minimum of the potential. The second situation 
(coherent oscillations) usually means that the 
friction term $3 H \dot{\vv}_0$ is negligible.

We conclude then that if the Universe
was initially dominated by a scalar field 
displaced from the zero of its potential, 
it would inflate exponentially until the friction term became small, and then
it would expand in a power-law just like in a matter-dominated Universe with 
$p_s=0$.

\section{Perturbations of FRW spacetimes}
\label{SecPeFRW}
 
The mosaic of hot and cold spots in the COBE map of the CMBR
shows that some regions of the early Universe were
slightly hotter (and therefore denser) than others.
Since energy curves space, this implies that the metric
was also largely homogeneous, with some small bumps
corresponding to the hot and cold spots. Call $\eps$ the adimensional
measure of the magnitude of these inhomogeneities. In the
case of our Universe at the time of decoupling, $\eps$ is, explicitly,

\be
\label{Eps}
\eps_{dec} = \left. \frac{\delta T}{T} \right|_{dec} \approx 10^{-5} \quad , 
\ee
where $\delta T$ denotes the fluctuations in the temperature field.
Metric perturbations can be preliminarily thought of as proportional to the
variation of the newtonian potential, which in turn is proportional to
the temperature fluctuations of the primordial plasma.
The constant $\eps$ thus regulates the size of both
matter and metric perturbations, and in fact can even be related to the
small parameter that enters when we consider gauge transformations (see
next section).

Although the background variables depend only on the coordinate $\eta$ (we 
use conformal time here for convenience)
the perturbed variables can vary along all the 4 coordinates. The scalar 
field, thus, will be written as

\be
\label{Sca_Ex}
\vv = \vv_0(\eta) + \eps \dv (\eta,\vec{x}) \quad ,
\ee
and the metric $\gmn$ as

\be
\label{Met_Ex}
\gmn = \ymn (\eta) + \eps \delta\gmn (\eta,\vec{x})
\ee
where $\ymn
$ is given in (\ref{Lin_El_Co}),

\be
\label{Met_Ba_Co}
\ymn = a^2(\eta) \left( \begin{array}{ll} 
			1   & 0 \\
			0  & -\delta_{ij} 
		\end{array}
       \right) \quad .
\ee

The metric perturbations $\dgmn$ of a homogeneous and isotropic background
(\ref{Met_Ba_Co}) can actually be classified into scalar, vector and
tensor, each category belonging to different ``multiplets'' of the group
of spatial coordinate transformations on a fixed time slice (for a review,
see \cite{RevPaper}). Scalar perturbations transform between each other
when the coordinate transformation can be expressed in terms of a
3-scalars; vector perturbations mix with other vector perturbations when
the transformation is due to a traceless 3-vector; and tensor
perturbations (also known as gravitational waves) are unchanged since
there is no way by which a genuine traceless, divergenceless 3-tensor can
produce a coordinate transformation. We will concern ourselves mostly with
scalar and tensor perturbations, since in most applications vector
perturbations decay rapidly.  As would be expected, each perturbative
``mode'' couple to different sectors of matter, in such a way that there
is no correlation between the 3 categories, and we can analyze them
separately\footnote{This is valid not only in linear theory, but even to
second order, as we will see later}. 

The {\bf scalar} perturbations can be constructed in terms of 4 scalar
quantities, $\phi$, $\psi$, $B$ and $E$: 

\be
\label{Met_Pe_Sc}
\dgmn^S = a^2(\eta) \left( \begin{array}{ll} 
			2\phi   & -B_{,i} \\
			-B_{,j}  & 2[\psi \delta_{ij} - E_{,ij}] 
		\end{array}
       \right) \quad .
\ee

{\bf Vector} perturbations can be written in terms of two transverse (or
solenoidal) 3-vectors
$S_i$ and $F_i$ (if they were not transverse we could extract scalars
- the divergences $S^i_{,i}$ and $F^i_{,i}$ - 
from them and construct new, transverse vectors)

\be
\label{Met_Pe_Ve}
\dgmn^V = - a^2(\eta) \left( \begin{array}{ll} 
			0   & -S_i \\
			-S_j  & F_{i,j}+F_{j,i} 
		\end{array}
       \right) \quad .
\ee

Finally, {\bf tensor} perturbations can be cast in terms of a traceless, transverse
3-tensor $h_{ij}$. Again, if the constraints $h^i_i=0$ or $h^i_{j,i}=0$ were not
satisfied, we could extract a scalar or a vector from $\hij$ and then construct
a new tensor that would be traceless and transverse. The metric perturbations
for gravitational waves are thus
\be
\label{Met_Pe_Te}
\dgmn^T = - a^2(\eta) \left( \begin{array}{ll} 
			0   & 0 \\
			0  & \hij 
		\end{array}
       \right) \quad .
\ee

The accounting for degrees of freedom is the following: 

\begin{quote}
4 d.o.f. for scalar modes (0 constraints);

4 d.o.f. for vector modes (2 constraints);

2 d.o.f. for tensor modes (4 constraints),
\end{quote}
\noindent giving a total of 10, which is the number of
independent components of the symmetric tensor $\dgmn$.

The splitting above, into scalar, vector and tensor modes, is specific to
the FRW background, and would not be valid if the 3-space was not
homogeneous and isotropic. In the case when the background is a
Schwarzschild geometry, which is symmetric on two of the variables (the
polar and azimuthal angles) but not on the radius, perturbations can be
separated into similar groups that transform according to rotations in
$S^2$.

\section{Diffeomorphism transformations}

General Relativity is a diffeomorphism-invariant theory, i.e., its action
and physical observables are symmetric under transformations that take any
one coordinate frame continuously into another. A general infinitesimal
coordinate transformation takes a system of coordinates $ \{ x^\mu \} $ of
a given manifold ${\cal M}$ into a different coordinate system $\{ \tx^\mu
\} $ in such a way that the structure of the manifold remains the same. 
This mathematical definition means that physically meaningful observables
such as the topology, the existence of singularities or the amplitude of
the quadrupole anisotropy in the CMBR are notions which are independent of
the coordinate frame we happen to choose.

In infinitesimal form, we can write a general gauge transformation as

\be
\label{Gen_Tr}
x^{\mu}_P \longrightarrow \tx^{\mu}_P = x^{\mu}_P + \veps \xi^\mu_P \quad ,
\ee
where $\veps$ is a small parameter (later $\veps$ will be identified with
the $\eps$ that measures the relative magnitude of perturbations)
and $\xi(x)$ is a general 4-vector\footnote{Whenever possible we will
drop tensorial indices, so here for example $\xi$ and $x$ mean clearly the
4-vectors $\xi^\mu$ and $x^\mu$. A generic tensor 
$q_{\mu \nu \cdots}^{\alpha \beta \cdots}$ will be denoted only
by $q$.}.
Here we present a simplified version of the treatment given in Chapter
5. For a fuller discussion of the issue of perturbations
of homogeneous spacetimes, see \cite{Sachs} and \cite{StewartWalker}

Now the question is, how do tensors transform under a
gauge transformation? In fact tensors by definition 
transform according to the law

\begin{equation}
\label{Tra_q}
q^{\mu \cdots }_{\nu \cdots } (x_P) \rightarrow 
{\tq}^{\mu \cdots}_{\nu \cdots } (\tx_P)=
{\partial \tilde x^\mu \overwithdelims() \partial x^\alpha }_P \cdots
{\partial x^\beta \overwithdelims() \partial \tilde x^\nu}_P \cdots
q^{\alpha \cdots}_{\beta \cdots}     (x_P)  \quad .
\end{equation}
Notice that on the l.h.s. of this equation the components $\tq$ are
evaluated
at the coordinate value $\tx^\mu_P = x^\mu_P + \veps \xi^\mu_P$.
Substituting this expression for $\tx$ into Eq. (\ref{Tra_q}) and expanding
both
the left and the right-hand-sides yields the definition of the Lie 
derivative

\be
\label{Lie_q}
\tq (x_P) = q(x_P)  - \veps \Lie q (x_P) + {\cal O}(\veps^2) \quad .
\ee

The Lie derivative can be interpreted as being the differential operator that
transforms tensors without changing the coordinate point at which the tensor
is evaluated\footnote{This is known as the ``passive'' interpretation. The
equivalent ``active'' picture is more geometric and prescinds of any
mention to coordinate frames (see, e.g., \cite{Wald,Stewart}). We will
discuss in
more detail the passive interpretation when we speak of Finite
Diffeomorphisms in Chapter 5.}.  We speak then of ``Lie dragging'' tensors
over real points whose coordinate values have been fixed, along a line with
tangent vector $\xi$. The difference between tensor components before and
after the dragging defines the Lie derivative. 

The algebraic definition of the Lie derivative is quite straightforward:  it
acts on 4-scalars as

\be
\label{Lie_Sc}
\Lie S  = S_{,\alpha} \xi^\alpha \quad ,
\ee
on vector fields as

\be
\label{Lie_Ve}
\left( \Lie v \right)^\mu = v^\mu_{,\alpha} \xi^\alpha - v^\alpha \xi^\mu_{, \alpha}
\quad ,
\ee
on 1-forms (covariant vectors) as

\be
\label{Lie_1f}
\left( \Lie u \right)_\mu = u_{\mu,\alpha} \xi^\alpha + u_\alpha \xi^\alpha_{, \mu}
\quad ,
\ee
and so forth for higher-ranking tensors. The case of the metric tensor $\gmn$ 
is particularly important:

\be
\label{Lie_Me}
\tgmn = \gmn - \veps \left( \Lie g \right)_{\mu\nu} 
	= \gmn - \veps g_{\mu\nu,\alpha} \xi^\alpha
	- \veps g_{\alpha\nu} \xi^{\alpha}_{,\mu} 
	- \veps g_{\mu\alpha} \xi^{\alpha}_{,\nu} \quad .
\ee
A coordinate transformation that leaves the metric unchanged is called an
isometry, and vector fields $\zeta$ (called Killing vectors) that generate
isometries obey the Killing equations\footnotesep=0.15in\footnote{It is
useful to note that
in the Lie derivatives all connections cancel, making the normal and the
covariant derivatives equivalent in those expressions.}\footnotesep=0.1in.

\be
\label{Kil_Eq}
g_{\mu\nu,\alpha} \zeta^\alpha
	- g_{\alpha\nu} \zeta^{\alpha}_{,\mu} 
	- g_{\mu\alpha} \zeta^{\alpha}_{,\nu} = - \zeta_{\mu ; \nu} - \zeta_{\nu ; \mu}
 	= 0 \quad .
\ee

We want to argue now that, to our purposes, the small parameter $\veps$ is
the same as the $\eps$ introduced before as the measure of the strength of
the perturbations in FRW spacetimes. As we discussed, we will be concerned
only with coordinate frames which describe the Universe as approximately
homogeneous and isotropic. This means that the background spacetime is
constrained to be described by the  metric (\ref{Lin_El_Co})
and the homogeneous scalar field $\vv_0(\eta)$, 
up to a trivial time reparametrization.
We can still allow, though, for coordinate transformations that change the
form of the matter and metric perturbations $\dv$ and $\dgmn$.
Equation (\ref{Lie_Me}) reads then

\beq
\nonumber
\tgmn 	&=& \tymn + \eps \tdgmn = \ymn + \eps \dgmn
	- \veps \left[ \Lie (\gamma + \eps \delta g) \right]_{\mu\nu} \\
	&=& \ymn +\eps \dgmn - \veps \gamma_{\mu\nu,\alpha} \xi^\alpha
	- \veps \gamma_{\alpha\nu} \xi^{\alpha}_{,\mu} 
	- \veps \gamma_{\mu\alpha} \xi^{\alpha}_{,\nu}  + 
	{\cal O}(\eps \veps) \quad ,
\label{Lie_Me_2}
\eeq
which after subtracting the background yields

\be
\eps \tdgmn (x) =  \eps \dgmn (x) - \veps \gamma_{\mu\nu,\alpha} \xi^\alpha(x)
	- \veps \gamma_{\alpha\nu} \xi^{\alpha}_{,\mu} (x) 
	- \veps \gamma_{\mu\alpha} \xi^{\alpha}_{,\nu} (x) + 
	{\cal O}(\eps \veps) \quad .
\label{Lie_Dg}
\ee
The matter perturbations change similarly, and it is easy to see that

\be 
\label{Lie_Dv} 
\eps \delta \tilde\vv (x) = \eps \dv - \veps \vv_{,\alpha}
\xi^\alpha (x) + {\cal O}(\eps \veps) \quad .  
\ee 
From the equations above we can conclude that if
the magnitude of perturbations is to be preserved under gauge
transformations, then $\veps \leq \eps$. In case of equality, the gauge
transformation can potentially redefine the perturbations; if $\veps \ll
\eps$ the gauge transformations do not alter significantly the perturbations,
and we can ignore their effect.  We will use the saturated bound $\veps =
\eps$ and drop their distinction altogether, since we are interested in gauge
transformations that are large enough that they actually affect the
perturbations.  Of course this upper bound constrains the class of gauge
transformations that are allowed, but the price of not going along these lines
is giving up a well-defined notion of background. In order not to make
notation too heavy, we drop any reference to $\eps$, and without ambiguity
assume it implicit in the definitions of $\xi$ and the perturbations.

If we separate the metric perturbations into the three categories
described in the end of the last section (scalar, vector and tensor),
gauge transformations do not destroy this classification. In a FRW
background spacetime any 4-vector $\xi (\vec{x},\eta)$ can be decomposed
in a 3+1 fashion into two 3-scalars and one transverse 3-vector,

\be
\label{Dec_Xi}
\xi^\mu = \left[ \xi^0 , \xi^{,i}_s + \xi^i_{T} \right] \quad,
\ee
where $\xi^0(\vec{x},\eta)$ and $\xi_s(\vec{x},\eta)$ are 3-scalars and 
$\xi^i_T (\vec{x},\eta)$ is a 3-vector such that $\xi^i_{T \, ,i} = 0$ .

Using the general coordinate transformations above, the metric
perturbations $\dgmn^S + \dgmn^V + \dgmn^T$ given in
(\ref{Met_Pe_Sc})-(\ref{Met_Pe_Te}) and the background metric
(\ref{Met_Ba_Co}) and substituting into Eq. (\ref{Lie_Dg}) we find that
metric perturbations change in the following way (a prime denotes
derivative with respect to conformal time, $a' = da/d\eta$): for scalar
perturbations,

\beq
\label{Tra_Sc_phi}
\phi \rightarrow \tilde{\phi} &=& \phi - \frac{a'}{a}\xi^0 - 
	{\xi^0}'  \quad , \\
\label{Tra_Sc_psi}
\psi \rightarrow \tilde{\psi} &=& \psi + \frac{a'}{a}\xi^0  \quad , \\
\label{Tra_Sc_B}
B \rightarrow \tilde{B} &=& B + \xi^0 - {\xi_s}' \quad , \\
\label{Tra_Sc_E}
E \rightarrow \tilde{E} &=& E -\xi_s \quad ; \\
\eeq
for the vector perturbations,

\beq
\label{Tra_Ve_S}
S^i \rightarrow \tilde{S}^i &=& S^i - {\xi^i_T}' \quad , \\
\label{Tra_Ve_F}
F^i \rightarrow \tilde{F}^i &=& F^i - \xi^i_T \quad ; \\
\eeq
finally, the tensor perturbations (gravity waves), as we noted, do not
transform,

\be
\label{Tra_Te}
h_{ij} \rightarrow \tilde{h}_{ij} = h_{ij} \quad . 
\ee

On the other hand, matter perturbations are transformed according to
their character: scalar field perturbations are simplest,
\be
\label{Tra_Dv}
\dv \rightarrow \delta \tilde{\varphi} = \dv - \vv_0' \xi^0 \quad .
\ee
Background 4-vector fields in homogeneous and isotropic 
spaces are such that their spatial
components vanish (a non-vanishing component $v^i_{(0)}$ would define
a preferred orientation and break isotropy), so 
$v^\mu_{(0)} = [ v^0_{(0)} (\eta) , 0 ]$.
Perturbations of vector fields in 3+1 can be
decomposed into a transverse
3-vector field plus two scalar fields, 
$\delta v^\mu = [ \delta v^0, \delta v^i_T + \delta v^{,i} ]$
and their components transform as

\beq
\label{Tra_Ve_1}
\delta v^0 &\rightarrow& \delta \tilde{v}^0 = \delta v^0
	- {v_{(0)}^0}' \xi^0 + v^0_{(0)} {\xi^0}' \quad , \\
\label{Tra_Ve_2}
\delta v &\rightarrow& \delta \tilde{v} = \delta v + 
	v^0_{(0)} {\xi^0}' \quad , \quad {\rm and} \\
\label{Tra_Ve_3}
\delta v^i_T &\rightarrow& \delta \tilde{v}^i_T = \delta v^i_T +
	v^0_{(0)} {\xi^i_T}' \quad .
\eeq
This separation can be carried out to tensors of arbitrary rank.

Finally, let us comment on the problem of the ``choice of gauge".
The metric tensor has 10 components, but due to the diffeomorphism
symmetry, only 6 are truly dynamical variables. In order to ``fix" 
a gauge we must impose constraints on the metric variables. From the 
definition of the diffeomorphism generator, Eq. (\ref{Dec_Xi}),
we draw that there are 2 constraints that can be imposed on the
4 scalar modes, and 2 that can be imposed on the 4 vector modes
(the tensor modes are already properly constrained by the TT
conditions). There are then 6 dynamical variables in the metric
tensor: 2 scalar, 2 vector and 2 tensor modes.

We must stress that the constraints on the metric perturbations should
completely remove the gauge freedom. Some gauge choices fail to do so (the
synchronous gauge defined by the
constraints $\delta g_{0\mu} = 0$ is a typical example), and the price to
pay for this shortcoming is having to deal with gauge modes that grow in
time and obscure the calculation. We will work later with a gauge
choice (longitudinal gauge) that eliminates completely the freedom under
coordinate transformations.

\section{Gauge invariant variables}

The transformation laws for scalar, vectors and tensor
perturbations show that they have distinct expressions
in distinct reference frames. Now that we know {\it how}
the gauge symmetry of gravity influences perturbations, we can try
to find physical observables that can describe those perturbations in an
invariant form. A quantity is said to be ``gauge invariant" if it
is independent of $\xi$ when the reference frame is transformed by
$x \rightarrow \tx = x + \xi$. Geometric quantities such as tensors
(scalars included) are {\it co}variant, not {\it in}variant, and
are not included in this category.

However, Bardeen\cite{Bardeen} was able to construct non-geometric
quantities defined with respect to a particular coordinate frame,
which take the same values in all nearby frames. An example of a
non-geometric quantity would be the sum of a scalar with the $0-0$
component of a mixed tensor of rank $(1,1)$.

In this monograph we will follow a slightly distinct but equivalent
construction than the one carried out in the standard references
\cite{RevPaper,Bardeen,Stewart90}. While this treatment is
less straightforward, it will prove much more useful when
we generalize gauge transformations in Chapter 5.

First, we note that $\xi$ is completely general, and
that it has the same order of magnitude as the
perturbations (the factor $\eps$ that we dropped earlier).
We can choose its components $\xi^0$, $\xi_s$ and $\xi^i_t$ to be
arbitrary functions in a particular frame. In particular we could
choose them to be linear combinations of the components of the metric
perturbations, and in fact that is what we should do if we
would like to go to a frame where, e.g.,  
$\tilde{E}=0$ : by Eq. (2.39) we easily see that we would have to
make a gauge transformation with $\xi_s = E$.

Call $X^\mu[\delta g]$ the special kind of ``vectors"
defined in terms of the perturbations such that, under a gauge
transformation,

\be
\label{Tra_X}
X^\mu [\delta g] \rightarrow \tilde{X}^\mu [\delta \tilde{g} ] =
	X^\mu [\delta g] + \xi^\mu \quad .
\ee
In other words, under a coordinate transformation generated by
$\xi$, the vectors $X$ transforms like the coordinates. What we
mean by (\ref{Tra_X}) is that $X^\mu$ are linear functional of the
components of the tensor $\dgmn$ which, under a
gauge transformation, change in such a way that $\tilde{X}^\mu$,
written in terms of $\tdgmn$, is related to $X^\mu$ by (\ref{Tra_X}).

Suppose now that we found at least
one vector $X$ that conforms to our specifications. 
Consider then the following
quantities $Q^{\mu \cdots}_{\nu \cdots}$ defined in terms of a generic
tensor $q^{\mu \cdots}_{\nu \cdots}$ with the help of $X$:

\be
Q \equiv q + \LieX q \quad .
\label{Def_Q}
\ee
If q has a perturbative expansion 
$q = q_0(\eta) + \delta q(\eta,\vec{x})$, then naturally
the background is the same, $Q_0 = q_0$, but not the perturbations:

\be
\label{Def_dQ}
\delta Q = \delta q + \LieX q_0 \quad .
\ee

The transformation law for $\delta Q$ can be calculated as follows:

\beq
\nonumber
\delta \tQ &=& \delta \tq + {\cal L}_{\tilde{X}} q_0 =
	\delta q - \Lie q_0 + {\cal L}_{X + \xi} q_0 \\
\label{Tra_dQ}
	&=& \delta q + \LieX q_0 = \delta Q \quad . 
\eeq
Therefore, the quantities $\delta Q = \delta \tilde{Q} $ are
gauge invariant variables! It is possible then to
define gauge invariant quantities related to the perturbations
of any background variables, provided that we can find a vector
$X$ that satisfies the transformation law (\ref{Tra_X}).

With the help of the transformation laws for the metric
perturbations it is easy to extract the components that
transform in the desired way. Since there is more than
one solution for each component $X^0$, $X_s$ and $X^i_t$,
we label them as follows:

\beq
\label{Def_X}
X^0_{(1)} &=& B-E' \quad , \\
\nonumber
X^0_{(2)} &=& \frac{a}{a'} \psi \quad , \\
\nonumber
X^0_{(3)} &=& - \int d \eta ( \phi + \frac{a'}{a} X^0_{(1,2)} )\quad , \\
\nonumber
X_{S(1)} &=& - E \quad , \\
\nonumber
X_{S(2)} &=& - \int d \eta ( B - X^0 ) \quad , \\
\nonumber
X^i_{T(1)} &=& -F^i \quad , \\
\nonumber
X^i_{T(2)} &=& - \int d \eta S^i \quad .
\eeq
In the definition of $X_{(3)}^0$ the last term can be any linear
combination of 
$X_{(1)}^0$ and
$X_{(2)}^0$ with norm 1 (see below).

For each one component we can actually use various definitions,
as long as they sum up to unity. For example, we could define

\be
\nonumber
X^0_{(\alpha)} = \frac{1}{4} X^0_{(1)} + \frac{3}{4} X^0_{(2)} \quad ,
\ee

or
\be
\nonumber
X_{S(\beta)} = 0.78 X_{S(1)} + 0.22 X_{S(2)} \quad .
\ee
There is an infinite number of definitions that we could choose,
corresponding to an infinite number of ways of defining gauge
invariant objects.
We would prefer the simplest possible definition, that in addition
is regular if we choose to switch off the expansion of the Universe and
is not nonlocal (does not include integrals).
We will also neglect the vector modes, since they do not
play a significant role neither in structure formation nor in
back reaction, which is the object of this work.

We are left with only one simple choice,

\be
\label{Def_X0}
X^0 = B-E' \quad ,
\ee
and

\be
\label{Def_X_s}
X_s \equiv X = - E 
\ee
and, of course, $X^i_T = 0$ since we are not going to allow for
vector modes in any reference frame.
With these definitions we can compute the explicit expressions for
gauge invariant metric components. Substituting in Eq. (\ref{Def_dQ})
and separating the components we have

\beq
\label{Def_Ph}
\phi &\stackrel{X}\rightarrow& \Phi = \phi + \frac{1}{a} [ a(B-E')]' \quad , \\
\label{Def_Ps}
\psi &\stackrel{X}\rightarrow& \Psi = \phi - \frac{a'}{a} (B-E') \quad , \\
\nonumber
B &\stackrel{X}\rightarrow& 0 \quad {\rm and} \\
\nonumber
E &\stackrel{X}\rightarrow& 0 \quad .
\eeq

The gauge invariant variable related to the scalar field perturbation is
also defined in Eq. (\ref{Def_dQ}),

\be
\label{Def_dV_gi}
\dv \quad \stackrel{X}\rightarrow \quad 
	\delta \vv_{gi} = \dv + {\vv_0}' (B-E') \quad . 
\ee

While the construction detailed above is undeniably more involved
than the one usually encountered in the literature, in the end of the
day it proves simpler. The classical construction consists of looking for
gauge invariant combinations of components of a given tensor with the metric
perturbations. In our construction, once the $X$ are properly identified 
and classified (a trivial procedure), it is a matter of simple
algebra to calculate the gauge invariant variables corresponding
to {\it any} tensor. It will also prove much simpler to deal with
once we go beyond the first order in perturbation theory in Chapter 
5.

So far we did not impose any constraints on the metric, therefore the
``gauge'' is completely general. Since we have excluded the vector
modes from the analyses (2 constraints), we still have to impose
2 conditions on the scalar functions. We could choose, for example,
the synchronous gauge $\phi =0$ and $ B = 0$,
but these 2 constraints are not sufficient to completely eliminate
the gauge freedom (see \cite{RevPaper}). We will use instead the
constraints that make definitions (\ref{Def_X0}) and (\ref{Def_X_s})
easiest to compute, i.e., $B=0$ and $E=0$, known as longitudinal gauge
(since the metric perturbations are diagonal). There are
three good reasons for this choice of gauge: one is that, as opposed to
synchronous gauge, longitudinal gauge does not carry any ``unphysical
modes'', therefore we do not have to deal with ``ghosts'' in the
calculation; two, $\phi$ can be connected to the gravitational potential
in the Newtonian limit, and it is desirable that we can
easily take that limit if we wish to; finally, in the gauge where $B=E=0$,
the
dynamical variables $\phi$ and $\psi$ actually coincide with the
gauge invariant variables $\Phi$ and $\Psi$, respectively.

In fact,
with the choice of the vector $X$ as in (\ref{Def_X0}) and
(\ref{Def_X_s}), 
{\it all} gauge-invariant variables are coincident with their
expressions in longitudinal gauge. This suggests an elegant way
of writing the equations for cosmological perturbations in a
completely gauge invariant manner: we derive the equations of motion
in longitudinal gauge, then substitute $\phi$, $\psi$ and $\dv$ for their
gauge invariant counterparts $\Phi$, $\Psi$ and $\dv_{gi}$.

Since we are going to use the longitudinal gauge for the remainder
of this paper whenever we speak of scalar perturbations, we can drop
the 
notational distinction between the gauge invariant variables
and their corresponding longitudinal gauge variables.
Thus from now on we restrict ourselves to the scalar metric 
perturbations given by 

\be
\label{Met_Lo}
\dgmn^S = a^2(\eta) 
	\left( 
		\begin{array}{ll} 
			2\phi   & 0 \\
			0       & 2 \psi \delta_{ij} 
		\end{array}
       \right) \quad .
\ee
It goes without saying that gravity waves do no interfere at all with
the discussion above, since they are gauge invariant to start with.

\section{Equations of motion for the perturbations}

The propagation of linearized perturbations in a background
spacetime is given by the Einstein field equations,
expanded in a Taylor series around the background metric $\ymn$ and
matter fields $\vv_0$, truncated at the second order. We will
use the following notation:

\beq
\nonumber
0 = \Gmn [g] - \Tmn[g,\vv] 
	&\equiv& \Pmn [g,\vv] \\
\label{EFE_1}
	&=& \left. \Pmn^{(0)} [\gamma, \vv_0] + 
	\delta \Pmn [\delta g, \dv] 
	\right|_{\gamma, \vv_0} \quad .
\eeq
The tensor $\Pmn^{(0)}$ is zero if the background equations 
$\Gmn^{(0)} = \Tmn^{(0)}$ are satisfied, which of course must be solved
prior to the linear equations $\delta \Gmn = \delta \Tmn$. It is
straightforward
but lengthy to derive the linearized equations of motion, and it
is widely available in the literature\cite{RevPaper}. We simply state here
the results for the cases of gravity waves (tensor perturbations) and
scalar perturbations. We neglect the vector modes, since they do not
appear in most models of structure formation.

Einstein's equations linearized around a FRW background spacetime
imply the following equations of motion for the tensor modes:

\be
\label{EOM_gw}
h_{ij}'' + 2 \H h_{ij}' - \nabla^2 h_{ij} = 0
\ee
where $\H = a'/a$ and a prime, as before, denotes a derivative with
respect to conformal time. Notice that the right-hand-side of
Eq. (\ref{EOM_gw}) does not contain any matter source, which is
the reason why gravity waves do not couple to matter.

It is useful to make the substitution

\be
\label{Sub_mu}
h_{ij} (\vec{x},\eta) = \frac{1}{a(\eta)} \int \frac{d^3k}{(2\pi)^{3/2}} 
	\left[ 
	\mu_k(\eta) \eps_{ij}(\vec{k}) 
	e^{i \vec{k} \cdot \vec{x} }  + c.c. 
	\right] \quad ,
\ee
in which case the equations of motion for $\mu(\eta)$ read
simply

\be
\label{EOM_mu}
\mu_k'' + \left( k^2 - \frac{a''}{a} \right) \mu_k = 0 \quad .
\ee

Due to the constraints on the tensor modes, $h^i_i=h^i_{j,i}=0$, the
polarization tensor $\eps_{ij}$ in (\ref{Sub_mu}) is a symmetric and
transverse-traceless (TT) tensor, $\eps^i_j = \eps_{ij} k^i = 0$. Consider
a gravity wave moving in the $z$ direction, that is, $\vec{k}=(0,0,k_z)$.
The constraints imply that the polarization
tensor for this plane wave is given in terms of the two variables
$\eps_{xx}=-\eps_{yy}$ and $\eps_{xy}=\eps_{yx}$. For each plane wave
there are then 2 degrees of freedom, transverse to the direction of
propagation of the wave.

When the background expands in time as $a \propto \eta^p$, the
equation can be solved in terms of (spherical) Bessel functions:

\be
\label{Sol_mu}
\mu(k,\eta) = \eta^{1/2} \left[ 
	C_1(k) J_{p-1/2}(k\eta) +
	C_2(k) Y_{p-1/2}(k\eta) \right] \quad.
\ee
These solutions can be easily expressed in cosmic time by using $a(\eta)
d\eta = \eta^p d\eta = dt$, therefore when $p\neq -1$ we have $\eta =
t^{1/(p+1)}$ and $a(t)=t^{p/(p+1)}$, while $\eta = e^{H_0t}$ 
and $a(t)=e^{H_0 t}$ if $p=-1$.

Scalar perturbations of the metric are much more complicated since they
couple to matter perturbations, which appear as a source term in the RHS
of the linearized Einstein equations. They are also much more interesting,
since they gave rise to the large-scale structure that we observe. The
metric for scalar perturbations in longitudinal gauge, given in Eq.
(\ref{Met_Lo}), can be simplified if we take into account that the
non-diagonal space-space ($i \neq j$) components of the EMT are
identically zero\footnote{This is also true of scalar hydrodynamical
perturbations when there is no anisotropic stress.}. This implies that
$\psi=\phi$, since the $i\neq j$ components of the Einstein tensor are
proportional to $\nabla_i \nabla_j (\phi - \psi)$.  With this
identification, we drop $\psi$ and the $0-0$, $0-i$ and $i=j$ equations
for the perturbations are,
respectively\footnotesep=0.15in\footnote{Conversion to cosmic
time $dt=ad\eta$ is easily done using the identities $f'=a\dot{f}$ and
$f''=a^2(\ddot{f} + \dot{a}\dot{f})$.}\footnotesep=0.1in,

\beq
\label{EOM_ph_00}
2 \nabla^2 \phi - 6 {\cal H} \phi' - 2 ( {\cal H}' + 2 {\H}^2 ) \phi 
	&=& \vv_0 ' \dv ' + a^2 V_{, \vv} \dv \quad , \\
\label{EOM_ph_0i}
2 \phi ' + 2 \H \phi 
	&=& \vv_0 ' \dv \quad , \\
\label{EOM_ph_ij}
2 \phi '' + 6 \H \phi' + 2 ( \H' + 2 \H^2) \phi 
	&=& \vv_0 ' \dv ' - a^2 V_{, \vv} \dv \quad ,
\eeq
where $\H = a'/a$.

Only two of the three equations above are
linearly independent, which can be verifyed by using the
background identity

\be
\label{EFE_con}
2(\H^2 - \H') = {\vv_0 '}^2 
\ee
and the equations of motion for the scalar field,

\be
\label{EOM_v0}
\vv_0 '' + 2 \H \vv_0' + a^2 V_{,\vv} = 0 \quad .
\ee

The rhs of Eq. (\ref{EOM_ph_00}) is in fact just twice the perturbed
energy density of the scalar field, $\delta \rho_s$. If we switch off the
expansion of the Universe ($\H,\H' \rightarrow 0$) this equation reduces to
Poisson's equation $\nabla^2 \phi = \delta \rho$, where the Newtonian
gravitational potential has been replaced by the the relativistic metric
perturbation $\phi$.

In addition to the Einstein equations,
the perturbed scalar field obeys the linearized Klein-Gordon equations in
a
curved FRW background. These can be obtained either
directly through the continuity equations $\nabla^\mu \Tmn = 0$, or
as the integrability condition on the system 
(\ref{EOM_ph_00})-(\ref{EOM_ph_ij}), since the
first equation is just a constraint and 
the second a true dynamical equation (second order in time derivatives).
In either case the equation for the evolution of $\dv$ is

\be
\label{EOM_dv}
\dv '' + 2 \H \dv' - 4 \vv_0 ' \phi' + 2 a^2 \Vl \phi + a^2 \Vll \dv - 
	\nabla^2 \dv = 0 \quad .
\ee

We can simplify the system even more by using the constraint
(\ref{EOM_ph_0i}) to express $\dv$ in terms of $\phi$. After
some algebra and the use of the background equations 
(\ref{EOM_v0}) and (\ref{EFE_con}) we get the surprisingly
simple wave equation

\be
\label{EOM_u}
u'' - \nabla^2 u - M_0^2 (\eta) u = 0 \quad ,
\ee
where we introduced the variable

\be
\label{Def_u}
u ( \eta, \vec{x}) = \frac{a}{\vv_0'} \phi (\eta, \vec{x})
\ee
and the auxiliary time-dependent ``mass''

\be
\label{Def_M0}
M_0^2 (\eta) = \frac{ \left[\H /a \vv_0' \right]''}{\H/a \vv_0'} \quad .
\ee

The main reason we use conformal time in this section is 
that Eq. (\ref{EOM_u}) looks very simple. In fact, in this 
form the problem of evolution
of perturbations turns out to be just scattering a wave off
a time-dependent potential. The technique to solve this problem is
self-evident: we will decompose the wave into a linear combination
of plane waves 

\be
\label{Def_u_k}
u(\eta, \vec{x}) = \int \frac{d^3 k}{(2\pi)^{3/2}}  
		u_{\vec{k}} (\eta) e^{i\vec{k} \vec{x}} \quad ,
\ee
where $\vec{x}=\vec{r}/a$ are comoving coordinates, and $\vec{k}$ is 
comoving momentum (the physical momenta $\vec{k}_{ph}$ are related
to comoving momenta by $\vec{k} = a \vec{k}_{ph}$).
We treat each $k$-mode separately,
since there is no coupling between modes to linear order.
In momentum space, the equation of motion for $u_k(\eta)$ is

\be
\label{EOM_u_k}
u_k'' +k^2 u_k - M_0^2 u_k = 0 \quad .
\ee

This equation can be solved in two asymptotic situations:  the
long wavelength (or low frequency, LF) limit, $k^2 \ll M_0^2$ and the
short wavelength (high frequency, HF) limit $k^2 \gg M_0^2$. The inverse
of the ``mass'' $M_0^{-1}$ is in fact a length scale proportional to the
radius of curvature of the Universe. For instance, when the dynamics of
the background is dominated by a scalar field potential and the Universe
is inflating then $M_0^{-1} \approx (a H)^{-1}$, that is, this ``mass''
scale is only the horizon scale\footnote{When the scale factor is
increasing exponentially the horizon size in physical coordinates,
$R_H=H^{-1}$, is approximately fixed. In comoving coordinates, though, a
fixed physical scale is exponentially smaller, $x_H=R_H/a \sim R_H
e^{-Ht}$.}. In this case, we speak of modes bigger (LF) and smaller (HF) 
than the horizon.

The HF limit is simplest, since we can neglect the time-dependent
mass $M_0$. Physically, it means that 
perturbations of very small wavelength do not feel the curvature of
spacetime (although they still feel the expansion, since their
physical wavelength is being stretched), and the solutions of
(\ref{EOM_u_k}) are 

\be
\label{u_HF}
u_k^{{\tiny (HF)} \pm} (\eta) = \hat{A}_k^{{\tiny (HF)} \pm} e^{\pm i k \eta}
\quad
.
\ee

If the expansion of the Universe is stretching the wavelengths faster
than its curvature radius is increasing, some modes will eventually
become larger than the horizon. During an inflationary period this
process becomes explosive, with modes crossing the horizon at an
exponential rate (remember that the horizon scale in comoving coordinates
is $R_H/a$).

For LF cosmological perturbations with $k^2 \ll M_0^2$, the solution
to Eq. (\ref{EOM_u_k}) is, after some algebra,

\be
\label{u_LF}
u_k^{\tiny (LF)} (\eta) = \hat{A}_k^{\tiny (LF)} \frac{1}{\vv_0'} 
	\left[ 
		\frac{1}{a} \int^\eta d \eta a^2(\eta) 
	\right]' \quad .
\ee
Notice that there is an additional integration constant implicit
in the integral in (\ref{u_LF}), so that both solutions 
$u_k^{\tiny (HF)}$ and $u_k^{\tiny (LF)}$ need to be
complemented by 2 ``boundary conditions" - typically the
metric perturbation and its time derivative on a given time slice. 
The general solution to second-order wave equations always involve
two linearly independent solutions. 
There is one dominant and one decaying solution for each mode, 
and we usually ignore the decaying one, except when making the transition
between two phases with different expansion regimes or when
connecting a HF mode that crosses the horizon to the LF solution
outside the horizon.

Now it is convenient to write the metric perturbations in terms of 
cosmic time $t$ (for time derivatives, use that $d\eta = dt/a$). 
The sub-horizon modes are

\be
\label{phi_HF}
\phi_k^{\tiny (HF)} = \dot{\vv}_0 
	\left[
		A_k^{+ {\tiny (HF)}} e^{+ik \int dt/a} +
		A_k^{- {\tiny (HF)}} e^{-ik \int dt/a} 
	\right] \quad ,
\ee
where the integration constants in the exponents can
be absorbed in the definitions of $A_k^{{\tiny (HF)} \pm}$ so that
there are really only 2 constants. 
Super-horizon modes behave like

\be
\label{phi_LF}
\phi_k^{\tiny (LF)} = A_k^{{\tiny (LF)} D} 
	\left(
		1 - \frac{H}{a} \int a dt
	\right) \quad ,
\ee
where the index ${\small D}$ will become clear in a minute.

To make the discussion more concrete, let us consider the case when the
background expands in a power-law, $a \sim t^p$ with $p>0$ and $p \neq 1$ (we
exclude the case $p=1$ for simplicity). There are two different regimes that
are possible with this type of scale factor, for which the deceleration
parameter $q \equiv \frac{\ddot{a}a}{\dot{a}^2} $ is bigger or less than zero:
the case $p<1$ gives $q<0$ and is decelerating, while the case $p>1$ is
accelerating, so $q>0$.

We can easily perform the integral in (\ref{phi_HF}), and the solution 
reduces to

\be
\label{phi_HF_p}
A_k^{\pm}e^{
	\pm \frac{i}{1-p} \frac{k}{k_0} 
	\left[ 
		\left(
			\frac{t}{t_0}
		\right)^{1-p} -1 
	\right] 
	} \quad ,
\ee
where $t_0 < t$ and $k_0$ are constants. In the case $p<1$ the
exponent increases with time, hence the perturbations
oscillate with a growing frequency and there is no dominant
solution. When the expansion is accelerating, $p>1$, the frequency
decays and the cosine mode becomes dominant, since the sine mode
becomes small as the argument in the exponential goes to zero.

LF modes in these backgrounds have the form

\be
\label{phi_LF_p}
A_k^D \left( 1 - \frac{p}{p+1} \right) + A_k^S \frac{1}{t^{1+p}} \quad .
\ee
In both cases, $p>1$ and $p<1$, the constant term is dominant, while the second
term fastly decays. This result is very important, and holds even when the
scale factor grows exponentially. We say that perturbations on scales bigger
than the horizon are ``frozen", i.e., the dominant mode in their amplitude is
constant in time.

The detailed description of spacetime in terms of perturbations is encrypted in
the ``spectrum" $\{ A_k \}$, and this description remains valid as long as the
interactions are approximately linear. For example, if we could deduce the
exact form of the spectrum at an early phase in the evolution of the Universe,
when nonlinear structures were still irrelevant, we would be able to determine
all the structure formed thereafter, including the times when nonlinearities of
a given scale became important. For the purposes of this work, knowledge of the
spectrum is crucial: it gives the information about how much amplitude there is
in all modes, and therefore it also carries the key to quantities like the
energy density and pressure carried by each of the modes.

Of course, the precise form of the spectrum for all $k$ cannot be deduced from
first principles without invoking transcendental powers, but we will see in
the next chapter that some physical processes in the early universe are
effective enough to create spectra with very distinct qualitative features. We
turn now to just such a mechanism: the inflationary Universe.


	\chapter{Inflationary Cosmology and Early Universe Physics}

%
%

There are several reasons to support the heterodox notion that the universe suffered
some sort of accelerated expansion fueled by exotic matter during a tiny fraction of
a second after the Big Bang.  The puzzles left unsolved by Big Bang cosmology are, in
order of importance\footnote{Personal taste might interfere with the order}:

\begin{quote}
$\bullet$ the cosmological constant problem; 

$\bullet$ the horizon or causality problem; 

$\bullet$ the problem of the origin of
structure; 

$\bullet$ the flatness (or entropy)  problem; 

$\bullet$ the problem of
primordial magnetic monopoles and domain walls..  
\end{quote}

\noindent As often happens historically, the solution to the least
important problem leads to an idea that elucidates other, more interesting questions.
This solution can still leave out the most interesting problems: 
the cosmological constant remains a mystery in cosmology,
although Inflation can solve all the other questions mentioned above.

As mentioned in the last chapter, inflation's account of the rich
structure of the universe is successful enough that this feature alone
should be sufficient to take the scenario seriously (although it is not
the only satisfactory mechanism on the market.) We discuss briefly the
other compelling reasons that make this theory so attractive.

\section{Puzzles of pre-inflationary cosmology}

\noindent $\bullet$ Horizon Problem 

\noindent 
A horizon at time $t_h$ is, generically, a
2-dimensional surface defined with respect to a given reference point
$P_h$ that separates the universe into 2 regions, a ``white" region
containing $P_h$ and a ``black" region. If a horizon at point $P_h$ exists
at time $t_h$, an observer at that point cannot receive information (light
signals, to be specific) from the black region.  If the information was
traveling from a distant point $P_{ph}$ in the remote past and could not
reach the point $P_h$ up to the time $t_{ph}$, we say that there is a
particle horizon.  If the information started to travel from a point
$P_{eh}$ at the time $t_{eh}$ but would never reach the point $P_h$, even
after an infinite amount of time, we have an event horizon. The black
regions in either cases are defined as the union of all points $P_{ph}$ or
$P_{eh}$, or, as the ``geometrical locus" of points satisfying the
respective causality conditions above.

A typical example is the Schwarzschild black hole spacetime, which possess
a horizon with respect to any point outside the Schwarzschild radius,
since the white external region is causally disconnected from the the
black interior of the hole. This case is unique, though, in that
a Schwarzschild black hole has always been there. Hence both particle and
event horizons can be defined, and they coincide in the ``black hole
horizon". The case of a collapsing star is more realistic, though, and in
that case only an event horizon is formed.

In Cosmology, FRW spacetimes usually have either one of the
two horizons. Consider the problem of propagation of light
in an expanding universe,

\be
\label{Lin_El_ho}
ds^2 = dt^2 - a(t)^2 d\vec{x}^2 = 0 \quad \Longrightarrow \quad 
	\Delta x = \int_{t_i}^{t_f} \frac{dt}{a} \quad ,
\ee
with the scale factor

\be
\label{a_p}
a(t) = t_0 \left( \frac{t}{t_0} \right)^p \quad ,  
\ee
where $t_0$ is the present time (or age of the universe) and
$p>0$ (remember that, for pressureless matter $p=2/3$, and for
radiation $p=1/2$.)

The comoving distance traveled by light signals between
the time interval $(t_i,t_f)$ is then given by

\be
\label{Dis_co}
\Delta x = \frac{1}{1-p} 
	\left[ 
		\left( \frac{t_f}{t_0} \right)^{1-p} -
		\left( \frac{t_i}{t_0} \right)^{1-p} 
	\right] \quad.
\ee
Let us answer first in which circumstances there is a particle horizon,
and why they are important.

If there is a finite maximal distance $\Delta
x_{ph}$ that a photon could have traveled if it was emitted at time
$t_i=0$ and received at any time $t_f$, then there is a particle horizon.
In this case, photons traveling from a spherical shell of radius $\Delta
x_{ph}$ towards the center of the sphere will just make it to the center,
with photons coming from outer shells not being able to get to the center
at the time $t_{ph}$.  From the expression to $\Delta x$ we see that
$t_i=0$ gives a finite radius only when $p<1$. For $p>1$ the expression
diverges and there is no particle horizon. Setting $t_f= t_{ph}$ we have
then

\be
\label{Par_ho}
\Delta x_{ph} (t_{ph}) = \frac{1}{1-p} \left( \frac{t_{ph}}{t_0} \right)^{1-p}
	\quad , \quad p < 1 \quad .
\ee

Particle horizons are important in Cosmology because, if one existed today,
CMB photons received from opposite directions would provide information about
two causally disconnected regions. To see this, compare the particle horizon
at the time of last scattering of the CMB (the time of decoupling,
$t_{dec}$) with the distance traveled by light from the surface to us:

\be
\label{Par_comp}
	\frac{\Delta x_{ph} (t_{dec})}{\Delta x (t_{dec} \rightarrow t_0)}
	\simeq 
	\left( \frac{t_{dec}}{t_0} \right)^{1/3} \simeq 10^{-2} \quad ,
\ee
where we have taken $p=2/3$ since the universe was very nearly matter
dominated since decoupling\footnote{The presence of a small
Cosmological Constant would not change that too much. We know
that, today, it could account for at most some 80\% of the energy
density in the universe, and as we go back in time this contribution
becomes negligible.}. 
We see that the particle horizon at
decoupling was a factor $1/100$ smaller than the particle horizon today.
In the CMB sky map, this fraction translates into a square of
a few degrees (the present particle horizon is the whole
$360^\circ$ sky). A few billion years from now this patch
would be even smaller, reflecting the fact the particle horizon
in the future has increased while the particle horizon at
decoupling remains fixed. That is, as we go into the
future, we gain access to more distant regions of the universe,
regions that did not ``know" about each other at $t_{dec}$.

A measurement of CMB radiation at the South pole and at the North pole would
be, therefore, a comparison between physical properties (such as temperature) 
of two regions that had never come into causal contact until the moment those
photons were observed. The fact that the CMB map shows a uniformity of 1 part
in $10^5$ despite the existence of particle horizons poses a
problem to the usual radiation- and matter-dominated cosmological
models. In order to explain this uniformity in the framework
of these models we have to invoke some acausal mechanism, like
preparing the initial state of the whole universe (or a good part of it)
right at the beginning of the Big Bang. We will see later how
inflation can solve this problem, at least for the patch of
the universe we happen to live in.

Event horizons, on the other hand, are also important,
although they do not lead immediately to problems with classical
cosmology. If 
there is a maximal distance $\Delta
x_{e}$ beyond which photons emitted on a spherical shell at a time $t_{eh}$
can {\it ever} reach the center of the sphere, then there is an event
horizon.  Photons emitted from the outer shells (with radii bigger than
$\Delta x_e $ ) at time $t_{eh}$ will never reach the center, because
the expansion overcomes the propagation of the photon. In other words,
the photon is redshifted to zero energy before it reaches the center
of the sphere. 

In formula (\ref{Dis_co}) we make the emission time
$t_i=t_{eh}$ and the observation time $t_f \rightarrow \infty$, and it
easy to see that for $p<1$ the expression diverges (the expansion
is not fast enough that it can redshift photons to zero.) For $p>1$ the
expansion is fast enough that event horizons are well defined,

\be
\label{Eve_ho}   
\Delta x_{eh} (t_{eh}) = \frac{1}{p-1} \left( \frac{t_0}{t_{eh}} \right)^{p-1}
        \quad , \quad p > 1 \quad .
\ee
In the special case when $a = a_0 e^{H_0 t}$, $H_0$ constant, the
event horizon can be easily calculated: it is, in comoving coordinates, 
$ \frac{e^{-H_0 t}}{H_0 a_0} $, and in physical coordinates simply
$H_0^{-1}$.

\noindent $\bullet$ The Structure Formation problem

\noindent
If the universe was perfectly homogeneous, no stars or galaxies
would have ever formed. We clearly need to introduce inhomogeneities
that can grow and become large-scale structure,
but in a purely classical theory these ``seeds" for structure can only
appear as {\it ad hoc} initial conditions. What explains the
scale-invariant pattern of the initial inhomogeneities and their
tiny initial amplitude? How did they arise?

\noindent $\bullet$ The Flatness problem

\noindent There is a critical energy density $\rho_c$ above which the
total mass of the universe is big enough to stop the present 
expansion and turn it into a contraction (``crunch''). 
If the density is smaller than this $\rho_c$,
the universe will expand forever, becoming colder and emptier.
If the universe has the density parameter $\Omega = \rho/\rho_c > 1$ 
it is said
to be closed; if $\Omega < 1$ it is said to be open; if
$\Omega$ equals precisely one, the universe is flat and it
will continue expanding forever, with an expansion rate that
vanishes asymptotically.

The flat case $\Omega = 1$ is clearly
unstable, just like the Einstein - de Sitter universe: any
deviation from flatness will increase rapidly with time.
In particular, the lifetime of a Planck-scale universe with energy density  
$\rho_{{pl}} \sim M_{pl}^4$ is of the order of a Planck time 
$t_{{pl}}^{-1} \simeq 10^{-44} s$. By lifetime here we mean the
time a closed universe would take to crunch or the time interval
that it takes for an open one to become effectively empty and void. 
Only if $| \Omega -1 | \sim 0$ a typical planckian universe can
survive a longer period of time without collapsing or turning into
vacuum.

The universe is presently very nearly flat, $0.1< \Omega_0 < 2$ , which
implies that in the past ($t \ll t_0$) it was much flatter than it is now,
$|1- \Omega (t)| \ll |1-\Omega(t_0)|$.  In fact, for the density parameter
to be so close to 1 today, one Planck time after the Big Bang the universe
would have to be flat to within one part in $10^{59}$, that is,
$|\Omega(t_{{pl}}) - 1| \leq 10^{-59}$ ! The question of why our universe
was so astonishingly flat soon after the Big Bang, or equivalently, why
it survived for so long, is known as the flatness problem.

\noindent $\bullet$  The Primordial Monopoles problem

\noindent As the temperature dropped in the early universe, symmetries
that existed at high temperatures are broken. For example,
the electroweak phase transition occurred at an energy
scale around $100$ GeV and gave mass to the originally massless
vector bosons of the Glashow-Salam-Weinberg theory.

During a symmetry-breaking phase transition, many strange objects
such as domain walls, cosmic strings and monopoles can be formed.
Most of these objects are stable, being topological defects,
and in the case of monopoles their annihilation rate is
too small and their creation rate too big to ignore them.
Sometimes the production
of monopoles during a phase transition is so copious that they 
may overpopulate
the universe. That is the problem with many grand unified theories,
which have t'Hooft-Polyakov magnetic monopoles
as a byproduct of symmetry-breaking phase transitions. 
The monopole creation rates are too high, and their
mass ($10^{16}$ times that of the proton) too big: if they existed
at all, the universe would have collapsed long ago.
The magnetic monopoles problem is how 
to reconcile grand unified theories
with the fact that the universe lived for so long and is so 
patently empty of monopoles. The same reasoning can be employed
to show that a particle theory with a phase transition that
breaks a $Z_2$ symmetry produces domain walls (regions interpolating
$|+\rangle$ and $|-\rangle$ vacua). There is no reason why electroweak
domain walls, for example, don't exist. However there is no 
evidence of that, on the contrary, their existence would have
catastrophic cosmological consequences and is a strong
indication that there is no discrete symmetry breaking in the
Higgs potential.

\section{The Inflationary Universe}
\label{SecInUn}

It was the last problem, that of topological defects, which precipitated the
discovery\cite{FirstInflation} that a brief period of exponential
expansion can explain not only the
absence of monopoles and domain walls, but also the regularity of the CMB
temperature map over the whole sky and the reason why the density
parameter was so close to the critical value 1.

The rationale is simple: an exponential expansion would reduce any initial
density of monopoles and domain walls (as it would of any preexisting
particles) exponentially. This is also known as the ``no-hair" theorem
of de Sitter space, analogous to the one for black holes: in de Sitter
space, all particles and inhomogeneities within the event horizon $H^{-1}$
will have left that sphere (or crossed the event horizon) by a time
of order $H^{-1}$. Events happening inside the horizon will not
be affected by events outside the horizon.

After this epoch of inflation, the
universe must be again filled with radiation and other sorts of
matter (but not monopoles or walls), hence the only visible effect of the
inflationary epoch would be to do away with the monopoles. This exit from
inflation, called reheating, is crucial to provide a fresh start to a
hot universe devoid of unwarranted creatures, but with a healthy
population of friendly animals like photons, baryons, leptons etc.

This exponential growth of the scale factor will also eliminate the
particle horizon problem.  A period of inflation in the early universe
means that between a time $t_1$ and a time $t_2$ the scale factor grows 
exponentially, $a \propto e^{H_0 t}$. After this period the universe
is reheated, or refilled with radiation, and its evolution is
once again given by $a \propto t^{1/2}$. After a bit of simple algebra
it is easy to obtain from (\ref{Lin_El_ho})
that the article horizon gets an exponential contribution.
If we set $t_2 = t_1 + N H_0^{-1}$ we have that, for any time
$t > t_2$, the particle horizon is

\be
\label{Par_ho_in}
\Delta x_{ph} (t) = 2 \left( \frac{t}{t_2} \right)^{1/2} +
	\frac{e^N}{N} \quad .
\ee
Clearly the exponential is going to dominate for $N$ large
enough, therefore this distance can be made much larger than the radius of
the present observable universe, if inflation lasts for a sufficient
number of ``e-folds" $N$. For the particle horizon at the time of
decoupling to be bigger than the radius of the observed universe at the
present time, $N \sim 50$. If $N \geq 50$ the horizon problem is
solved: the CMB is so nearly isotropic because the distant regions
of the sky we observe now were in fact microscopically close at times
$t < t_i$ in the early universe. Their present cosmological distance is
due to an inflationary phase that took place after $t_i$ and
exponentially stretched all physical distances for a time 
$\geq 50 H_0^{-1}$.

It is clear then that an inflationary phase in the early universe,
with a sufficient number of e-foldings $N \geq 50$, is highly
desirable. Our task now is to construct a realistic model with
these features. Below we describe the simplest and most successful
of those models, the so-called Chaotic Inflation scenario
\cite{LindeSlava,LindeBook}. We do not
discuss Old, New, Double or String inflation in this monograph (for
reviews, see \cite{LindeBook,InflationBook,RevRobert}.)

Consider a scalar field $\vv$ with the potential

\be
\label{Sca_Po}
V(\vv) = \frac{m^2 \Mpl^2}{n} \left( \frac{\vv}{M_{pl}} \right)^n 
       = \frac{m^2}{n} \vv^n 
\ee
where $n>0$, $0< m \ll \Mpl $ and we have also introduced Planck units,
$M_{pl}=1$. The energy density $\rho=T_{00}$ of this field is given by
(\ref{EMT_Sc}),

\be
\label{Ene_Sc_FRW}
\rho^S (\vec{x},t) = \frac{\dot{\vv}^2}{2} + \frac{(\nabla\vv)^2}{2 a^2} 
			+ V(\vv) \quad ,
\ee
and its pressure $p = -T^i_i/3$ is

\be
\label{Pre_Sc_FRW}
p^S (\vec{x},t) = \frac{\dot{\vv}^2}{2} - \frac{(\nabla\vv)^2}{6 a^2}   
                        - V(\vv) \quad .
\ee
It is not necessary to assume that the scalar field is the {\it only}
component in the very early universe: spin $1/2$ and $1$ fields could
also be present but their influence will be vanishingly small, as
we show below.

One Planck time after the Big Bang, the typical energy density
of a Planck-size region of the universe is, by virtue of the Heinsenberg
uncertainty principle, of order $1 [M_{pl}^4]$. Focus on the 3 different
components of the energy density
and pressure, the kinetic terms $\dot\vv ^2$, the gradient terms
$(\nabla\vv)^2$ and the potential term $V(\vv)$. We assume that each one
of those terms and, in addition, the curvature $R^2 \sim H^4$ should
assume values in the vicinity of $1 [M_{pl}^4]$. These conditions are
then,

\beq
\label{Pla_Ki}
\dot\vv^2 		&\sim& {\cal O}(1) \quad , \\
\label{Pla_Gr}
(\nabla\vv)^2 		&\sim& {\cal O}(1) \quad , \\
\label{Pla_Po}
\frac{m^2}{n} \vv^n 	&\sim& {\cal O}(1) \quad , \\
\label{Pla_H}
H	 		&\sim& {\cal O}(1) \quad ,
\eeq
The orders of magnitude above are all consistent with
the Klein-Gordon equations for the scalar field and
with the Einstein equations that relate the curvature
to the energy and pressure density of $\vv$.
It is easy to see
that the value of the scalar
field $\vv$ in a Planck-size region $H^{-1} \sim 1$ is given most simply
by the second to last equation, 

\be
\label{Sca_Pl}
\vv_{{pl}} \sim m^{-2/n} \gg 1 \quad .
\ee
The remaining conditions, (\ref{Pla_Ki}) and (\ref{Pla_Gr})
impose constraints on the time derivatives and gradients of
the scalar field. These are actually upper bounds, since
if any one of the terms is bigger than the Planck scale,
it will dominate over the others but then the universe does not
have a classical description, since quantum fluctuations
of the metric would destroy the classical notions of spacetime.
Let us focus on a region of the universe that allows a classical
description, and in which the scalar field is relatively homogeneous.
To be precise, consider a region bigger than one Hubble horizon
$H^{-1}$, in which the scalar field have an approximately homogeneous
value over a fixed time slice, 
$\left| \frac{\nabla^2\vv}{\vv} \right| \ll |H|^{-2}$.
We assume, in addition, that the scalar field
varies slowly in time in that region, 
$\left| \frac{\dot\vv}{\vv} \right| \ll |H| $, although this condition
can be relaxed, since the friction term $3H\dot\vv$ in the Klein-Gordon
equation supresses any initial time derivatives of the homogeneous field.

With those assumptions, the energy and pressure of the scalar field
are approximately given by 

\beq \label{Ene_Sc_In} \rho^S &\simeq& V(\vv) \quad , \\ \label{Pre_Sc_In}
p^S &\simeq& -V(\vv) \quad , \eeq which, by the continuity equations
$\dot\rho + 3 H (\rho+p) = 0$ means that ${\dot\rho}^S \simeq 0$. The
homogeneous scalar field, thus, contribute an approximately constant
energy density, while all other particles (such as photons, electrons and
monopoles) contribute an energy density which decays in time like
radiation, $\rho_r \propto a^{-4}(t)$.  It is easy to conclude that the
scalar field, just like a cosmological constant, quickly dominates the
evolution of the universe over the other matter fields, which are
exponentially dissolved or ``redshifted". The scalar in question is not
necessarily a fundamental scalar, it could be for example a 0-spin
condensate of a fermionic or bosonic field.  That the future evolution of
this patch of the universe is independent of the non-scalar spectator
particles is a consequence of the ``no-hair" theorem for de Sitter.

Summarizing, if there is a domain (``bubble") in the very early universe of
sufficiently large size (bigger than $H^{-1}$) in which an interacting
scalar field assumes an approximately homogeneous value, this patch will
become dominated by the scalar field in a time lapse of the order
$H^{-1}$. Since the main contributions of the scalar field to the energy
and pressure come from the scalar potential $V(\vv)$, the universe
inflates - the scale factor increases exponentially. We call the inflaton
the scalar field responsible for cosmological inflation.

Clearly, regions outside of the inflating bubbles become irrelevant, since
the inflating domains occupy an exponentially larger share of the global
volume of the early universe. According to the chaotic scenario, we live in
one such bubble that inflated, reheated and cooled off to eventually become
the dust-dominated universe that is visible to us today. As we shall see
later, though, not all bubbles begin or end the inflationary cycle at the
same time. In fact, new inflating bubbles are continuously generated in the
chaotic scenario, a phenomenon due to long wavelength quantum fluctuations
of the inflaton field that occur when the scalar field is above the
so-called self-reproduction scale $\vv_{sr} = m^{-2/(n+2)} < \vv <
\vv_{{pl}}$. A bubble in which the scalar field has fallen to a value
smaller than $\vv_{sr}$ is not affected by macroscopic quantum fluctuations
of the scalar field, and is said to ``evolve classically" (see Section
3.3 for a detailed discussion.)

Let us consider for now the fate of one such classically evolving bubble
in the case where the bubble have no boundaries, that is, as if it
occupied the whole universe (we know this approximation is valid because of
the no-hair
theorem for de Sitter.)  For simplicity, assume that at time $t_i$ there
is a bubble of radius $H^{-1}$ in which a homogeneous scalar
field\footnote{This assumption is natural since a quasi-homogeneous bubble
expands to a radius $\propto 3 H^{-1}$ in a time $H^{-1}$ while during
this
interval the inhomogeneities grow at a much slower rate. Therefore, if we
consider as initial time $t_i + H^{-1}$, we can chose another bubble of
radius $H^{-1}$ inside the original bubble that is homogeneous enough for
our purposes.} takes the value $1 \ll \vv_i < \vv_{sr}$.

The Klein-Gordon equation dictating the evolution of the scalar field
is, without approximations and in a FRW homogeneous background,

\be
\label{Kle_Go_FRW}
\dalamb \vv + V'(\vv) 
	= \ddot\vv + 3 H \dot\vv - \frac{1}{a^2} \nabla^2 \vv + V'(\vv) 
	= 0 \quad ,
\ee
where $V' = dV(\vv)/d\vv $ is the derivative of the potential
(\ref{Sca_Po}).
Immediately after $t_i$ this equation can be approximated by

\be
\label{EOM_vv_In}
3H\dot\vv + V' \simeq 0 \quad \Longrightarrow \quad 
	\dot\vv \simeq - \frac{V'}{3H} \quad .
\ee
The scale factor, on the other hand, is determined by one of
Eintein's equations in curved 3-space sections,

\be
\label{Ein_Eq_Cu}
3 H^2 + 3 \frac{k}{a^2} = \rho^S \quad ,
\ee
where $k = \pm 1 , 0$ corresponds to a closed, open or flat universe.
The term $k/a^2$ should be small at $t_i$, since we assumed that
curvature, likewise radiation and other matter components, is
not predominant at this early post-planckian time. As the rapid
expansion begins (exponential or power-law with $t^q$ and $q\gg1$),
the curvature term becomes negligible as compared to $H^2$.
In the exponential case, $H^2$ remains constant while  
$k/a^2 \propto e^{-2Ht}$, and in the
power-law case $H^2$ falls as $t^{-2}$ while the curvature
term falls as $t^{-2q}$.

At this point a major success of inflation is already clear:  if the
curvature term in the Einstein's equation is not big enough to rapidly
crunch or empty the universe, inflation will make it vanishingly small as
compared to the energy density of the scalar field. Therefore, open and
closed universes that go through a period of inflation asymptotically
resemble a flat universe. After inflation is over and the scale factor
evolves as $t^q$ with $q<1$ the curvature term decays less fast than the
energy density of matter, and the instability that we discussed in the
second chapter starts to magnify the curvature again. This scenario can
then explain why we don't observe a large curvature today: it was
made
exponentially small in the inflationary epoch.

We neglect the curvature term from now on, since the discussion above 
shows that it is irrelevant during inflation.

In the simplest but typical case which we consider 
here, $n=2$ in the scalar potential, the
maximal value for $\vv_i$ is $\vv_{sr}=m^{-1/2}$, and the value above
which the classical description brakes down is $\vv_{{pl}}=m^{-1}$ (see
Fig. 3.1.) The scalar field ``rolls" down the potential after
the scalar field drops below $\vv_{sr}$, until it reaches $\vv \simeq 1$
and the inflationary epoch ends. 

\begin{figure} 
\centerline{\epsfig{file=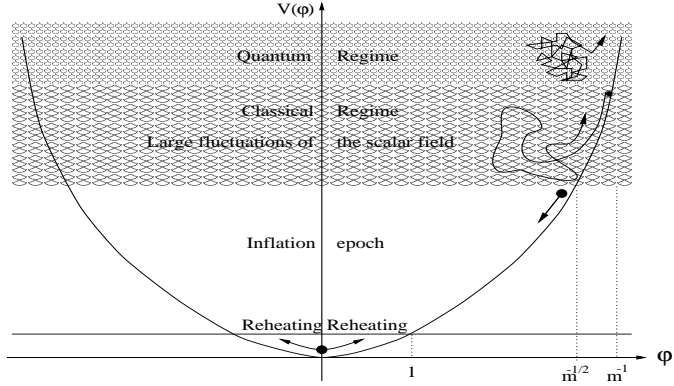,height=2.0in, width=3.5in}}
\vspace{10pt}
\caption{\small Potential energy $V(\vv)=m^2 \vv^2/2$. Above the energy
scale
$1[M_{pl}^4]$, quantum gravitational effects are important and the
classical interpretation of spacetime brakes down. Above the scale
$\vv \sim m^{-1/2}$, quantum fluctuations of the scalar field are
the dominant driving force and can force the field to go both up
and down the potential hill. Below this scale, classical evolution
is unaffected by quantum fluctuations and the scalar field slowly rolls
down the potential. After the field drops below $\vv \sim 1[M_{pl}]$,
inflation ends and the reheating period starts, where oscillations of the
scalar field lead to particle creation.}
\label{Fig_Ch_Po}
\end{figure}

The solution to Eqs. (\ref{EOM_vv_In}) and
(\ref{Ein_Eq_Cu}) for $n=2$ and $k=0$ are straightforward:

\be
\label{H_vv_m2}
H \simeq \frac{m}{\sqrt{6}} \vv
\ee
and

\be
\label{Vv_t_m2}
\vv(t) - \vv_i \simeq - m \sqrt{\frac{2}{3}} (t-t_i) \quad ,
\ee
where $\vv_i$ is the value of the scalar field at the time $t_i$.
The Hubble parameter as a function of time is then

\be
\label{H_t_m2}
H(t) \simeq H_i - \frac{m^2}{3} (t-t_i) \quad ,
\ee
with $H_i = \frac{m}{\sqrt{6}} \vv_i$. The scale factor can be calculated
from this expression,

\be 
\label{a_t_m2}
a(t) \simeq a_i e^{H_i (t-t_i) - \frac{m^2}{6}(t-t_i)^2} \quad .
\ee
It is sometimes useful to use the homogeneous scalar field as a time
parameter, and in that case we have the simple expression

\be 
\label{a_vv_m2}
a(t) \simeq a_i e^{ \frac{1}{4} (\vv_i^2 - \vv^2) } \quad .
\ee

Notice that, since $m \ll 1$, the time derivative of the Hubble parameter,
$\dot{H} = -m^2/3$, is much smaller than $H_i \leq m^{1/2}/\sqrt{6} $,
that is, it is valid to approximate the Hubble parameter as a constant
during inflation. It can also be verified that the kinetic term $\dot\vv^2
\simeq 2m^2/3$ is much smaller than the potential term $m^2 \vv^2/2$, so
that it can be neglected in the expressions for the energy density and
pressure.  In conclusion, the solutions found are consistent with the
approximations, and inflation only makes the solutions more accurate.

During inflation, quantum fluctuations of the scalar field are
coupled to the expansion of the universe, which amplifies these
fluctuations. After their wavelength crosses the horizon size $H^{-1}$,
the fluctuations freeze (their amplitude becomes constant) 
and quantum corrections cease to be important. These cosmological
perturbations from inflation reenter the horizon once
inflation is over, and they provide the basis for the inflationary
model for structure formation.

We will discuss the generation and evolution of perturbations at greater
length in the next section, but for now it suffices to say that the scale
at which the fluctuations are generated is approximately constant during
inflation. Fluctuations of the size of the Hubble radius have
approximately the same amplitude during inflation.  After a perturbation
becomes larger than the horizon, its amplitude freezes, although its
wavelength continues to be redshifted as $\lambda_{ph} \propto a(t)$.

When inflation is over, the Hubble radius increases much faster than
$a(t)$, and scales outside the horizon start to cross back into the
horizon. Provided the reheating period (see below) is much smaller than
the inflationary epoch, the perturbations generated during inflation with
same amplitude reenter the Hubble radius $H^{-1}$ after reheating.  Since
their amplitude has been frozen from the moment they crossed out of the
horizon, when they cross back into the horizon their amplitudes are
obviously approximately equal. This scale-invariant spectrum of
perturbations obtained after the end of inflation is verified by the
analysis of the CMB, and constitutes another major success of the
inflationary models (but not one exclusive of these models.)

When do inflation end? From Eqs. (\ref{Ene_Sc_FRW}) and (\ref{Pre_Sc_FRW}) 
we see that the kinetic term $\dot\vv^2/2 \simeq 2m^2/3$ becomes
comparable to $m^2 \vv^2/2$ when $\vv = \vv_{reh} \simeq 1$. At this
point the second derivative in the Klein-Gordon equation
(\ref{Kle_Go_FRW}), $\ddot\vv$, cannot be neglected anymore. In fact, when
the scalar field reaches $\vv_{reh}$ it accelerates down the potential,
since the friction term $3H\dot\vv$ has become much smaller. Another
way of seeing this is noting that the natural time scale of the
inflaton, $m^{-1}$, becomes equal to the cosmological time scale
$H^{-1}$ when $\vv \simeq 1$, and after the field falls even further,
the friction term becomes much smaller, ending the ``slow-roll" 
(see Fig. 3.1.)

In the new
regime the scalar field oscillates fastly between two maxima
of the potential with very little friction. The Klein-Gordon equation
can be approximated by
\be
\label{Kle_Go_Re}
\ddot\vv + m^2 \vv \simeq 0 \quad ,
\ee
whose solution is straightforward,

\be
\label{Vv_t_re}
\vv(t) \simeq \vv_* \sin{(mt + \theta)} \quad .
\ee
Notice that energy density and pressure oscillate together with the scalar
field, but if we average over a cycle $m^{-1}$ the result is

\be
\label{Ene_Pre}
\langle \rho \rangle_{m^{-1}} = \frac{m^2 \vv_*^2}{2} \quad , \quad 
\langle p \rangle_{m^{-1}} = 0  \quad .
\ee
This is the equation of state of pressureless dust, $p=0$, and we know
that in that case the continuity equations $\dot\rho + 3 H (\rho+p) =0$
imply that the energy density falls like $a^{-3}$. This time dependence
is hidden in $\vv_*$, which in fact decays slowly, 
while the scale factor falls as a power-law so that the approximation 
is consistent.

This period of fast oscillations (with period $m^{-1} \gg H^{-1}$)  marks
the end of cosmological inflation, since the scale factor now grows as a
power-law, $a \propto t^{2/3}$. At this point the couplings of the
inflaton field to other matter fields becomes also important, and particle
production occurs through the mechanism of parametric
resonance\cite{Reheating}. This period is known as
``reheating", and is the process by which the empty post-inflationary
universe is replenished of the hot particles that we know were present at
very early times.  Of course, the temperature of the plasma during
reheating should not be bigger than the GUT scale, otherwise the problem
of monopoles would strike again.

The inflaton field eventually decouples from the other
matter sectors, and usually it is assumed that it settles
at the minimum of its potential. At the end of reheating
virtually all the energy contained in the scalar field
at the end of inflation has been converted into
ultra-relativistic particles (radiation). Sometimes the
inflaton field leaves remnants, and it has been suggested that
it could even provide some of the dark or ``missing" matter
predicted by primordial nucleosynthesis.

\section{Cosmological perturbations from inflation}

One of the main applications of inflationary models is
the origin of large-scale structure.
Inflation provides a causal mechanism whereby
microscopic quantum fluctuations of the scalar field 
evolve into the classical 
cosmological perturbations that originated the structure
of the universe.

During inflation, the analysis at the end of Chapter 2 shows
that the metric perturbations have a time dependence given by
Eqs. (\ref{phi_HF}) and (\ref{phi_LF}),

\be
\nonumber
\phi_k^{\tiny (HF)} = \dot{\vv}_0
        \left[
                A_k^{+ {\tiny (HF)}} e^{+ik \int dt/a} +
                A_k^{- {\tiny (HF)}} e^{-ik \int dt/a}
        \right] \quad ,
\ee
when the perturbation is smaller than the horizon, and

\be
\nonumber
\phi_k^{\tiny (LF)} = A_k^{{\tiny (LF)}}
        \left(
                1 - \frac{H}{a} \int a dt
        \right) \quad ,
\ee
when it is bigger than the horizon.

The factors $A_k$ are as yet undetermined, and the classical theory of
perturbations does not provide any clue to what their value is. Of course,
they are fixed by the initial conditions. When there is no theory
available to provide these initial conditions, we can only determine the
$A_k$'s through the observed inhomogeneities in the present-day universe,
and use the theory of cosmological perturbations to deduce what the
``seeds" might have looked like at very early times.  Modern attempts at
explaining large-scale structure such as inflation and topological
defects, try to provide physical mechanisms that could generate a spectrum
of $A_k$ that is consistent with the observed map of the CMBR from a
minimal set of parameters and initial conditions. 
Chaotic Inflation is particularly minimalistic,
and necessitates only the self-couplings of the inflaton and its initial
value on a homogeneous time slice.

Clearly, the theory about the origins of the seeds are outside
the scope of the classical theory of cosmological perturbations.
In the inflationary scenario, the mechanism stems from the inherently
quantum fluctuations of a scalar field in flat
spacetime. In the case of topological defects,
the origin of perturbations is a ``classical" effect 
in the sense that the perturbations are created by the motions of 
macroscopic solitonic objects.

On small wavelengths, fluctuations of the scalar field around a
homogeneous background $\vv_0$ can be described by the usual canonical
quantum operators acting on a Fock space.  When their wavelength crosses
the Hubble radius, the fluctuations become squeezed in states with a
very high occupancy number, and are thus best
described by functions with a statistical distribution. Fluctuations
bigger than the horizon during inflation are said to be
classical perturbations, in the sense that they follow the classical
equations of motion, $\delta \Gmn = \delta \Tmn$, without quantum
corrections.

Fluctuations of the homogeneous scalar field inside the horizon can be
treated by quantum field theory. 
Outside the horizon the expectation value of the field excitations is
very high, and can be replaced by a sum over momentum modes of classical
fields with amplitudes obeying a Gaussian statistical distribution. The
spatial average $\langle \delta f \rangle$ of linear perturbations yield
zero identically, just as happens with the expectation value of a linear
quantum fluctuation ($\langle 0| \delta f^+ a^+ + \delta f^- a^- |0\rangle
=0$ at all times). The notion of spatial average here is defined on the
biggest possible scale, that is, on the scale of the physical radius of
the entire inflating bubble, $H^{-1}(t_i) a(t)/a(t_i) \gg H^{-1}(t)$,
where $t_i$ is the time when inflation started in that bubble.  Of course,
fluctuations around the homogeneous value are by definition zero on this
largest scale, although they might be nonzero on smaller domains inside
the initial bubble.

The two-point functions of fluctuations, as opposed to linear
functions of the perturbations, do not vanish in general,
$\langle 0 | \delta \hat{f}_k \delta \hat{f}_{k'} | 0 \rangle \propto
\delta_{k'k}$. 
Knowledge of the correlation functions of
field operators at the horizon scale ($|k| \sim H$) 
gives the amplitudes of modes $|\delta f_k|^2$
with a classical interpretation and Gaussian statistics.

The reasoning used below is as follows: we first determine the amplitude
of the modes of the perturbations when they cross outside the horizon. 
That fixes the amplitudes of all modes created during inflation that are
of cosmological interest. We then just follow the classical evolution of
the modes to obtain the spectrum of the perturbations,
both during inflation when their wavelength is exponentially bigger
than the horizon $H^{-1}$, and after inflation ends when they start
to reenter the horizon.

We start by presenting a non-rigorous derivation\cite{LindeSlava} of the
``spectrum" $\phi_k$, based on the Hawking temperature associated with a
de Sitter spacetime\cite{HawkingGibbons}.  Quantum fluctuations of the
scalar field on scales of order the Hubble radius $H^{-1}$ should have an
amplitude proportional to the Hawking temperature of the thermal
fluctuations of de Sitter. Their root mean square value on a sphere with a
radius of the size of the Hubble horizon is then

\be
\label{Flu_Ha}
|\dv_H| \sim T_H = \frac{H}{2\pi} \quad.
\ee
Since $\vv(t) \gg 1[M_{pl}]$ and $H \sim m\vv_0(t)$ during inflation,
these perturbations are small, as expected: $\dv/\vv_0 \ll 1$.
The energy density of the universe, during inflation, is dominated
by the potential energy $V(\vv)$, and thus 
fluctuations with scales close to the Hubble horizon should also be
proportional to the fluctuations of the scalar field,
$\delta\rho_H/\rho \sim \delta V/V$. Using the model with
$V=m^2 \vv^2/2$ and Eq. (\ref{Flu_Ha}), as well as the 
identity\cite{RevPaper}
$\left. \frac{\delta\rho}{\rho} \right|_{k(t)} = -2 \phi|_{k(t)}$
valid when the perturbation crosses the horizon, $k(t)=H$, we 
get that, during inflation, the amplitude of the modes for the scalar
perturbation $\phi_H$ is (up to some numerical coefficients),

\be
\label{Spe_easy}
|\phi_H| \simeq m \quad .
\ee
This amplitude corresponds to a perturbation on the scale of
the Hubble radius, and real perturbations are expected to be
positive or negative, that is, the fluctuations either
raise or lower the energy density inside the horizon.

We are primarily interested in the spectrum of
the momentum modes, so we need to perform a Fourier transformation
starting from the fluctuations $\phi_H(r)$. This quantity reflects
net perturbations inside a sphere of radius $r$, and its value can
assume both negative and positive values. The best we can do then is to
evaluate the square of this quantity on spheres of radius $r$. The
integration over volume inside the sphere introduces a factor of $r^3 \sim
k^{-3}(r)$, so we find the result 

\be
\label{Fou_ph}
\phi_{k(r)}^2 \simeq k^{-3}(r) \left| \phi_H \right|_r^2 
	\quad \rightarrow \quad 
	| \phi_{k(r)} | \simeq k^{-3/2}(r) \left| \phi_H \right|_r \quad .
\ee
$\left| \phi \right|_{k(r)}^2$ is usually referred to as the
{\it power spectrum} of the field $\phi$ at the scale k(r). We finally
find that, for quantum fluctuations
with size comparable to the Hubble radius, $k \sim H$, 

\be
\label{Spe_ph_easy}
|\phi|_k \simeq k^{-3/2} m \quad .
\ee

Due to the special time-translation invariance properties
of de Sitter spacetime, we can generalize this result
to other scales. As soon as the
physical wavelength of a fluctuation becomes bigger than the horizon, the
time dependence of its amplitude $\phi_k$ is given by the classical
equations of motion and, as a consequence, its amplitude freezes 
[actually it evolves very slowly with time, $\dot\phi_k/\phi_k \ll H$, see
Eq. (\ref{phi_LF})]. What that
means is that we can assume that the amplitudes (\ref{Spe_easy}) carries
on for a long time after the perturbation leaves the
horizon, and, as long as inflation goes on, this amplitude is
approximately constant. Therefore, during inflation and for
modes with wavelengths bigger than the horizon, the spectrum is 
given by (\ref{Spe_easy}), where the $k$'s correspond to scales
bigger than the horizon, $k > H$.

The argument above relied on a separation between the classical
and the quantum regime that was criticized for being artificial. The full
quantum treatment is slightly more complicated, but the end result,
Eq. (\ref{Spe_ph_easy}) during inflation, is essentially the same.

We now proceed to the full quantum treatment of the creation of
perturbations\cite{RevPaper}.  Our aim is to quantize, using the canonical
scheme, the coupled scalar and metric perturbations, and calculate their
two-point correlation functions $\langle \delta \hat{f}^2 \rangle$. 
In order to use the variational principles (find the canonically
conjugated momenta and the Hamiltonian), the action must be
determined to second order in the perturbations. We start from the
fundamental action

\be
\label{Tot_Ac}
S = \int \sqrt{-g} d^4 x \left\{ -\meio R + {\cal L}_m \right\} \quad,
\ee
where $R$ is the Ricci scalar curvature and ${\cal L}_m$ is the Lagrange
density function for the matter sector. We are interested in a scalar
field, for which

\be
\label{Lag_Sc}
{\cal L}_\vv = \meio D^\mu \vv D_\mu \vv - V(\vv) \quad .
\ee

We should now expand the action (\ref{Tot_Ac}) to second order in the
scalar perturbations $\phi$ and $\delta \vv$, around the flat FRW
background with $a(t)$ and $\vv_0(t)$. The $0^{th}$-order term gives
the equations for the backgrounds, while the $1^{st}$-order term is
identically zero (its variation $\delta S$ is proportional to the
background equations of motion).

It turns out that, due to the
constraint (\ref{EOM_ph_0i}), there is only one dynamical variable
expressing scalar perturbations, the gauge-invariant potential ${\cv}$.
We shall find explicit solutions for the modes $\cv_k$, 
then relate those to the modes $\dv_k$ and $\phi_k$. Finally,
we give explicit expressions for the spectra of the scalar
perturbations.

The action in
terms of $\cv$ is most easily expressed in conformal time, and we
simply state the result here (for a detailed discussion, see Ref.
\cite{RevPaper}): 

\be
\label{del_S}
\delta^2 S = \meio \int d^4 x 
	\left\{ \cv'^2 - (\nabla \cv)^2 + \frac{z''}{z} \cv^2 \right\}
	\quad ,
\ee
where we have discarded several total derivatives
that do not affect the quantization procedure or the classical evolution.
The potential $\cv$ has dimensions of mass, and is defined as

\beq
\label{Def_cv}
\cv (\eta,\vec{x}) &=& a \dv(\eta, \vec{x}) + z \phi (\eta \vec{x}) 
	\quad , \\
\nonumber
z (\eta)  &=& a {\vv_0}' /\H \quad , 
\eeq
where $\H=a'/a$.
The action (\ref{del_S}) describes a scalar field $\cv$ with
a time-dependent mass $M^2 = - z''/z$, in analogy with the
equation of motion (\ref{EOM_u}) for the perturbation $u$ defined
in Section 2.6 . This mass is, as before, a function of
the background quantities $\vv_0(\eta)$ and $\H$. When the
scalar field is not rolling down the potential, ${\vv_0}'=0=z$, 
the perturbations are not amplified and as a consequence there is no
creation of particles. The quantization procedure used here is also
analogous to the quantization of scalar fields in Minkowski space-time in
the presence of external fields.

The second step in canonical quantization is to determine the
canonical momenta associated with $\cv$, and impose the
commutation relations between the field operators and the
canonically conjugated momenta. The momenta are defined
with the help of (\ref{del_S}) as

\be
\label{Def_Mo}
\pi (\eta,\vec{x}) = \frac{\partial {\cal L}^{(2)} }{ \partial \cv'} =
\cv'
\quad .
\ee
In the quantum theory, fields and momenta $\cv$ and $\pi$ are raised to
the category of Heisenberg operators $\hat\cv$ and $\hat\pi$, and we
impose the canonical same-time commutation relations

\be
\label{Com_Rel}
[ \hat\cv(\eta,\vec{x}), \hat\cv(\eta,\vec{x}') ] = 
[ \hat\pi(\eta,\vec{x}), \hat\pi(\eta,\vec{x}') ] =  0
\ee
and

\be
\label{Com_Rel_2}
[ \hat\cv(\eta,\vec{x}), \hat\pi(\eta,\vec{x}') ] =  
	i \delta (\vec{x}-\vec{x}') \quad .
\ee

The dynamics of the field operators is given by the hamiltonian equations
of motion

\be
\label{Ham_EOM}
	i \hat\cv' = [\hat\cv, \hat H] 
		\quad , \quad i 
	\hat\pi' = [\hat\pi, \hat H] \quad ,
\ee
where the Hamiltonian can be easily reconstructed from
the second-order action $\delta^2 S$,

\be
\label{Ham}
\hat H = \int d^3 x (\cv'\pi - {\cal L}) =
	\meio \int d^3 x
        \left\{ \pi^2 + (\nabla \cv)^2 - \frac{z''}{z} \cv^2 \right\}
        \quad .
\ee

The equation of motion for $\hat\cv$ can be obtained from the
hamiltonian system (\ref{Ham_EOM}), or equivalently by varying the 
action $\delta^2 S$ with respect to the field $\cv$:

\be
\label{EOM_cv}
\hat\cv'' - \nabla^2 \hat\cv - \frac{z''}{z} \hat\cv = 0 \quad .
\ee

The best way to solve this equation is, clearly, to expand
the field in plane waves (generalized spherical harmonics in the case
of open and closed spaces):

\be
\label{Exp_cv}
	\cv(\eta,\vec{x}) 
	= \frac{1}{\sqrt{2}} \int \frac{d^3 k}{(2\pi)^3/2}
	\left[
		\cv_k^\ast e^{i \vec{k} \vec{x}}  a^{-}_k +
		\cv_k      e^{-i \vec{k} \vec{x}} a^{+}_k
	\right] \quad ,
\ee
where the operators $\hat{a}$ will be interpreted as raising and
lowering (creation and annihilation) operators acting on the vacuum
$|0\rangle$ of the Fock space representation (we use the Heisenberg
representation). All time dependence is contained in the
functions $\cv_k(\eta)$, which now satisfy the equations

\be
\label{EOM_cv_k}
{\cv_k}'' + (k^2 - {z''}/{z}) \cv_k = 0 \quad .
\ee

In terms of the creation and annihilation operators, the commutation
relations are the ones for a simple harmonic oscillator,

\be
\label{Com_a}
[a_k^-,a_{k'}^-]=[a_k^+,a_{k'}^+]=0 \quad , \quad 
[a_k^-,a_{k'}^+]=\delta (\vec{k}-\vec{k'}) \quad .
\ee
The mode functions $\cv_k(\eta)$ must obey the following
normalization conditions so that the last commutation relation
is equivalent to $[\cv(x),\pi(x')]=i\delta(x-x')$:

\be
\label{Nor_cv_k}
\cv_k' \cv_k^\ast - {\cv_k^\ast}' \cv_k = 2i \quad .
\ee

Usually, modes associated with the annihilation operator $\hat{a}_k$ have
a positive frequency $w^2(k) = k^2 + M^2$, while the modes corresponding
to $\hat{a}^\dagger$ have negative frequency. In our case the mass term
$M^2 = -z''/z$ is time-dependent, leading to a mixing between the
positive- and negative-frequency modes associated with $a^-_k$ and
$a^+_k$. Hence, even if we define a vacuum state for the Fock
representation at some initial conformal time $|0,\eta_1\rangle$, this
vacuum will
appear to have a non-vanishing particle content at a later time, $a^-_k
(\eta_2 > \eta_1) |0, \eta_1 \rangle \neq 0$. Mathematically, it is
possible to
define eigenfunctions $\psi_k(\eta_1)$ of the creation and annihilation
operators at time $\eta_1$, but these cannot be expressed in terms of
eigenfunctions at a later time $\eta_2$, $\psi_k(\eta_2)$.  The number
operator
at time $\eta_1$, $N_k^1 =a_k^+ (\eta_1) a_k^-(\eta_1)$, vanishes for
the vacuum $|0, \eta_1\rangle$, but is not zero if the operator is
evaluated at a later time,

\be
\label{Num_op}
\langle N_k^2 \rangle_{\eta_1} = \langle 0,\eta_1 | N_k (\eta_2) |0,
\eta_1\rangle \neq 0
\quad .
\ee

Conversely, it is equivalent to say that the creation and annihilation
operators at different times mix with each other. With that
perspective, it makes no sense to define a unique set of eigenfunctions
that are annihilated by {\it all} operators $a_k^-(t)$.
The linear transformations that relates
creation and annihilation operators at different times are called
Bogoliubov transformations:

\be
\label{Bog_tr}
\left( \begin{array}{c} 
	a_k^- (\eta_2) \\ a_k^+ (\eta_2) 
\end{array} \right) =
\left( \begin{array}{cc} 
	\alpha_k (\eta_1,\eta_2) &
	\beta_k^\ast (\eta_1,\eta_2) \\
	\beta_k (\eta_1,\eta_2) &
	\alpha_k^\ast (\eta_1,\eta_2) 
\end{array} \right)
\left( \begin{array}{c} 
	a_k^- (\eta_1) \\ a_k^+ (\eta_1) 
\end{array} \right)  \quad ,
\ee
where unitarity implies that $|\alpha_k|^2 - |\beta_k|^2 = 1$.

If we define the number operator at the time $\eta_2$ in terms of the
vacuum at $\eta_2$, it is clear
that, at an earlier time $\eta_1$, the occupation
number is given by

\beq
\label{Num_21}
\langle N_k^2 \rangle_{\eta_1} 	&=&
\langle 0, \eta_1 | a_k^+(\eta_2) a_k^- (\eta_2) | 0, \eta_1 \rangle  \\
\nonumber
&=& 
\langle 0, \eta_1 | a_k^-(\eta_1) \beta_k (\eta_1,\eta_2) \beta_k^\ast
(\eta_1,\eta_2) a_k^+(\eta_1) | 0, \eta_1 \rangle 
= |\beta_k (\eta_1, \eta_2)|^2 \quad,
\eeq
and conversely, if we define the vacuum at $\eta_2$, the number
operator defined in terms of the vacuum at $\eta_1$ has the expectation
value

\be
\label{Num_12}
\langle N_k^1 \rangle_{\eta_2} = 
\langle 0, \eta_2 | a_k^+(\eta_1) a_k^- (\eta_1) | 0, \eta_2 \rangle =
|\beta_k (\eta_2, \eta_1)|^2 \quad.
\ee
The ``forward" ($\eta_1 \rightarrow \eta_2$) and ``backward" 
($\eta_1 \leftarrow \eta_2$) Bogoliubov coefficients 
can be related by inverting the matrix (\ref{Bog_tr}) and taking into
account the unitarity constraint,

\be
\label{Bog_in}
\alpha_k (\eta_2,\eta_1) = \alpha^\ast (\eta_1,\eta_2) \quad , \quad
\beta_k (\eta_2,\eta_1) = - \beta (\eta_1,\eta_2) \quad ,
\ee
therefore it is a consequence of unitarity that
$\langle N_k^1 \rangle_{\eta_2} = \langle N_k^2 \rangle_{\eta_1}$.

To summarize, an observer at $\eta_2$ will measure a nonvanishing number
of particles $N^2_k(\eta_1)$ in the ``vacuum" state $|0,\eta_1\rangle$ as
long as the Bogoliubov transformations are not trivial, i.e., as long
as $\beta_k \neq 0$. This ``particle production" from an initial vacuum
state is a consequence of the time-dependent background external fields,
and is the mechanism responsible for the origin of cosmological
perturbations in inflationary models.

The ambiguity in chosing a vacuum is related to the freedom in choosing
the precise initial conditions for the mode functions $\cv_k$ and $\cv_k'$
at some initial time. Since the equations of motion that determine the
$\cv_k$'s is second-order in time, there are two integration constants
that must be fixed. The normalization constraints (\ref{Nor_cv_k})
reduce the number to one free parameter, which is
essentially the choice of an initial time $\eta_0$ such that the vacuum $|
0, \eta_0\rangle$ is empty at $\eta_0$.

During the inflating period, high frequency modes are redshifted by the
expansion of the universe, and eventually become low frequency. We will
define the vacuum states in the limit $k \rightarrow \infty$, when the
perturbations are well inside the horizon and we can
consider the background to be approximated by flat Minkowski, therefore
the ambiguity in the definition of a particle that occurs for
long wavelength modes in curved spacetimes is absent\cite{RevPaper}. 
The equations of motion (\ref{EOM_cv_k}) and the normalization conditions
(\ref{Nor_cv_k}) imply that the mode functions are

\be
\label{Sol_cv_k}
\cv_k (\eta_i) \simeq M(k\eta_i) k^{-1/2} e^{+ik\eta_i} \quad , \quad 
\cv_k'(\eta_i)  \simeq iN(k\eta_i) k^{1/2} e^{+ik\eta_i} \quad  \quad
(k\eta_i \gg 1) \quad ,
\ee
where $|M(k\eta_i)| \rightarrow 1$ and $|N(k\eta_i)| \rightarrow 1$
when $k\eta_i \gg 1$ and the normalization condition
is

\be
\label{Nor_MN}
MN^\ast + M^\ast N = 2 \quad .
\ee
The time $\eta_i$ can be chosen to be the time when the inflationary
epoch begins, so that the vacuum is defined with respect to this
initial moment. If there are particles or inhomogeneities present at that
moment such that vacuum would not be the appropriate definition,
this ``initial time" can be chosen a few e-folds later, when
the expansion of the universe will have diluted the initial
deformities.

The choice of initial time is akin to the choice of
initial conditions on a big enough patch of the universe. Since the
observed universe originated inside the Hubble radius at $N\simeq 50$
e-folds before the end of inflation, we could chose this initial
time as being any moment from the beginning of inflation 
up to the time $\eta_{60}$ when there were $60$ e-folds left to
the end of inflation. Although the choice of initial vacuum in this
time interval does not alter the results of calculations of the CMBR, 
it will play a significant role when we discuss
the global features of the chaotic scenario, that is, features
of the spacetime on scales of the particle horizon
$H^{-1}(t_i) a(t_0)/a(t_i)$ and larger ($t_0$ is the time now and
$t_i$, as before, is the time when inflation started.)
Notice that if inflation occurred for more than $60$ $e$-folds, 
the particle horizon could be many orders of magnitude
bigger than the presently observable universe.

We now wish to go back to the scalar perturbations $\phi$ and
$\dv$. First, remember that the two are related by the
constraint (\ref{EOM_ph_0i}),

\be
\nonumber
\phi ' + \H \phi = \meio \vv_0 ' \dv \quad ,
\ee
so that all we have to do is relate $\cv$ to $\phi$.
From the definition of $\cv$, (\ref{Def_cv}), and Eq. (\ref{EOM_ph_ij}),
it is straightforward to obtain, after some juggling with algebra, that

\be
\label{phi_cv}
\phi_k(\eta) = - \frac{1}{k^2} \frac{\vv_0'^2}{\H} 
	\left( \frac{\cv_k}{z} \right)' \quad
\ee
where, as before, $z=a\vv_0'/\H$.

The reason why we used the potential $\cv$ instead of the
variable $\phi$ in the quantization procedure is that
the action in terms of $\phi$ is much more complicated,
and thus the quantization much more involved when
written in terms of that variable.

The equation obeyed by $\phi_k$ have been derived before,
see Eq. (\ref{EOM_u}), 

\be
\label{EOM_u_2}
u_k'' + k^2 u_k - \frac{(1/z)''}{1/z} u_k = 0 \quad ,
\ee
where the variable $u_k = a\phi_k / \vv_0'$, introduced in Eq.
(\ref{Def_u}), is obviously related to $\cv_k$ by

\be
\label{u_cv}
u_k = - \frac{1}{k^2} \frac{z}{2} 
	\left( \frac{\cv_k}{z} \right)' \quad .
\ee

We will use now explicit solutions to $u_k$ in the short
and long wavelength limits, and subject them to the
constraints (\ref{Sol_cv_k}) and (\ref{Nor_MN}), found
through the quantization of the $\cv$-modes. 
First, in the short wavelength limit of Eqs. (\ref{EOM_u_2}), 
$k^2 \gg (1/z)''/(1/z)$,  there are clearly oscillating solutions with
frequency $k$. We impose on these solutions the constraints found upon
quantization, (\ref{Sol_cv_k}), which, using the relations above, are

\beq
\label{uk_eta_i}
u_k(\eta_i) 
	&=& -\meio 
	\left[ 
		\frac{i}{k^{3/2}} N(k\eta_i) -
		\frac{z'(\eta_i)}{z(\eta_i)} \frac{1}{k^{5/2}} M(k\eta_i)
	\right] 
	\quad , \\
\nonumber
u_k'(\eta_i) 
	&=& -\meio 
	\left[ 
		\frac{1}{k^{3/2}} M(k\eta_i) +
		3 \frac{z'(\eta_i)}{z(\eta_i)}
		\left( \frac{i}{k^{3/2}} N (k\eta_i) -
			\frac{z'(\eta_i)}{z(\eta_i)} \frac{1}{k^{5/2}}
			M (k\eta_i)
		\right)
	\right] 
	\quad . \\
\label{uk'_eta_i}
\eeq
The solutions in the high frequency, small wavelength limit
are then written as

\be
\label{Sol_u_HF}
u_k(\eta) = u_k(\eta_i)  \cos{[ k (\eta-\eta_i)]} +
	    \frac{u_k'(\eta_i)}{k} \sin{[ k (\eta-\eta_i)]} \quad .
\ee

The low frequency, long wavelength solutions [when
$ k^2 \ll (1/z)''/(1/z) $ ] are

\be
\nonumber
u_k = A_k \frac{1}{z} \int d\eta z^2 \quad ,
\ee
but we can use the background equations inside the integral,
integrate by parts and the final result is

\be
\label{Sol_u_LF}
u_k (\eta) = A_k \frac{1}{\vv_0'} 
	\left( 
		\frac{1}{a} \int d \eta a^2 
	\right)' \quad .
\ee

So far the constants $A_k$ (which, as we discussed before, determine
the spectrum of the long wavelength perturbations) are undetermined.
But the spectrum $u_k(\eta_i)$ for small wavelengths has been
fixed by quantization. Therefore, what we have to do to determine
the $A_k$'s is follow a short wavelength mode with physical
wavelength $\lambda_{ph} = 2\pi a/k$ until it reaches
the limiting wavelength 
$2\pi a k^{-1}_{\eta_{_\H}} \sim \sqrt{ (1/z)/(1/z)'' }$, and
then use the junction conditions to fix the amplitude of that mode for
all later times.

A rigorous treatment would forcibly involve the approximate solutions
when $k^2 \sim (1/z)''/(1/z)$, which are very difficult to obtain.
There is a much simpler way, though, which involves taking
into account the relation between the $\cv_k$ and $u_k$ given in Eq.
(\ref{u_cv}). We know that for the $\cv_k$ modes
the scale above which the fluctuations are high frequency is
$k_\cv^2 = z''/z$. For the $u_k$ modes, though,
that scale is $k_u^2 = (1/z)''/(1/z)$. Therefore, if these scales
overlap over a wide enough band, both the oscillatory solutions
(\ref{Sol_u_HF}) and the large-wavelength solutions (\ref{Sol_u_LF})
must be good approximations to the exact solution.
During an inflationary period, this interval is indeed large, in fact

\be
\nonumber
z \left( \frac{1}{z} \right)'' \simeq m^2 a^2(\eta) \quad \ll \quad
\frac{z''}{z} \simeq \H^2(\eta) \simeq 
	\frac{m^2}{6} \vv_0^2 (\eta) a^2 (\eta)
	\quad ,
\ee
since $\vv_0 \gg 1$ during inflation. Therefore, 
for scales such that $ma(\eta) < k < \H(\eta) $, both
solutions are approximately valid. We can then 
express the approximate
solutions in this interval as

\be
\label{Jun_u_k}
u_k(\eta) = \frac{	u_k(\eta_i)\cos{(k\eta_i)}
			- [u_k'(\eta_i)/k]\sin{(k\eta_i)}
		}{	\{ (1/\vv_0') [ (1/a) \int d\eta a^2 ]'
			\}_{\eta_\H(k)}
		}
	    \frac{1}{\vv_0'} \left( \frac{1}{a} \int d\eta a^2 \right)'
	\quad ,
\ee
where the junction time $\eta_\H(k)$
represents the time when the scale $k=\H(\eta_\H(k))$ crosses
the Hubble radius, that is, when the physical wavelength
of the mode $k$, $\lambda_{ph}= 2\pi a(\eta)/k$, becomes bigger than the
horizon $H^{-1}= a \H{-1}$.

Therefore, formulas (\ref{Jun_u_k}) and (\ref{Sol_u_HF})
are, respectively, valid for long wavelength and short wavelength
perturbations in an inflationary universe\footnote{ In fact, the formula
for the long wavelength perturbations is valid for
perturbations outside the horizon even after inflation
has ended (see \cite{RevPaper} for a detailed analysis).}.
We will use these formulas to calculate the amplitude in each
mode of the field $\phi_k$, for $k$'s representing both
perturbations bigger and smaller than the horizon. 
As noted after Eq. (\ref{EOM_u_2}), the modes $\phi_k$ can be
expressed in terms of the modes $u_k$ by

\be
\label{phi_u}
\phi_k (\eta) = \frac{\vv_0'}{a} u_k(\eta) \quad .
\ee

For the simple case when the inflaton potential is $V= m^2 \vv^2/2$,
the expressions for long and short waves simplify substantially, and
after we take into account the approximate solutions for the
scale factor in (\ref{a_vv_m2}), approximation (\ref{H_vv_m2})
and the useful indentity $\H \simeq -2 \vv_0' \vv_0$ the result is

\be
\label{Spe_ph}
|\phi_k| (t) = k^{-3/2} \frac{m}{2\sqrt{6}\pi} 
	\times 
	\left\{ 
	\begin{array}{ll} 
		1 \\
		\left[
			1 - \frac{4}{\vv_0^2(t) - \vv_0^2(t_r)} 
			\log{ \left( \frac{k}{H(t)a(t)} \right) }
		\right] \, ,
	\end{array}
	\right. 
\ee
with $k > H(t)a(t)$ in the former case and 
$ \quad H_i a(t_i) < k < H(t)a(t)$ in the latter.
In this expression, $t_i$ is the time when inflation started, and
$t_r$ is the time when it ends. To the reader concerned with units, we
note that had we retained units in the expression above, we would get
$m/M_{pl}$ (or $\sqrt{G} m$)  instead of $m$. We also notice that we
have retrieved the result of Eq. (\ref{Spe_ph_easy}), with a
logarithmic correction that comes from the fact that the Hubble parameter
is only approximately constant during inflation, such that the
Hawking temperature at the beginning of inflation is bigger
than that at the end.

We can also calculate the amplitude of perturbations after inflation is
over. Assuming that the universe behaves essentially as if it was
dominated by radiation, the scale factor has the time dependence $a
\propto t^{1/2}$ (this is a good approximation because during the brief
period of reheating, $a \propto t^{2/3}$). The result is very simple,

\be
\label{ph_rad}
\phi_k (t>t_r) \simeq k^{-3/2} m 
	\log{( \frac{\lambda_{ph}}{\lambda_\gamma} )} \quad ,
\ee
where $\lambda_{ph} = 2\pi a(t)/k$ is the physical wavelength of
the perturbation with comoving wavenumber $k$, and $\lambda_{\gamma}$
is the characteristic wavelength of the cosmic background radiation.

In observational cosmology the object of interest is the
power spectrum, which, as mentioned at the beginning of this section, is
defined as the adimensional measure of a classical fluctuation moded out
by a volume factor, e.g., 

\be
\label{Pow_Sp_ph}
\left| \delta^{\phi}_k \right| \equiv 
	k^{3/2} \left| \phi_k \right| \quad
\ee 
and

\be
\label{Pow_Sp_dv}
\left| \delta^{\delta \vv}_k \right| \equiv
        k^{3/2} \left| \frac{\delta\vv_k}{\vv_0} \right| \quad .
\ee

From Eqs. (\ref{Spe_ph}) and (\ref{ph_rad})  we see that the power
spectrum of the scalar perturbations created during inflation is nearly
{\it scale invariant}, with only logarithmic corrections that are not very
important on scales that are relevant cosmologically. The power spectrum
is also proportional to the square root of the self-coupling of the scalar
field ($\sqrt{m^2}$ in our case of a massive inflaton). The fluctuations
in the temperature seen in the CMBR are roughly proportional to the
spectrum calculated above, and the amplitude measured is about $\delta T/T
\sim 5\times 10^{-6}$.  This implies that the mass of the inflaton has to
be very small, $m \sim 10^{-6}[M_{pl}]$, so that the perturbations created
by inflation have the correct amplitude. That the mass of the inflaton
happens to be so
small is referred to as the ``fine-tuning problem".

Summarizing, we saw that inflation provides a causal mechanism that
generates perturbations with near identical amplitudes on cosmological
scales. The power spectrum of perturbations is nearly scale-invariant, and
is proportional to the mass of the inflaton. These perturbations are
generated continuously during inflation, inside the causal Hubble horizon. 
They leave during inflation (their physical wavelengths become bigger than
the horizon) and reenter later, after inflation ends.

\section{Global structure of Chaotic Inflation}
\label{SecGlSt}

We have mentioned in Section 3.2 that, above a certain
scale, large quantum fluctuations of the scalar field might
give rise to a process that generates new inflating bubbles
\footnote{As we have indicated before, a bubble is a horizon-sized
spherical domain
in the exponentially expanding universe. After one $e$-fold, there
are $e^3$ ``new" bubbles inside the initial bubble.}.
We will see later that the back reaction of scalar perturbations created
during inflation is important on scales where the
process of self-reproduction is active.
In this section we explain how the chaotic inflationary model can
regenerate itself, and investigate
some features of the universe on very large
scales\cite{LindeSlava}.

It was shown in the last section that small wavelength quantum
fluctuations originated inside the Hubble horizon have their amplitude
magnified, and
their wavelength redshifted, as inflation proceeds.  If perturbations have
a small amplitude, they don't have a noticeable effect on the evolution of
bubbles. On the other hand, if a perturbation in the energy density inside
an initial bubble is so large that after it leaves the horizon the
background energy density in ``new" bubbles is affected, then the
perturbations have to be accounted for in the dynamics of these
bubbles\cite{Linde}.

We will show below that above a certain energy scale (the so-called
``self-reproducing scale"), quantum fluctuations of the scalar field are so
large that they are dominant in the dynamics of the bubbles, and actually
drive the scalar field {\it up} its potential hill in most regions of the
universe\cite{Starob}. 
Since the expansion rate is proportional to the potential energy
of the scalar field, regions in which the scalar field is higher up its
potential expand faster, and as a result these bubbles dominate the
ultra-large scale structure of the universe.  A bubble in which the scalar
field happened to fall below the self-reproducing scale (called also a
``mini-universe") evolves classically, in the sense that the classical
equation of motion for the scalar field (\ref{EOM_vv}) is obeyed, without
quantum corrections. The expansion rates of these classical mini-universes
are much slower than elsewhere, therefore they occupy an ever decreasing
volume fraction in the universe.

In bubbles where the energy density is above the self-reproducing scale,
quantum fluctuations drive the scalar field both up and down the
potential. As a result, after a few e-folds an initially homogeneous
bubble would turn into a rather inhomogeneous collection of bubbles, some
with higher, some with lower energy density (see Fig. 3.2). 
From a homogeneous bubble, after a few e-folds we would get many different
bubbles with different energy densities, expansion rates etc. In
particular, the process could drive the energy density in some regions to
such high values that the symmetries of the fundamental Grand Unifyed
Theory would be restored. If a domain happens to cool down and come out of
this symmetry-restored region to become a FRW universe, the physical
properties of this ``baby universe" could be radically different from
ours, since its GUT vacuum is not necessarily the same as ours. Hence the
names ``self-reproduction" and "mini-universes": according to the chaotic
scenario, there is a continuous creation of inflating bubbles, each of
which might have completely different physics.

\begin{figure} 
\centerline{\epsfig{file=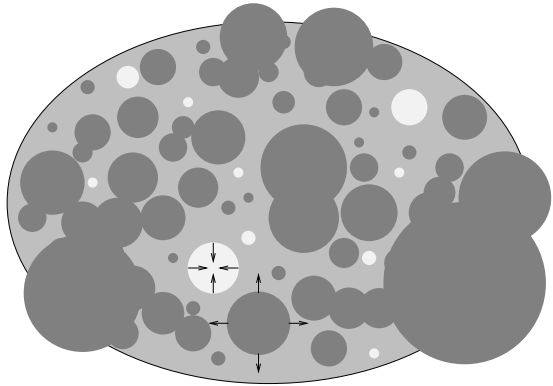,height=2.0in}}
\vspace{10pt}
\caption{\small A region of constant comoving volume with an
average value of the scalar field $\bar\vv>\vv_{sr}$ (``initially
homogeneous bubble"). Dark bubbles correspond to
domains where the scalar field fluctuated to a value bigger than
$\bar\vv$, and in the
light bubbles it fluctuated down to $\vv < \vv_{sr}$. The dark
bubbles occupy an exponentially high comoving volume
of the initial bubble, since their Hubble parameter is bigger than the
Hubble parameter of the initial bubble.}
\label{Fig_Se_Re}
\end{figure}

In all fairness, the global structure of the chaotic scenario bears no
immediate concerns for astronomers preoccupied with the observable
universe, since all of it would have come from an entirely classical region
some $60$ $e$-folds before reheating. However, it is crucial when we
consider, for example, the nature of the Big Bang singularity, the initial 
conditions for the universe and its topology, and as a means to
justify the anthropic principle in inflationary cosmology.

A rigorous analysis of the global structure involves the diffusion
equation for the probability distribution of the value of the
scalar field, $P(\vv,t)$. The final result, though, is the same
as the one obtained by a much simpler argument explained
below\cite{LindeSlava}.

A classical description of spacetime is only possible when the energy
density of the universe has fallen below $\rho \sim V(\vv)  \sim
1 [M_{pl}^4]$. The value of the scalar field, in the massive case, is by
assumption smaller than $\vv < \vv_{pl} = m^{-1}$, with the mass
constrained by the observed CMB anisotropies to be $m \sim < 10^{-6}$. A
microscopic domain of the size of a Hubble radius $H_{pl}^-1 \sim l_{pl}=
10^{-44}$ cm would have inflated to a radius today of about

\be
\label{Rad_pl}
R(t_0) \sim l_{pl} e^{-\frac{1}{4} [\vv^2(t_0) - \vv_{pl}^2] }
	\sim e^{10^{10}} {\rm cm} \quad ,
\ee
quite a few orders of magnitude bigger than the observable universe
today. Notice that the estimate above is actually a lower bound, since
we know that quantum fluctuations should make most regions inside
the initial planck-size bubble inflate for much longer than that.

The generation of long wavelength perturbations is completely
analogous to the mechanism described in the previous chapter:
quantum perturbations of the scalar field are redshifted,
while their amplitude is approximately frozen once they
become larger than the horizon. The difference is the
{\it amplitude} of the quantum fluctuations in a region of the universe in
which the scalar field is above the self-reproducing scale.
As we have noted before, quantum fluctuations
during inflation have a power spectrum given by 

\be
\label{Del_vv_Qu}
|\Delta^{Q} \vv| \sim \frac{H}{2\pi} \quad ,
\ee 
which can be interpreted as meaning that the scalar field takes a step of
size $H/2\pi$ every time lapse $\Delta t = H^{-1}$.

At the same time, the approximate equation of motion during
inflation, Eq. (\ref{EOM_vv_In}), implies that after the same
time lapse $H^{-1}$ the change in the scalar field from the force
$V_{,\vv}$ will be

\be
\label{Del_vv_Cl}
\Delta^{Cl} \vv \simeq \frac{V_{,\vv}}{V} H^{-1} \simeq \frac{2}{\vv}
	\quad .
\ee
By comparing the two values we see that if we require that quantum
fluctuations be unimportant, then

\be
\label{Com_De}
\vv \ll \vv_{sr} = m^{-1/2} \quad ,
\ee
and the converse if the classical motion of the scalar field is
negligible when compared to the quantum fluctuations.
Notice that even though the quantum fluctuations are important in
this limit, we still have $|\dv/\vv| \ll 1$ so that the linear perturbative
approach is still valid.

The behavior of a domain with $\vv < \vv_{sr}$ is very simple, and
is given by the familiar picture of a scalar field rolling down
the hill of the potential. Regions in which $\vv > \vv_{sr}$ are much
more interesting, since they can show a multitude of distinct behaviors,
including mini-universes with radically different physical properties.
The continuous creation of inflating bubbles from other inflating
bubbles means that spacetime has a fractal structure, maybe even a
self-similar one. Given the virtually infinite variety of mini-universes
that can be created, the chaotic scenario implies that some universes
must be capable of supporting life.

The main result from this section is Eq. (\ref{Com_De}):  above the
self-reproducing scale, quantum fluctuations dominate the evolution of the
scalar field. In a mini-universe like ours, the value of the scalar field
at the beginning of inflation was $\vv_{sr}=m^{-1/2}$, and it rolled down
from there. The total size of our mini-universe is then roughly $L_0 \sim
10^{10^5}$ cm, still a number for which units are quite meaningless.

The main point of this monograph is to show that the influence of the
quantum creation of perturbations is not restricted to the
effect to linear order caused by fluctuations in the self-reproducing
epoch. In the next chapter we discuss how
cosmological perturbations for which the linear effect is negligible
can still have a decisive effect on the dynamics of the background.


	\chapter{The Effective Energy-momentum Tensor}

Gravitational waves curve the space in which they propagate. But how can
the gravitational field be at the same time the cause and effect of the
gravitational field? What is the difference between the gravitons that make
up a gravity wave and the gravitons that constitute a background metric
in which the gravity waves propagate? How do they interact?

Due to the nonlinearities of Einstein's equations, the notions of
gravitational fields and sources are not as markedly distinct as in other
field theories like electromagnetism. Indeed, if gravity can be described
by anything that looks like a quantum field theory, then in some classical
limit of this theory there is a spin-2 massless particle, the graviton.
The propagation of gravitons in flat spacetime can be described in terms
of transversal plane wave solutions to the classical (usually Einstein's) 
theory. These waves, parametrized by a wavenumber $\vec{k}$, clearly carry
some energy and momentum proportional to $k$, and as such possess a
``gravitating mass", in analogy with non-abelian Gauge Theories where
the field strength can carry color charge.

The situation is more complicated when the background in which the
gravitons are propagating is a curved one. Nevertheless, in the
so-called Geometrical Optics limit, when
the wavelength of the graviton is much smaller than the curvature radius,
the former result is still valid: gravitational waves have energy and
momentum and, therefore, are themselves capable of curving spacetime.
The back reaction from gravitons in the high frequency limit has been
explored in many applications, most notably with respect to
nucleosynthesis\cite{Nucleo}. There, an excess of gravity waves can speed
up the cooling rate of the universe and increase the fraction of
helium, leading to constraints on the energy density of gravity waves at
the time of nucleosynthesis from the observation of the abundances of
primordial elements.

These ideas are old, and have been pushed forward since the late 50's and
60's, by the work of Wheeler\cite{Geometrodynamics}, Brill and
Hartle\cite{BrillHartle}, Brill\cite{Brill}, and culminating with an 
important result by Isaacson\cite{Isaacson}, who showed that the
energy-momentum tensor of small wavelength and high frequency (HF)
gravitational waves is ``gauge invariant", i.e., has the same value in any
reference frame. This result was crucial in that it put to rest any fears
that the energy density and momentum of gravitational perturbations are a
gauge effect.

Some questions, though, remain open, namely, what happens
outside the limit of HF perturbations? How to apply the results to
cosmology and, in particular, to inflationary cosmology?

\section{Historical remarks}

The first person to think of using gravitational waves as a means to curve
a homogeneous spacetime was Wheeler\cite{Geometrodynamics}.  Consider
geometries in which a background changes smoothly, due to the presence of
perturbations with high frequency but small amplitude. Wheeler showed
that, as long as the perturbations had small amplitudes, the metric could
be regarded as a smooth background ($\ymn$) with small ripples ($\hmn$),
and that the Eintein's equations made sense perturbatively, even when the
perturbations were HF.

In the absence of matter, the Einstein's equations for a background
metric with perturbations can be expanded in the usual Taylor series,

\be
\label{EFE_vac}
\Gmn = 0 \quad \Rightarrow \quad \Gmn^{(0)} + \Gmn^{(1)} + \Gmn^{(2)} +
\cdots = 0
	\quad ,
\ee
where the superscripts indicate the order of the term in the
perturbations. Assume that the dynamics of the background is determined
by the
superposition of gravity waves $\hmn$, but somewhat independent of the
dynamics of one given mode (wave). In that case, Einstein's equations
decouple and read

\beq
\label{EFE_vac_1}
\Gmn^{(1)} [h] &=& 0 \quad , \\
\label{EFE_vac_02}
\Gmn^{(0)} [\gamma] &=& \taumn^{(2)} 
	\equiv - \Gmn^{(2)}[h^2] \quad ,
\eeq
where now it is implied that the first equation is to be solved
for each mode separately, and
that the RHS term on the second equation involves the superposition of all
the wave modes
(the $0^{th}$-order term does not involve any perturbations
at all). Both $\Gmn^{(1)}[h]$ and $\Gmn^{(2)}[h^2]$ are evaluated at
the background $\ymn$, which means simply that both the propagation
of the linear modes and their quadratic interactions are mediated by the
background spacetime.

We could interpret the RHS of (\ref{EFE_vac_02}) as a source, a
second-order {\it
effective} energy-momentum tensor $\taumn$ for the gravity waves $\hmn$
in the background $\ymn$, although it is in reality nothing but the
second-order term of the Taylor expansion of $\Gmn$. In this form, it is
also known as Landau's pseudotensor of energy-momentum of
gravity\cite{Landau}.

This effective energy-momentum tensor (EEMT) is conserved,
${\Gmn^{(2)}}^{|\mu} =0$, as it should be given that the Bianchi
identities are geometric identities for the exact metric ($|\mu$ is the
covariant derivative with respect to the background metric). The
nonlinearities here are evident: the source of the gravitational field is
a conserved current which is a function of the gravitational field
itself. In Maxwell's equations, the source is a conserved, charged
current, but the electric and magnetic fields do not carry any charge. The
gravitational field, on the other hand, has a ``gravitational charge",
namely, mass.

Some comments are in order at this point. First, the typical examples of
backgrounds that we should have in mind are Friedman-Robertson-Walker
spacetimes (homogeneous space) and black hole spacetimes ($S^2$
symmetry). Perturbations around these spacetimes can only be defined with
the help of an average with respect to the symmetric background
coordinates [$x^3$ for cosmology, angles $(\theta,\phi)$ for black
holes] in a given patch of the spacetime. The
average of a perturbation $\delta f$ should yield
zero {\it by definition}, otherwise we could re-define the background and
perturbation as $f_0 \rightarrow f_0 + \langle \delta f \rangle$ and
$\delta f \rightarrow \delta f - \langle \delta f \rangle$, where
$\langle \rangle$ denotes the average,

\be
\label{Ave_dv}
\langle \delta f \rangle =
	\left\{
	\begin{array}{ll} 
	\frac{1}{V} \int_V \, d^3x \, \delta f (\vec{x},t)
	\quad & {\rm for} \, {\rm cosmology} \\
	\frac{1}{4\pi} \int \, d\theta \, d\phi \, 
	\delta f [(\theta,\phi),(r,t)]
	\quad & {\rm for} \, {\rm black} \, {\rm holes} .
	\end{array}
	\right.
\ee

Second, all perturbations $\delta f$ in these symmetric spacetimes can be
decomposed into {\it modes}, which are conventionally chosen to be
orthonormalized solutions of Laplace's equations in the background of
choice. In flat FRW the modes are simple plane waves $e^{\pm
i\vec{k}\vec{x}}$ labeled by the wavenumber $\vec{k}$, while in
Schwarzschild the modes are spherical harmonics $Y_{lm}(\theta,\phi)$
labeled by the angular momentum numbers $(l,m)$. We can use the
orthonormality conditions to simplify integrals of quadratic
combinations of the perturbations and turn them into sums over the
labels, so for example in cosmology we have 

\be
\label{Int_ex}
\langle \delta f(x) \, \delta g(x) \rangle (t)
	= \frac{1}{V} \, \int d^3 x \, \delta f(x) \, \delta g(x)
	= \int d^3 k  \,  \delta f_k(t) \, \delta g_k(t) \quad .
\ee
In the case of black holes, we would have a sum over the numbers $(l,m)$
in the RHS of (\ref{Int_ex}) instead of the integral (sum) over $k$.
The conclusion is that the averaging of quadratic combinations
of the perturbations is equal to a sum over the modes of the
perturbations. A superposition of modes is, therefore, the phase space 
equivalent of a spatial average. In cosmology, given that the
amplitudes of the modes
have a Gaussian distribution, the sum could also be interpreted
as an ensemble average.

This suggest a rigorous way of deriving the system of equations
accounting for back reaction: we start with the Taylor expansion of
Einstein's tensor, Eq. (\ref{EFE_vac}), and perform an average.
The term $\langle \Gmn^{(1)} \rangle$ vanishes, since it is linear in
the perturbations, and we are left with

\be
\label{EFE_av}
\Gmn^{(0)} = - \langle \Gmn^{(2)} \rangle \quad ,
\ee
where, of course, $\langle \Gmn^{(0)} \rangle = \Gmn^{(0)}$ since the
background metric depends only on the variable $t$, in the case of
cosmology, and on $r$ and $t$ in the case of black holes. The EEMT can
then be interpreted
in either one of two equivalent ways: 1) it is the correction
to the homogeneous component coming from inhomogeneities, or 2) it
is the sum of the stresses and energies in each momentum mode. 
The remaining equation for the perturbations, $\Gmn^{(1)}=0$, follows
naturally from (\ref{EFE_vac}), since it decouples from the other
terms and thus must be zero by itself.

The first explicit construction of a consistent solutions to Einstein's
equations in the form above was the gravitational geon, due to Brill and
Hartle\cite{BrillHartle}\footnote{Wheeler, who suggested the problem, had
for some time been constructing electromagnetic and neutrino
geons\cite{WheelerGeon}, with the intent of examining the problem of
gravitational collapse in black hole physics.}. They were looking for
stable, spherically symmetric superpositions of small amplitude
gravitational waves which were held together by their own gravitational
attraction.

Clearly, if this configuration is to be stable and
nondispersive, the background must be highly curved, and therefore it is
necessary to demand a lot of energy in the gravity waves, that is, their
frequency $\omega \propto E$ must be extremely high.

Although the gravitational geon model ended up depending on an unrealistic
limit\footnotesep=0.15in\footnote{The gravity waves had to be all
propagating over an
infinitely thin spherical shell, and their wavelength had to be infinitely
small compared to the radius of the shell.}\footnotesep=0.1in, they found
that the gravity
waves carried a mass and produced an exterior Schwarzschild spacetime
outside of the geon. That work established also that the problem of
propagation of gravitational waves in curved spacetimes need not be
treated through the full nonlinear Einstein equations, at least as far as
the system (\ref{EFE_vac_1})-(\ref{EFE_vac_02}) is concerned.

At the same time Brill\cite{Brill} completed an investigation
of gravitational waves in closed universes, and he concluded
that gravitational waves in an empty, closed universe would
cause this universe to collapse. Furthermore, the time development of a
universe filled with HF gravity waves is identical to that of
a universe filled with radiation, that is, $p_r = \rho_r/3$.
This was further evidence that gravitons, at least in the HF
(or geometrical
optics) limit, are just massless particles with 2 transversal degrees
of freedom, like the photon.

After the ADM formalism\cite{MTW} was developed in 1964, a
series of papers\cite{BrillDeser} found that the
mass-energy of any nonsingular gravitational wave configuration
that is bounded by flat spacetime is positive definite (this result
was later generalized to any asymptotically flat gravitating systems).
Interestingly, in his 1964 paper Brill also formulated situations in which
the interaction energy between distinct bundles of gravity waves is
negative. As we will se in later chapters, the energy density in some
cosmological perturbations corresponds to interaction energies, and
have a negative sign as well.

\section{The geometrical optics limit and gauge invariance of the EEMT}

One of the difficulties that arose during those years is the gauge
problem\footnote{See \cite{Fock} for a brief discussion.}:  small
transformations of the coordinate system induce gauge transformations of
the
metric perturbations - see Eq. (\ref{Lie_Dg}) - and since the EEMT is
defined in terms of the metric perturbations, it should then also change
under a gauge transformation, $\taumn[h^2] \rightarrow
\tilde{\tau}_{\mu\nu}[\tilde{h}^2] \neq \taumn[h^2]$. Therefore a change
in the coordinate frame would induce a change in the EEMT. This implies,
e.g., that the mass of the geon as calculated by Brill and Hartle would
not have an invariant meaning, assuming different values in the different
frames used to describe the gravitational waves.

In 1968 Isaacson\cite{Isaacson} (see also
\cite{Choquet,TaubMacCallum,Burnett,Efroimsky}) came up with the answer
for this important question: he showed that in the limit of HF
gravitational waves,
the effective energy-momentum tensor averaged over space {\it and} time is
gauge invariant.  In other words, assume that the gravity waves are HF,
and calculate their effective energy-momentum tensor, $\Gmn^{(2)}$. The
wavelengths and periods of the waves, $l$, are much smaller than any
corresponding background scales $L$ by assumption, so we can proceed to
averaging over the space and time coordinates on scales bigger than $l$
but smaller than $L$ (see Fig. 4.1). Isaacson showed that
the result of this calculation is independent of coordinatization, under a
few weak conditions\cite{Isaacson,Choquet,TaubMacCallum,Burnett}. 

\begin{figure} 
\centerline{\epsfig{file=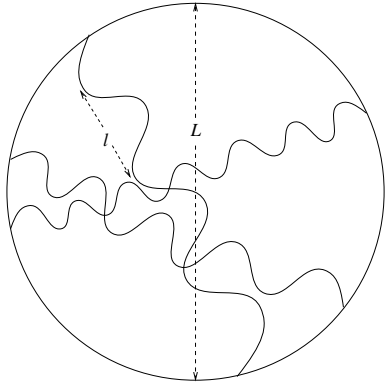,height=2.0in}}
\vspace{10pt}
\caption{\small A background spacetime with perturbations superimposed. In
the
geometrical optics limit, the wavelength $l$ of perturbations
is much smaller than the curvature radius of the background, $L$. } 
\label{Fig_Ba_Pe}
\end{figure}

We will now provide a new proof of this theorem, which does not rely on
the extensive algebraic manipulations needed in previous proofs. In order
to do that we introduce some formalism borrowed from functional analysis
by De Witt \cite{DeWitt} that greatly simplifies this and other
calculations (see Appendix A). To facilitate the discussion we also drop
tensorial indices whenever unambiguous, so $\Gmn \rightarrow G$, $\gmn
\rightarrow g$ etc.

We will be interpreting the functions
$w(x)=w[\partial /\partial x,g(x)]$ (Einstein tensor or Riemann
curvature, for example) defined on the manifold ${\cal {M}}$ as
the parametrized set of functionals $w_x$ defined on the space of
functions $g(x^{\prime })$ according to the formula

\begin{equation}
w_x \equiv w[\partial /\partial x,g(x)]=\int w[\partial /\partial
x^{\prime},g(x^{\prime })]\delta (x-x^{\prime })dx^{\prime }  \quad ,
\label{f}
\end{equation}
where $\delta (x-x^{\prime })$ is the Dirac delta function. Then the
functional derivative $\delta w_x/\delta g(x^{\prime })$ can be 
defined in the
standard way

\begin{equation}
\delta w_x=\int \frac{\delta w_x}{\delta g(x')} 
	\delta g(x') dx^{\prime } \quad ,  
\label{f1}
\end{equation}
where $\delta w_x$ is the change of the functional $w_x$ under
an infinitesimal
variation of $g(x^{\prime })$: $g(x^{\prime })\rightarrow g(x^{\prime
})+\delta g(x^{\prime }).$
 
If, for instance, $w_x=g(x),$ then
 
\begin{equation}
\delta w_x=\delta g(x)=\int \delta (x-x^{\prime })\delta g(x^{\prime
})dx^{\prime } \quad , 
\label{f3}
\end{equation}
and comparing this formula with (\ref{f1}) we deduce that
 
\begin{equation}
\frac{\delta w_x}{\delta g(x^{\prime })}=\delta (x-x^{\prime }) \,\, .
\label{f2}
\end{equation}
As another example, consider $w_x=\partial^2g(x)/\partial x^2$. Using
the definitions (\ref{f}) and (\ref{f1}) one gets
 
\begin{equation}
\frac{\delta w_x}{\delta g(x^{\prime })}=\frac{\partial ^2}{\partial
x^{\prime 2}}\delta (x-x^{\prime }) \quad .
\label{f4}
\end{equation}
 
In the following, the functional derivative $F_{xx^{\prime }}=\delta
G_x/\delta g(x^{\prime })$ will be treated as an operator which acts on
the function $f(x^{\prime })$ according to the rule

\begin{equation}
F_{xx^{\prime }} \cdot f(x^{\prime })=\int F_{xx^{\prime }}f(x^{\prime
})dx^{\prime }  
\label{f5}
\end{equation}
We will also use DeWitt's condensed notation \cite{DeWitt} and assume that
continuous variables $(t,x^i)$ are included in the indexes, e.g.,
$A^\alpha (x^{i},t)=A^{(\alpha ,x^{i },t)}=A^{m},$
where $m$ is used as the collective variable (``field index") to denote
$(\alpha, x^{i}, t).$ In addition we adopt as a natural extension of the
Einstein
summation rule that ``summation'' over repeated indexes
also includes integration over appropriate continuous variables, e.g.,
 
\begin{equation}
A^m B_m=A^{(\alpha ,x,t)}B_{(\alpha ,x,t)}=\sum_\alpha \int A^\alpha
(x,t)B_\alpha (x,t)dxdt  
\label{f6}
\end{equation}

We shall write functional derivatives using the following
short hand notation:

\begin{equation}
\frac{\delta w_{x}}{\delta g(x')} \equiv \frac{\delta w}
{\delta g^{m'}} \equiv w_{,m'}
\label{f7}
\end{equation}
For instance, the formula that we are interested in for the proof
of gauge invariance of the EEMT of HF gravity waves,

\begin{equation}
\Lie w(x)=\int d^4x^{\prime }
\frac{ \delta w(x) }{ \delta g(x^{\prime}) }
\Lie g(x^{\prime }) \quad  ,
\label{Lie4}
\end{equation}
in condensed notation takes the form  

\begin{equation}
\Lie w = w_{,m} \cdot (\Lie g)^m \quad .
\label{Lie_identity}
\end{equation}  
We will se later (Chapter 5) that this formula is a trivial consequence of
the diffeomorphism invariance of the theory. It expresses the geometrical
fact that the infinitesimal transformation induced on a tensor
functional $w[g]$ by a gauge transformation can always be expressed in
terms of the infinitesimal change in the functions $g$.

As further examples of the formalism, the Taylor expansion of the Einstein
tensor in the metric perturbations $h_{\mu\nu}$ can be written in this
formalism as simply

\be
\label{Def_G1_Fu}
G^{(1)}[h] \, = \, \left. G_{,m} \right|_{\gamma} \cdot h^m  
\ee
and

\be
\label{Def_G2_Fu}
G^{(2)}[h^2] \, = \, \meio \left. G_{,mn} 
		\right|_{\gamma} \cdot h^m h^n \quad .
\ee

We now proceed to prove that the EEMT of HF gravity waves is
gauge invariant. First, we note that 4-divergences can be integrated
out under a space-time average, i.e., they do not contribute to the
effective energy and stress $\langle \taumn \rangle$ of the gravity
waves\cite{Isaacson}. For cosmological perturbations
which are small wavelength but low frequency the average in time
cannot be done, but the average in space guarantees that we can still
integrate by parts expressions containing spatial derivatives.

We will now show that under a gauge transformation that takes the metric
perturbations $h \rightarrow h - (\Lie \gamma)$ [see Eq. (\ref{Lie_Dg})]
the corrections to the EMT are of the form of a total derivative and thus,
in the HF limit, can be dropped in expressions containing the averaged
(effective) EMT.

For the metric $g = \gamma + h$, the EEMT
is given by Eq. (\ref{Def_G2_Fu}).
If it is the gauge-transformed metric $\tg = \gamma + h - (\Lie \gamma)$
that we are considering, the expression is now simply

\be
\label{tau_Is_Tr}
\tilde\tau = \tau + 
	\meio G_{,mn} \cdot \left[ - 2 h^m (\Lie \cdot \gamma)^n +  
	(\Lie \cdot \gamma)^m (\Lie \cdot \gamma)^n \right] \quad .
\ee
Using the identity (\ref{Lie_identity}) we can integrate by parts
the $\xi$-dependent correction terms and get

\beq
\tilde\tau - \tau &=& \meio \Lie \cdot \left[ 
		G_{,m} \cdot (\Lie \cdot \gamma)^m \right]
		- \meio G_{,m} \cdot (\Lie \cdot \Lie \cdot \gamma)^m  \\
\label{tau_diff}
		& & - \Lie \cdot \left[ G_{,m} \cdot h^m \right] +
		G_{,m} \cdot (\Lie \cdot h)^m \quad .
\eeq
Now the expression between square brackets in the first term is just
$\Lie G[\gamma]$, which vanishes by the equations of motion for the
background $G[\gamma]=0$. The expression inside the square brackets
in the third term is just the equations of
motion for the perturbations $h$, therefore it also vanishes. We are left
then with the expression

\be
\label{tau_diff_final}
\tilde\tau - \tau = G_{,m} \cdot \left[ (\Lie \cdot h)^m - 
	\meio (\Lie \cdot \Lie \cdot \gamma)^m \right] \quad .
\ee
Now all that is left is to show that this is a total derivative. This is
easy to do if we notice that $G_{\mu\nu \, ,m}h^m =
R_{\mu\nu \, ,m} h^m - 1/2 \ymn R_{,m} h^m - 1/2 \hmn R[\gamma]$,
but for gravity waves in vacuum $R[\gamma]=R_{,m}h^m=0$, and therefore 
the Einstein tensor reduces to the Ricci tensor. Now, the
Ricci tensor for a general metric perturbation $\delta g$ is given by

\beq
R_{\mu\nu \, ,m} \delta g^m 
	&=& 	\dalamb \, \dgmn + 
		\delta g^\alpha_{\,\, \alpha|\mu \nu} -
		\delta g^\alpha_{\,\, \mu|\nu\alpha} - 
		\delta g^\alpha_{\,\, \nu|\mu\alpha} \\
\label{Ricci}
	&=&	\left[ 
		\delta g^\beta_{\,\, \beta|\mu} \delta^\alpha_\nu +
		\delta g_{\mu\nu}^{\quad |\alpha} -
		\delta g^\alpha_{\,\, \mu|\nu} - 
		\delta g^\alpha_{\,\, \nu|\mu} 
		\right]_{|\alpha} \quad ,
\eeq
that is, a 4-divergence. Therefore, from Eq. 
(\ref{tau_diff_final}) we obtain for the metric perturbation
$\delta g = \Lie \cdot ( h - \Lie \cdot \gamma )$ that

\be
\label{EEMT_Is_gi}
\ll {\tilde\tau}_{\mu\nu} - \taumn \gg = 0 \quad ,
\ee
where $\ll \cdots \gg$ denotes an average over
both space and time, over scales bigger than the wavelengths and
periods of the HF gravity waves. This completes the proof of the gauge
invariance of the effective EMT of high frequency gravitational
perturbations. Notice that the proof is only valid for HF perturbations
because we used integration by parts in time, which is not justified when
the perturbations are low frequency. We also remark that if the
EMT of some matter fields $f^m$ is such that $T[f]_{,m} \delta f^m$ can
also be expressed as a total derivative, then the proof above is valid for
a non-vacuum spacetime with that matter.

The physical content of the proof is simple: HF gravitational waves
correspond to highly localized gravitons that do not ``feel" the
background curvature, and therefore their short-time dynamics is
essentially equivalent to that of gravitons propagating in a flat
background spacetime. It is not surprising then that in this limit, called
the geometrical optics limit, the energy and momenta of the gravitons have
an invariant meaning in the class of reference systems that describe the
background. Only in cases where coordinate transformations can actually
fuzzy the localizability of the gravitons, or the notion of their energy,
we are going to find difficulties with the interpretation of the EEMT.

A crucial drawback of this proof of gauge invariance of the EEMT is
that it does not apply when matter is present, being valid only for vacuum
spacetimes with gravitational waves. Scalar field and metric perturbations
are then excluded from the proof, although they are the driving force
behind structure formation and actually contribute much more than
gravitational waves to the energy density of the universe.

Besides this weakness, there is at least one situation of interest in
which the geometrical optics limit fails to apply: cosmology. As we saw in
the theory of cosmological perturbations, there are physically relevant
perturbations with a gauge invariant meaning on scales bigger than the
horizon, i.e., with wavelengths bigger than the curvature radius $H^{-1}$. 
The amplitude of these perturbations is virtually frozen, which means that
their period is also much bigger than the time scale of the background,
$H^{-1}$.

The reason why we want to study the back reaction from perturbations which
are bigger than the horizon during inflation is that the scales outside
the horizon during inflation actually correspond to the present scales of
the apparent horizon of the universe and beyond. As we saw earlier, the
Hubble radius when there were $50$ $e$-folds left before the end of
inflation correspond to the present apparent horizon. In other words, the
region of the universe that emitted the CMBR that we measure today
originated all from inside a bubble of physical radius $H^{-1}(t_{50})$ at
time $t_{50}$, expanded to become a region of radius $H{-1}(t_{50}) \times
e^{50}$ containing some $e^{3 \times 50}$ Hubble-size volumes, then after
inflation ended the horizon increased to eventually encompass all these
$e^{150}$ regions. If we focus on a bubble at a time $t_b < t_{50}$, we
would be looking later at a region of spacetime which is some orders
of magnitude bigger than the present radius of the observable universe.

When we set out to solve the equations for the background in one of these
bubbles, we simply assume the fact that the description in terms of a
scale factor and a homogeneous background scalar field is valid inside the
entire bubble of radius $R_b (t) = H^{-1}(t_b) e^{t-t_b}$, for all future
times $t>t_b$. We wish to go beyond this calculation, and take into
account the contributions from all perturbations inside the bubble.  
Whatever goes on outside the inflating bubble is of exponentially
small consequence to processes inside the bubble, since the rate of
expansion of the bubble creates physical space at a rate that quickly
obliterates any outside processes and restrict their effects to the region
close to the boundary $R_b(t)$ of the bubble. We say then that the
physical radius of the bubble, $R_b(t)$,
is also a particle horizon for that bubble.

The perturbations whose wavelengths are smaller than $H^{-1}(t)$ we call
small-wavelength and HF (high frequency), while the ones which are bigger
than that [but smaller than $R_b(t)$] are called long wavelength and
low frequency (LF) (see Fig. 4.2).  The problem of back
reaction of HF perturbations should also include the back reaction of
quantum fluctuations and the renormalization of the stress-energy tensor,
which is are much more complicated issues that we do not consider here. On
the other hand, the back reaction of LF perturbations can be treated as a
completely classical problem (although their generation is an inherently
quantum process), and this is in fact the subject of the present work.

\begin{figure} 
\centerline{\epsfig{file=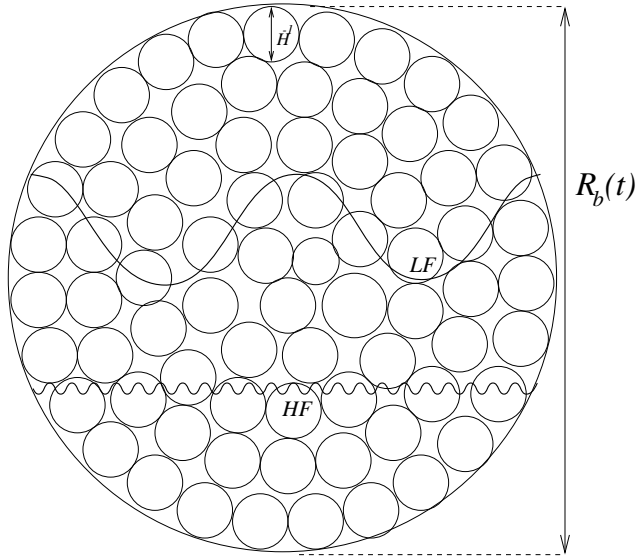,height=2.9in}}
\vspace{10pt}
\caption{\small After a time corresponding to $n$ $e$-folds of the scale
factor,
a bubble of initial radius
$R_b (t_b) = H^{-1}(t_b)$ has inflated to the size 
$R_b(t) \sim H^{-1}(t_b) e^{n(t)}$.
The small circles correspond to horizon-sized spheres of
radii $H^{-1}(t)$. Perturbations that exited the horizon at a time
$t_{H_k}$ have a wavelength $l$ much smaller
than $R_b(t)$ but much bigger than the Hubble radius $H^{-1}(t)$.}
\label{Fig_Ba_Pe_Ch}
\end{figure}

The perturbations created during inflation have a wavelength much smaller
than the particle horizon $R_b(t)$ of the bubble, but the ones that we are
interested in have been redshifted out of the Hubble horizon and thus
have, obviously, a wavelength bigger than the Hubble horizon. When we talk
later about the back reaction of scalar perturbations during inflation in
the chaotic inflationary model, it is with respect to these LF, classical
perturbations.

A large class of cosmological perturbations are then not HF, which
destroys the purpose of the integral over time in the space-time average.
Without the integral over time, the proof of gauge invariance of
the EEMT is invalid.
Besides, when there is matter filling up the universe,
there is again no guarantee that the EEMT is gauge invariant, even for
HF perturbations.
But without a simple gauge-invariant meaning to the
EEMT, how can we make sense of the problem of back reaction? This will be
the topic of the next chapter. Now we review some recent attempts
at back reaction.

\section{The averaging problem}

The back reaction of perturbations on smooth backgrounds is a small set of
a more generic class of problems, known as the ``averaging"  or ``fitting"
problem\cite{Ellis,Buchert}. Locally the universe is quite inhomogeneous,
with the largest structures that we see on scales of $100$ Mpc, compared
to $3000$
Mpc for the radius of the observable universe. We assume that if the
inhomogeneities were smoothed out evenly throughout the whole universe,
the time evolution of this homogeneous spacetime obtained by averaging
would not be very different from the true, inhomogeneous universe. 
However, nobody has ever provided an explicit covariant procedure whereby
an inhomogeneous universe is mapped into a homogeneous one.

The more basic problem is, whether and how a general relativistic system
can be split at all into background and perturbations, and
what dynamics the averaged fields follow. 
It can also be formulated
as the loose statement that the averaged dynamics is different from the
dynamics of the average.

In the context of cosmology, the assumption is that on a sufficiently
large scale the universe is isotropic and homogeneous, therefore on these
scales the split into background and perturbations is justified. This is
true as long as the density contrast $\delta\rho/\rho$ and the metric
perturbations (or, equivalently, the newtonian potential) are much smaller
than one. However, since very early times in the history of the universe
there are structures which have gone into the nonlinear regime, i.e., the
perturbations on the region around those structures are bigger than one.
Therefore, in cosmology the averaged variables used to describe
homogeneous Friedmann universes are obtained by construction rather than
by derivation.

The averaging problem has been explored also in Newtonian
cosmology\cite{Ehlers}, where the situation is much simpler than in the
relativistic case. For instance the gauge freedom of the perturbations,
which makes the analysis involved in the relativistic case, is substituted
by the symmetry under a galilean transformation, $\vec{x} \rightarrow
\vec{x} + \vec{v}t$, where $\vec{v}$ is a constant velocity field.  The
dynamical variables of newtonian cosmology are the eulerian fluid
variables: mass density $\rho_N(\vec{x},t)>0$, fluid velocity
$\vec{v}(\vec{x},t)$ and gravitational acceleration $\vec{g}(\vec{x},t)$. 
The back reaction of perturbations in newtonian cosmology turns out to
depend only on the rate of expansion $\vec{\nabla} \cdot \vec{v}$ and
rotation $\vec{\nabla} \times \vec{v}$ of the fluid, which are clearly
invariant under galilean transformations.

Another interesting result from the newtonian theory is that for closed
spacetimes the perturbations do not influence in the overall
expansion\cite{Ehlers}. For flat spacetimes, the perturbations can
influence the expansion on arbitrarily large but finite scales, and the
average motions are approximated well by a Friedmann model on scales much
larger than the scale on which perturbations have an effect on the
expansion.

In general relativity the problem of averaging is much more involved than
in newtonian cosmology, because of the facts that 1) there is no preferred
time-slicing, and 2) the metric itself is a dynamical variable, which
should be averaged over.  A solution to the averaging problem in general
relativity would require an as yet unspecified ``macroscopic description" 
(see, for example, Ref.  \cite{Zalaletdinov}), in analogy to the
macroscopic formulation of Maxwell's electromagnetism inside dielectric
media.

Another suggestion that has been made\cite{Carfora,CarforaPiot} is that a
inhomogeneous spacetime manifold with initial Cauchy data $({\cal M},
g_{ij},K_{ij})$ can be continously deformed into a
homogeneous spacetime supported by the manifold, $({\cal M},
\bar{g}_{ij}, \bar{K}_{ij})$ (see Fig. 4.3). The procedure
is similar to the renormalization group approach: the metric on a
point of the inhomogeneous spacetime is rescaled according to the
curvature at that point\cite{Hamilton}. The end result of this flow is a
homogeneous metric and curvature, corresponding to a FRW
universe\footnote{The procedure is only applicable to closed spacetimes
which are topologically $S^3/\Gamma$. The dominant energy condition
$ |{p}| \leq {\rho} $ must also be
satisfied on the manifold.}. The procedure, though,
seems too complicated to apply in any realistic situation

\begin{figure} 
\centerline{\epsfig{file=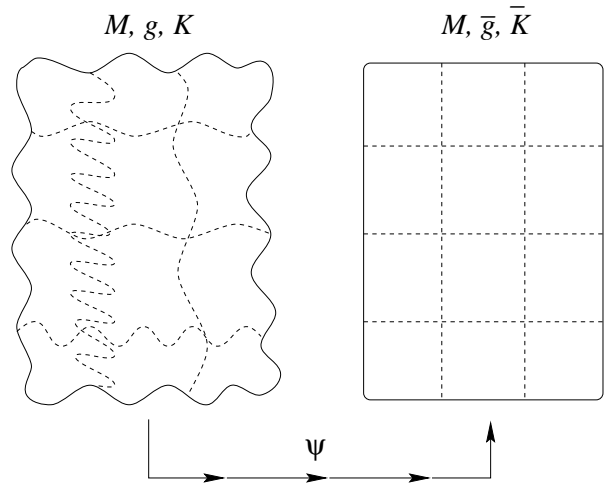,height=2.5in}}
\vspace{10pt}
\caption{\small A manifold (${\cal M},g,K$) corresponding to an
inhomogeneous spacetime can always be mapped by a continuous
deformation $\Psi$ onto another manifold which corresponds to a homogeneous
spacetime ($\bar{{\cal M}},\bar{g},\bar{K}$).} 
\label{Fig_Tr_Ma}
\end{figure}

Most recently the problem of back reaction has been addressed by a
number of authors in connection with the effect of inhomogeneities
on the expansion of the universe.

Futamase\cite{Futamase} analysed the corrections to
the expansion law of the universe using the ADM formalism.
He did not solve the dynamical problem of back reaction, rather
he focused on the question of local observables such as the expansion rate
of the universe. 
Futamase and Bildhauer\cite{Bildhauer} have looked into the problem of
back reaction of hydrodynamic perturbations in dust-filled universes, and
concluded that the influence of high frequency perturbations increases the
expansion rate of the universe, but that this effect decays rapidly with
time. In their treatment, they allow for
nonlinear density perturbations, as long as the metric perturbations
remain linear.  Their results are reinforced by some numerical work done 
by Shibata {\it et al.}\cite{Shibata} on the effect of axisymmetric
gravitational waves on a de Sitter background spacetime, which shows
that unless the gravity waves form a black-hole singularity, they are
redshifted and the de Sitter spacetime quickly becomes empty.

Also related to the back reaction program is the subject of local
observables in an inhomogeneous universe\cite{Uros}. One of the
most interesting questions that have been addressed lately is the
variations in local measurements of the Hubble constant, and their
relation to a globally defined expansion parameter\cite{Turner}. This
might be of crucial importance when comparing with the Hubble parameter
which will be inferred from the maps of the CMBR anisotropy.


	\chapter{Finite Diffeomorphisms}


We saw in the last chapter that when the perturbations
have a time period that is bigger than the expansion time $H^{-1}$,
the effective (averaged) energy-momentum tensor (EEMT) is
not gauge-invariant, that is, it assumes different
values in different coordinate frames. This would
put in jeopardy any attempts to tackle the back reaction
of low frequency (LF) perturbations on the background, because the problem
itself would be ill-defined: which gauge should one choose
to perform the back reaction calculations?

The root of the problem, as it stood up to now, was that the second-order
problem of back reaction was not formulated in a covariant manner up to
second order.  The perturbative expansion of the Einstein tensor was done
to second order [see Eq. (\ref{EFE_av})], but the gauge transformations
obeyed by objects in this perturbed spacetime were still regarded as
first-order, linear gauge transformations.

Clearly, a consistent, covariant theory of second-order perturbations
includes gauge transformations that are not truncated at first
order.
In fact, we will be able to formulate the whole problem of a
perturbative expansion of a diffeomorphism-invariant theory in
an explicitly covariant manner\cite{MAB96,ABM97}.

Calculating corrections to arbitrary orders is equivalent to regarding
perturbations as finite corrections, which can be improved in successive
orders of magnitude in perturbation theory.  The parameter $\eps$ of the
second chapter, which expresses the strength of the perturbations ($\eps
\sim \delta\rho/\rho$), is a finite number in any interesting perturbative
theory. Furthermore, as we have argued in Chapter 2, $\eps$ is
indistinguishable from $\varepsilon$, which is the small parameter that
sets the size of the shift vector $\eps\xi$. Hence the parameter
$\veps=\eps$ that regulates the diffeomorphisms also have to be regarded
as a finite number.  This leads us to consider the theory of {\it finite
diffeomorphisms}, an issue that was apparently ignored by most authors in
General Relativity. The formalism can be applied to any relativistic
system, and relies only on the assumption that there is an unambiguous
background spacetime with perturbations.

\section{Finite gauge transformations}

A diffeomorphism of a manifold is a transformation on the map $\Theta_x: P
\in {\cal M} \rightarrow x^\mu \in \Re^4$ of identification of the points
of the manifold. We call it a gauge transformation $\Psi(\lambda)$ if the
initial map $\Theta_x$ can be continuously
deformed\footnote{Transformations such as the one that takes a cartesian
$(x,y,z,t)$ system into a spherical $(r,\theta,\phi,t)$ one are obviously
not included in this category.} into the map $\Theta_{\tx}$ by
taking the affine parameter $\lambda$ away from zero. In terms of the
labels $x$ and $\tx$ we have

\beq
\nonumber
P \stackrel{\Theta_x} \longrightarrow &x^\mu_P& \\
\nonumber
&\downarrow& \Psi(\lambda)\\
\nonumber
P \stackrel{\Theta_{\tx}} \longrightarrow &\tilde{x}^\mu_P& 
\eeq
By the expression above we mean that the same point $P$ of the manifold,
with coordinate label $x_P$ in the original frame, has a label $\tx_P$ in
the tilde (transformed) frame.

We can construct explicit coordinate transformations for every point $P$
of the manifold ${\cal M}$ by considering a congruence of vectors
$\chi^a = [\lambda, \chi^\mu(\lambda)]$
parametrized by the affine parameter $\lambda$ in the following
way\cite{Stewart90}. 
First, fix $\chi^\mu (P; \lambda = 0) = x^{\mu}_P$ and $\chi^\mu (P;
\lambda = \eps) = \tx^{\mu}_P$, that is, set the small number
$\Delta\lambda=\eps$ as determining how ``close'' the two coordinate
systems are along the curves $\chi$ (see Fig. 5.1).  

\begin{figure} 
\centerline{\epsfig{file=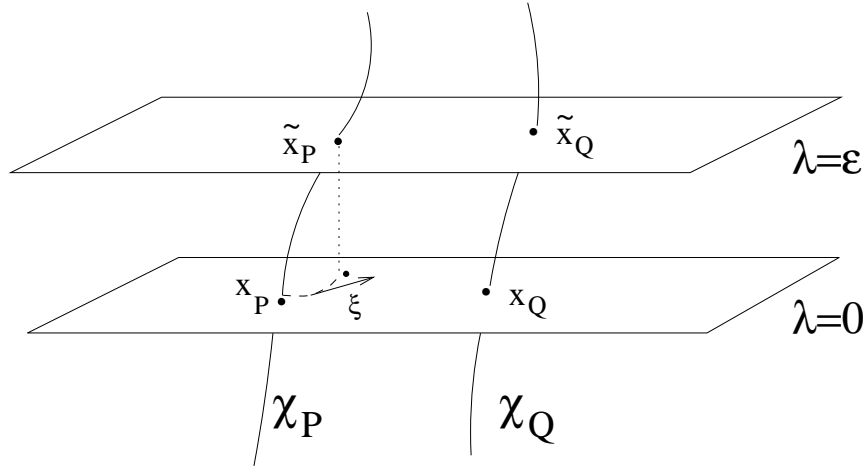,width=4.5in}}
\vspace{10pt}
\caption{\small Two charts, $x$ and $\tx$, related by a diffeomorphism
transformation generated by the congruence of vectors $\chi$. Every point P
of the manifold has a representation $x_P$ and $\tx_P$ in both charts, 
which are linked by the vectors $\chi$. To specify a gauge
transformation is equivalent to
specifying either the congruence $\chi^a$ or its tangent vector $\xi^\mu$ 
in the direction of the hypersurfaces of constant $\lambda$.}
\label{Fig_Fl}
\end{figure}

The affine parameter defines a foliation of charts along $\chi$, and the
tangent vector field along the hypersurfaces of constant $\lambda$ is
given by

\be
\label{Def_Xi}
\frac{d\chi ^\mu (P;\lambda )}{d\lambda } = \xi ^\mu (P;\lambda )
\ee
Now transport the coordinate system $x$ along the curve $\chi$ from
$\lambda=0$ up to $\lambda = \eps$, that is, solve Eq. (\ref{Def_Xi}) 
with initial conditions $\chi^\mu (P; \lambda=0) = x^\mu_P$. The solution
is given in terms of the Taylor series,

\begin{eqnarray}
\label{Flo_Xi}
\nonumber
\tx^\mu_P &=& \chi^\mu (P; \lambda = \eps) 
	= \chi^\mu (P; \lambda = 0) +
	\eps \left. \frac{\partial \chi^\mu}{\partial \lambda} 
	           \right|_{\lambda = 0} +
	\meio \eps^2  \left. \frac{\partial^2\chi^\mu}{\partial \lambda^2} 
		         \right|_{\lambda = 0} + {\cal O}(\eps^3) \\ 
	&=& x^\mu_P + \eps \xi^\mu_P + 
		\meio \eps^2 \xi^\mu_{,\nu} \xi^\nu + {\cal O}(\eps^3) 
\end{eqnarray}
where in the second line we used the identity

\be
\left. \frac{d^2 \chi^\mu }{d\lambda^2} \right|_{P; \lambda=0} 
=	\left( \frac{\partial \xi^\mu}{\partial \chi^\nu} 
	\frac{\partial \chi^\nu}{\partial \lambda} \right)_{P; \lambda=0} 
=	\xi^\mu_{,\nu} \xi^\nu  \quad .
\ee
Equation (\ref{Flo_Xi}) can be written in the compact form

\begin{equation}
\chi_P^\alpha (P;\lambda =\eps)=(e^{\xi ^\beta \frac \partial
	{\partial x^\beta }}x^\alpha )_P \quad ,  
\label{ct2}
\end{equation}
where we have included the small parameter $\eps$ in the definition
of $\xi$. Thus the general coordinate transformation
$x\stackrel{\xi }{\rightarrow } \tilde x$
on ${\cal {M}}$ generated by the vector field $\xi ^\alpha (x)$
can be written as

\begin{equation}
x^\alpha \rightarrow \tilde x^\alpha \, = 
	\, \Psi(\lambda = \eps) \cdot x^\alpha \, = \,
	e^{\xi ^\beta\frac{\partial}{\partial x^\beta }}x^\alpha\quad.
\label{coor} 
\end{equation} 
Conversely, given any two coordinate systems
$x^\alpha $ and $\tilde x ^\alpha $ on ${\cal {M}}$ which are not too
distant, we can find the vector field $\xi ^\alpha (x)$ which generates
the coordinate transformations $ x\rightarrow \tilde x$ in the sense
(\ref{coor}). This class of coordinate transformations is suitable for our
purposes, although it does not cover the more generic family of ``knight
diffeomorphisms" of Bruni {\it et al.}\cite{Bruni}\footnote{The cited
authors use a family of
congruences $\chi_n$, instead of a single congruence $\chi$ to generate
the diffeomorphisms. Since the
order in which the congruences are used to transform the charts does matter
(at least in curved spacetime), the family of diffeomorphism generated in
this manner appears to be more general than the class of diffeomorphism
generated by a single congruence of vectors.}. To our knowledge, the
first to consider gauge transformations to second order was
Taub\cite{Taub}.

The variables which describe physics on the manifold ${\cal {M}}$ are
tensor fields $q$. As discussed in section 2.4, under coordinate
transformations $x\rightarrow \tilde x$ the value of $q$ at the given
point $P$ of the manifold transforms according to Eq. (\ref{Tra_q}),

\begin{equation}
q(x_P)\rightarrow {\tq}(\tx_P)=
{\partial \tilde x \overwithdelims() \partial x}_P \cdots
{\partial x \overwithdelims() \partial \tilde x}_P q(x_P)  
\label{Tra_q_4}
\end{equation}
Note that both sides in this expression refer to the same point of
manifold which has different coordinate values in different coordinate
frames, that is ${\tx}_P\neq x_P.$ The question about the transformation
law for the tensor field $q$ can be formulated in a different way. Namely,
given
two different points ${\cal P}$ and ${\tP}$, which have the same
coordinate values in different coordinate frames, that is
${\tx}_{\tP}=x_{\cal P}$, we could ask how to express the components of
the tensor in the coordinate frame $\tilde x$ at the point $\tP$
(denoted by ${\tq}_{\tP}$) in terms of $q$
and its derivatives given in the frame $x$ at the point ${\cal P}$.
The answer to
this question is found with the help of the Lie derivative [see Eq.
(\ref{Lie_q})] of the tensor $q$ with respect to
the vector field $\xi$ which generates the appropriate coordinate
transformation $x\rightarrow \tilde x$ according to (\ref{coor}) :

\begin{equation}
{\tq}(\tx_{\tP}=x_0) \, = \, [1 - \Lie ] \cdot q (x_P=x_0) + O(\xi ^2)
\quad .
\label{Lie1} 
\end{equation}
We wish now to generalize this expression and write a formula for
the exact (to all orders) diffeomorphism generated by a small but
finite vector field $\xi$.

It is useful to notice that the operator $\Lie$ is the generator of the
algebra of the group of diffeomorphisms, in the same manner as the angular
momentum operators $\hat{J}$ are the generators of rotations in 3-D space. 
A quantum mechanical state $|\psi(\theta)\rangle$ can be rotated by a
finite angle $\Delta\theta$ through the action of the rotation operator
$e^{-i \, \hat{J} \, \Delta\theta } | \psi(\theta) \rangle =
|\psi(\theta+\Delta\theta)\rangle$. Similarly, the operator realising
finite diffeomorphisms can be expressed as an exponential of the
differential operator $\Lie$, where the parameter $\eps=\Delta\lambda$
(included for simplicity in the vector field $\xi$) plays the role of the
angle $\Delta\theta$.  This operator can be found in some
advanced books on differential geometry\cite{Schouten} as well as in some
recent papers on the subject of finite diffeomorphisms\cite{Bruni}. Let us
call this exponential of the Lie derivative the Lie differential operator,

\begin{eqnarray}
q(x) \rightarrow \tq (x) &=& (e^ {-\Lie} \cdot q)(x)  \nonumber \\
	&=&
	q(x)-(\Lie \cdot q)(x) + \meio (\Lie \cdot \Lie \cdot q)(x)
		+{\cal O}(\xi ^3) \quad.
\label{Lie_Ge}
\end{eqnarray}
Equations (\ref{Lie1}) and (\ref{Lie_Ge}) are tensor equations, where for
notational convenience, tensor indices have been omitted. Note that,
in spite of the fact that the transformation law (\ref{Lie_Ge}) is the
consequence of transformation law (\ref{Tra_q_4}), they are different in
the following respect: transformation law (\ref{Tra_q_4}) is well defined
for any tensor given only at the point $P$, while (\ref{Lie_Ge}) is
defined
only for tensor {\it fields.}

With the Lie differential operator we have a simple
prescription for the gauge transformation of all tensor fields
of the manifold. It is valid, in particular, for tensor fields
which are functionals of other tensor fields, e.g. $A_{\mu\nu} [B]
(x) = B_\mu(x) B_\nu(x)$. Under a gauge transformation both $A$ and
$B$ are transformed by the Lie operator, 

\beq
\nonumber
A \, &\longrightarrow& \, \tilde{A} 
	= e^{-\Lie} A \quad ,\\
\nonumber
B \, &\longrightarrow& \, \tilde{B} =
	e^{-\Lie} B \quad .
\eeq
However, the tensor field $\tilde{A}$ is the same functional of $\tilde{B}$
as the tensor field $A$ was a functional of $B$, therefore we conclude that

\be
\label{Lie_Ge_A}
\tilde{A} = e^{-\Lie} A[B] = A[e^{-\Lie}B] \quad .
\ee
Clearly, for this nontrivial identity to be true the Lie derivative has to
obey all the properties of the usual partial derivatives when applied to
tensor fields and functionals. For local tensors obtained by sum
($A_\mu=B_\mu+C_\mu$), multiplication ($A_{\mu\nu} = B_\mu C_\nu$),
contraction ($A_\mu = B^\alpha_{\, \alpha \mu}$) or any combination
thereof, the composition rule (\ref{Lie_Ge_A}) is a consequence of
linearity and of Leibniz's rule,

\beq
\label{Pro_Li_Li}
\Lie (\alpha A + \beta B) &=& 
	\alpha \Lie  A + \beta \Lie  B \quad , \\
\label{Pro_Li_Le}
\Lie (A B) &=& (\Lie  A) B + A (\Lie  B) \quad ,
\eeq
where $\alpha$ and $\beta$ are constants and $A$ and $B$
tensor fields on the manifold. A key consequence of these
properties is that the Lie operator obeys the same composition rule as
differential operators, $d f(g) = \frac{df}{dg} dg$.
Therefore, all local tensor fields which are functionals of other
tensor fields constructed in the ways described above follow the
composition rule,

\be
\label{Pro_Li_Fu}
\Lie  A [B] (x) = \int \, d^4x' \, \frac{\delta A(x)}{\delta B(x')}
			\Lie  B(x') 
		= A_{,m}  (\Lie B)^m (x) \quad .
\ee
(see section 4.2 for notational conventions). For the example
$A_{\mu\nu}=B_\mu B_\nu$ this relation can be easily verified
by using that

\be
\nonumber
\frac{ \delta A_{\mu\nu}(x) }{\delta B_\alpha (x')} = 
	\delta^\alpha_\mu \delta (x-x') B_\nu(x) +
	B_\mu (x) \delta^\alpha_\nu \delta (x-x')  \quad ,
\ee
where $\delta(x-x')$ is the 4-dimensional Dirac delta function.

Now we will show that it follows from the composition law
(\ref{Pro_Li_Fu}) for the 
tensors built by sum, multiplication or contraction of other tensor
fields, that the Lie operator obeys 
equation (\ref{Lie_Ge_A}):

\beq
\nonumber
e^{-\Lie} A[B] &=& 
	\left( 1- \Lie + \meio \Lie \Lie  + \ldots \right) A[B] \\
\nonumber
	&=& A - A_{,m} (\Lie B)^a + \meio A_{,mn} (\Lie B)^m (\Lie B)^n
		+ \meio A_{,m} ({\cal L}_\xi^2 B)^m + \ldots \quad \\ 
\label{Lie_Explicit}
	&=& A[ (1-\Lie + \meio \Lie \Lie + \ldots ) B] = A[ e^{-\Lie}B ] 
	\quad ,
\eeq
where $B$ can be any collection of tensors labeled by the indices $a$.
The two last terms on the second line follow from the identity

\be
\label{Com_2}
\Lie [ A_{,m} (\Lie B)^m] = A_{,mn} (\Lie B)^m (\Lie B)^n + 
A_{,m} (\Lie \Lie B)^m \quad .
\ee
Notice that this last equality follows from the Leibniz rule, since
$\Lie \cdot (A \cdot C) = (\Lie \cdot A) \cdot C + A \cdot (\Lie\cdot C)$ 
and $C=\Lie \cdot B$.
Therefore the composition law to first order together with
Leibniz's rule implies that the composition law to {\it all orders}, Eq.
(\ref{Lie_Explicit}), is satisfied. Since the composition law to
first order is itself a consequence of Leibniz's rule [see
Eq. (\ref{Pro_Li_Fu})], we can say that everything hangs on the Leibniz
rule: if it is valid for a tensor functional, then relation
(\ref{Lie_Explicit}) is also valid.

We wish now to generalize this result for more generic tensor functionals,
in particular the ones built with the help of covariant derivatives, e.g. 
$A^\mu_{\, \nu} = B^\mu_{\,\, ; \nu}$. In order to do that we need to
verify Leibniz's rule for the covariant derivative, that is, we must show
explicitly that the metric tensor transforms in such a way that the
covariant derivative $D_\alpha [\partial/\partial x,
\Gamma^\sigma_{\mu\nu}]$ acting on any tensor field transforms in the same
way as a covariant vector $B_\alpha$ multiplying that tensor field.

As before, we only need to verify this to first order, since the result
for higher orders follows from Leibniz's rule. This long
calculation is done in Appendix B, and the end result, as expected, is
that covariant derivatives indeed obey the rule,

\be
\label{Lei_D}
 \Lie \cdot ( D[\frac{\partial}{\partial x} , \Gamma ] \cdot q) =
	D [\frac{\partial}{\partial x} , \Lie' \Gamma ] \cdot q +
	D [\frac{\partial}{\partial x} , \Gamma] \cdot (\Lie \cdot q)
	\quad ,
\ee
where $(\Lie' \Gamma)^\alpha_{\mu\nu} = (\Lie \Gamma)^\alpha_{\mu\nu} +
\xi^\alpha_{,\mu\nu}$ is the appropriate differential operator acting on
the non-tensorial connection.
This fundamental relation can also be
extended to higher orders, and expressed as

\be
\label{Lie_Cov}
e^{-\Lie} D_\alpha \left[ \frac{\partial}{\partial x}, \gmn \right] = 
	D_\alpha \left[ 
	\frac{\partial}{\partial x}, (e^{-\Lie} g)_{\mu\nu} \right] \quad .
\ee

In sum: the Lie operator passes through the differential operator
$\partial/\partial x$ and acts only on the metric in expressions which
involve a covariant derivative. Since we have proved the composition
relation for tensors which are sums, multiplications, contractions
and covariant derivatives of other tensor fields, the result is
automatically proven for any combination of these operations. In this
manner we cover all the objects which we will be interested in
for the purposes of General Relativity and classical Field Theory.

In particular, the composition rule (\ref{Lie_Cov}) implies that the
Riemann tensor, defined as $R^\alpha_{\beta\mu\nu} v^\beta = [D_\mu,D_\nu]
v^\alpha$ for any vector $v$ transforms in the same way. If we regard the
Riemann tensor as a functional of the metric, we have that

\beq
\label{Riemann}
e^{-\Lie} \{ R^\alpha_{\, \beta\mu\nu}[g] v^\beta \} &=& 
	(e^{-\Lie} R)^\alpha_{\, \beta\mu\nu}[g] (e^{-\Lie} v)^\beta \\
\nonumber
	&=& [ D_\mu (e^{-\Lie} g) , D_\nu(e^{-\Lie} g) ]
		(e^{-\Lie}v)^\beta \\
\nonumber
	&=& R^\alpha_{\, \beta\mu\nu}(e^{-\Lie}g) (e^{-\Lie}v)^\beta
\eeq
from whence we deduce that 

\be
\label{Ope_Riemann}
e^{-\Lie}R[g] = R[e^{-\Lie}g] \quad ,
\ee 
a very nontrivial result for a nonlinear, differential functional 
of the metric such as the Riemann tensor. Of course, contractions of the 
Riemann tensor such as the Ricci curvature and the Ricci scalar clearly
follow the same composition rule (for an explicit
proof of the composition law for the Ricci tensor, see
Appendix B.)

It is clear now that the composition rule applies to all riemannian
tensors, the Ricci tensor, the Einstein tensor, the stress-energy
of a scalar field and so on. Indeed, any of the usual covariant
expressions $A$ which are expressed as a function of other tensors $B$
follow the rule $e^{-\Lie} A[B]=A[e^{-\Lie}B]$. We will se in
the next section that this rule is the base for a covariant
formulation of the back reaction problem.

\section{Finite gauge transformations and general covariance}

Let us consider now what are the implications of the conclusions of
the last section to the problem of back reaction of perturbations
on a given background spacetime. Let us first reformulate the problem,
to include not only vacuum but also matter spacetimes. The Einstein
field equations can be written in the form\footnote{It is a trivial but
sometimes tedious task to verify that these equations can be written
equivalently in the covariant, mixed or contravariant form.}.

\be
\label{Def_Pi}
\Gmn[g] - \Tmn[g,\vv] \, \equiv \, \Pmn[g,\vv] \, = \, 0 \quad ,
\ee
for the exact metric $\gmn$ and the exact matter fields $\vv$.
Let us denote the matter and metric variables collectively as
$q^m$, the background variables by $b^m$ and the perturbations by
$p^m$ ($b$ would be, for example, the homogeneous scalar field $\vv_0$
and the metric $\ymn$ of
Chapter 2, and $p$ the scalar field perturbations $\dv$ and metric
perturbations $\dgmn$, in the case of scalar field matter in a FRW
background). In addition we include the second-order corrections
since the beginning, and denote them by $r^m$. We have then,
with factors of $\eps$ explicit for clarity,

\be
\label{Def_bpr}
q^m(\vec{x},t) = b^m(t) + \eps p^m(\vec{x},t) + \eps^2 r^m \quad ,
\ee
where we are keeping in sight that we ultimately wish to attack the
problem of back reaction in cosmology, where the background variables
depend only on the time coordinate. For the back reaction problem in black
hole physics, the background variables would depend on the radius $r$
from the singularity. The zero mode of the variables $r^m$ will be
regarded as the back reaction corrections $r^m_o(t)$ to the background
$b^m(t)$.

With the perturbative expansion (\ref{Def_bpr}) we can also
expand the Einstein field equations,

\be
\label{EFE_pert}
\Pi [ q^m] = \Pi [ b^m](t) + \eps \Pi_{,m} p^m (\vec{x},t) + 
	\meio \eps^2 \Pi_{,mn} p^m p^n + 
	\eps^2 \Pi_{,m} r^m + {\cal O}(\eps^3) \quad .
\ee
To lowest order, we have the background equations 

\be
\label{Pi_b}
\Pi[b] = G[b]-T[b]= G^{(0)} - T^{(0)} =0 \quad , 
\ee
which in the case of
cosmology can be solved for the
homogeneous background metric (scale factor) and matter fields on
hypersurfaces of constant time.

The next order in perturbation theory is the equations of motion for
the perturbations, 

\be
\label{Pi_p}
\Pi_{,m} p^m = G_{,m}p^m - T_{,m} p^m = G^{(1)} - T^{(1)} = 0 \quad ,
\ee
which, again in the case of cosmology, gives the time
dependence of the $k$-modes (see Chapter 2). It is important to notice
that
the equations (\ref{Pi_p}) do depend on the background solution
obtained from Eqs. (\ref{Pi_b}). This is simply the statement that
perturbations
propagate differently in different background curvatures, which is
obvious but should be stressed in this context.

To second order in perturbation theory we have that the equations
of motion are

\beq
\nonumber
\meio \Pi_{,mn} p^m p^n + \Pi_{,m} r^m &=& 
	 \meio \left( G_{,mn} p^m p^n - T_{,mn} p^m p^n \right) \\
\label{Pi_r}
	&&+ \, G_{,m} r^m - T_{,m} r^m = 0 \quad .
\eeq
Notice that the terms between brackets are simply the EMT for
perturbations defined in the previous chapter.
Notice also that in second order in perturbation theory we require that
both the $0^{\rm th}$ and $1^{\rm st}$-order dynamics have been solved,
that is, that the background and linear perturbations are known
to within a level of accuracy of order $\eps$. This sequence can be
carried to arbitrary orders in perturbation theory, with the caveat
that at each order it is necessary to introduce additional degrees of
freedom for the metric and matter fields.

We will show now that these equations have a covariant meaning, that is,
that a gauge transformation does not change the content of Eq.
(\ref{EFE_pert}). Under a gauge transformation, the fields $q$ transform
as

\be
\label{Gau_q}
q^m \rightarrow \tilde{q}^m = (e^{-\Lie} q)^m  \quad .
\ee
The Einstein equations in the coordinate frame $\tx$ are the same
tensors, written in terms of the metric and matter fields in that
coordinate frame,
\be
\label{Pi_tq}
\tilde{\Pi} = \Pi [ e^{-\Lie}q] = e^{-\Lie} \Pi[q] = 0 \quad ,
\ee
where in the last equality sign we have used the equations of motion
written in the coordinate frame $x$, $\Pi[q]=0$. We also made use of the
``composition rule" derived in the last chapter to pull out the Lie
differential operator. The conclusion is: in both coordinate frames $x$
and $\tx$, the content of the Einstein equations is the same for a given
exact spacetime. In other words, we have verified explicitly the general
covariance of Einstein's General Relativity and found simple formulas
that relate the form of these equations
in each coordinate frame. Since the same reasoning
applies to any equations written in terms of tensors on the manifold, this
result applies to any covariant theory with or without matter, such as
higher derivative ($R^2$) theories of gravity, scalar-tensor theories,
dilaton gravity etc.

To make matters more clear, let us verify explicitly that, to second order,
the Einstein equations (\ref{EFE_pert}) are a covariant set of equations.
The gauge transformation of the field variables
$q$ is, expanding (\ref{Gau_q}),

\be
\label{Gau_q_pert}
(e^{-\Lie} q)^m = \left( 1-\Lie+\meio \Lie^2 \right) 
	(b + p + r)^m + {\cal O}(\eps^3) \quad ,
\ee
where we assume that $\xi$ and $p$ carry a factor of $\eps$ and
$r$ carries a factor of $\eps^2$. Moreover, we assume that, for our
purposes at least, the vector $\xi$ averages
to zero (this will be explained shortly).
Sorting the above equation by orders
of $\eps$ we have

\beq
\label{Gau_b}
b^m &\rightarrow& \tilde{b}^m = b^m \\
\label{Gau_p}
p^m &\rightarrow& \tilde{p}^m = p^m - (\Lie b)^m \\
\label{Gau_r}
r^m &\rightarrow& \tilde{r}^m = r^m - (\Lie p)^m + \meio (\Lie^2 b)^m 
	\quad .
\eeq
Notice that the background metric and matter fields $b(t)$ are unchanged
by the gauge transformation (this is actually one of the basic assumptions
- see Chapter 2). The perturbations $p(\vec{x},t)$ are transformed in the
usual way, through the Lie derivative of the background.
Since one of the key assumptions is that the linear perturbations average
to zero in all
coordinate systems that we wish to consider (see, again, Chapter 2), it is
clear
that the vector $\xi$ also averages to zero, $\langle \xi \rangle = 0$. 
The novelty comes from Eq. (\ref{Gau_r}):
the second-order variables $r$, which
will be used to
describe the feedback from the perturbations, are changed in a nontrivial
manner that involves Lie derivatives of both the perturbation and the
background variables.

It is obvious that the zero and first order equations are covariant (see,
e.g., Section 2.5). Let us see how the second-order equations of motion
are rewritten. In the transformed reference frame we have, to order
$\eps^3$,

\beq
\nonumber
\meio \Pi_{,mn} \tp^m \tp^n &+& \Pi_{,m} \tr^m \\
\nonumber
&=& 
	\meio \Pi_{,mn} (p^m - \Lie b^m) (p^n - \Lie b^n) + 
	\Pi_{,m} \left( r^m - \Lie p^m + \meio \Lie^2 b^m \right) \\
\nonumber
	&=& 
	\meio \Pi_{,mn} p^m p^n + \Pi_{,m} r^m  -
		\Lie ( \Pi_{,m} p^m ) + \Pi_{,m} (\Lie p)^m \\
\nonumber
	&& + \meio \Lie [ \Pi_{,m} (\Lie b)^m ] -
		\meio \Pi_{,m} (\Lie^2 b)^m - 
	\Pi_{,m} (\Lie p)^m + \meio \Pi_{,m} (\Lie^2 b)^m  \\
\label{Pi_r_tr}
	&=& 
	\meio \Pi_{,mn} p^m p^n + \Pi_{,m} r^m  = 0 \quad,
\eeq
where we have used the equations of motion for the background
$\Lie \Pi[b]= \Pi[b] = 0$
and for the perturbations $\Pi_{,m} p^m=0$ to cancel some terms. We have
also used integration by parts, e.g. 
$\Lie \Pi_{,m}p^m = \Pi_{,mn}p^m (\Lie b)^n + \Pi_{,m} (\Lie p)^m$ . 
We see then that the content of the equations in one reference frame,
(\ref{Pi_r}), is the same as the equation in another frame,
(\ref{Pi_r_tr}). This completes the proof of general covariance of the
back reaction equations.

\section{Background variables, averaging and covariance}

So far we have only established that the general problem of feedback of
perturbations is well defined, at least before any average is taken. As we
showed in the last chapter, averages are fundamental to the problem of
back reaction. However, it is conceivable that the average might break the
covariance of the problem, since it might pick a preferred time-slicing.

Equation (\ref{Pi_r}), as it stands, involves all quadratic
nonlinearities, from the back reaction on the homogeneous background to
the second-order corrections to linear perturbations coming from mode-mode
($kk'$) couplings. Although the subject of second order perturbations is a
very interesting one (see, e.g., Ref. \cite{Matarrese}), it is not covered
here. We are interested instead in how the perturbations can affect the
so-called zero modes (that is, the variables that describe the homogeneous
component of the spacetime.)

The second order perturbations can be separated into two categories:  the
zero modes $r_o^m$, which do not vanish under spatial average, and the
inhomogeneous k-modes $r_k^m$ (k comoving), which can actually be regarded
as corrections to the perturbative modes $p_k^m$. We are not interested in
the modes $r_k^m$, as we said in the last paragraph, so we concentrate on
the homogeneous modes $r_o^m(t)=\langle r^m \rangle$. We want to write
down the equation for the zero modes. Clearly, an average performed over
(\ref{Pi_r}) will pick all the homogeneous contributions, and that should
be regarded as the dynamical equation for the zero mode,

\be
\label{Pi_r_av}
	\langle \Pi_{,m} r^m \rangle = \Pi_{,m} r^m_o \, = \,  
		- \meio \langle \Pi_{,mn} p^m p^n \rangle = \tau
		\quad , 
\ee
where we have retrieved the notation that the term on the right is the
{\it effective} (averaged) energy-momentum tensor (EEMT) of the
perturbations $p^m$. We also used the fact that $\Pi_{,m}r^m$ on the LHS 
is linear in $r^m$, therefore the average of everything but the zero modes
vanish.

In the literature, the terms on the left of Eq. (\ref{Pi_r_av}) are
usually taken as simply the background equations, so that
$G^{(0)}_{\mu\nu} = T^{(0)}_{\mu\nu} + \tau^{(2)}_{\mu\nu}$ and the EEMT
can be interpreted as a homogeneous stress-energy contribution from the
perturbations. However, when solving for back reaction using these
equations we assume that the background has already been solved and that
back reaction will at most contribute with small deviations from this
background. Therefore even in this simplified version the assumption is
that we will eventually solve for the back reaction equations
perturbatively, i.e. as in the format of Eq.  (\ref{Pi_r_av}).
Furthermore, we have no {\it a priori} reason to believe that the back
reaction variables $r_o^m$ have the same symmetry structure and degrees of
freedom as the background variables $b_o^m$. The typical example would be
a de Sitter spacetime background with an EEMT that breaks time translation
invariance.  The perturbative format of the back reaction equations is
then the most useful and well defined.

Notice that gauge transformations have a surprising effect on the zero
modes $r_o$, which can be retrieved from formula (\ref{Gau_r}),

\be
\label{Gau_r0}
\tr_o^m = r_o^m - \langle (\Lie p)^m \rangle + 
	\meio \langle (\Lie^2 b)^m \rangle \quad ,
\ee
where the average picks the zero modes from
the second and third terms in the formula. This is a fundamental
change with respect to previous treatments of back reaction: the 
realization that quantities with respect to the homogeneous mode have
different expressions in different coordinate systems. The
$r_o^m$ are in fact corrections to the background variables $b_o^m(t)$, 
hence an accurate description of the background is attained only with
$b_o^m(t) + (\eps^2) r_o^m(t)$. If the $r_o$'s change under a gauge
transformation, it means that the background itself is redefined at second
order by a gauge transformation,

\be
\label{Gau_back}
b^m(t) + r_o^m(t) \quad \longrightarrow \quad b^m(t) +
\tr_o^m(t) \quad .
\ee
Even if $r_o^m=0$ initially, after a finite gauge transformation
it is no longer zero.

An interesting example is the case of a coordinate frame
${x}$ the observers of which see the universe as an exactly
homogeneous spacetime, $p^m=r^m=\ldots =0$. Suppose now
that we perform a gauge transformation, that is, change the set
of basic observers at $x$ for whom the universe is perfectly
homogeneous, to a set of observers at $\bar{x}^\mu = {x}^\mu +
\zeta^\mu$ who now see ``gauged" perturbations

\be
\label{Per_gauge}
\bar{p}^m = p^m 
- ({\cal L}_\zeta b)^m =  - ({\cal L}_\zeta b)^m \quad ,
\ee
and $b^m$ are the homogeneous background variables measured by the
observers at $x$. Now, do these ``gauged" perturbations have an effect
on the (inhomogeneous) background in which they propagate? If the
``true" spacetime does not have any perturbations, how could there
be any back reaction at all?

Consider the back reaction equations in
each reference frame. In the original frame there are no
back reaction equations, since the variables are all zero,
except the background ones. In the barred frame, though, the 
equations are

\be
\label{BR_bx}
\Pi_{,m} \bar{r}^m = - \meio \Pi_{,mn} \bar{p}^m \bar{p}^n = 0 \quad
\Longrightarrow \quad \bar{r}^m \neq 0 \quad .
\ee
From this equation we find that there must be a non-vanishing 
back reaction effect in the barred frame. In fact, it is easy to see from
formula (\ref{Gau_r}) that the back reaction in the barred frame is given
by $r^m = \meio ({\cal L}_\zeta^2 b)^m $. This expression
indeed solves exactly the back reaction equations of motion,

\be
\label{Sol_BR}
\Pi_{,m} \meio ({\cal L}_\zeta^2 b)^m  = 
	-\meio \Pi_{,mn} ({\cal L}_\zeta b)^m ({\cal L}_\zeta b)^n
	\quad .
\ee
That the equation above is an identity can be seen by integrating the Lie
derivative by parts and using the equations of motion for the background,
$\Pi_{,m} ({\cal L}_\zeta b)^m = {\cal L}_\zeta \Pi[b] = 0$.

Of course, if we did not know that the LHS of the back reaction equations
change under a gauge transformation, we could never make sense of the fact
that the EEMT assumes different values on different reference frames.  The
back reaction variables $\bar{r}^m = \meio ({\cal L}_\zeta^2 b)^m $ are in
fact indispensable to guarantee that the perturbed spacetime seen by
observers in the frame $\bar{x}$ corresponds to a homogeneous spacetime in
the original frame $x$ at all later times.

It can be checked that the Einstein equations in each perturbative order
are zero identically (i.e., they are void of dynamical content), provided
that the referential spacetime was perfectly homogeneous. We could also
ask the converse question, namely: given a set of perturbative solutions
to Einstein's equations to all orders, is it possible to find a
homogeneous spacetime corresponding to it?  Moreover, is this homogeneous
spacetime unique (up to a trivial time reparametrization), and what is its
significance? We are not going to answer these broad questions here (see
\cite{Carfora}), although by the end of this chapter the problems
connected with the formulation of the back reaction equations will be
clarified.

In this spirit, let us verify now that the back reaction equations for
the zero mode are covariant. This is not a trivial matter, as it might
appear at first sight. We want to prove that 

\be
\label{Pi_rr}
\Pi_{,m} r_o^m = - \meio \langle \Pi_{,mn} p^m p^n \rangle 
\quad \stackrel{?}\Longrightarrow \quad 
\Pi_{,m} \tr_o^m = - \meio \langle \Pi_{,mn} \tp^m \tp^n \rangle  \quad ,
\ee
with $\tr_o^m$ given by formula (\ref{Gau_r0}) and $p^m$ by (\ref{Gau_p}).
Substituting these expressions into the equation on the right of
(\ref{Pi_rr}) we have

\beq
\label{Pi_rr2}
 \Pi_{,m} \tr_o^m &+& 
	\meio \langle \Pi_{,mn} \tp^m \tp^n \rangle =
	\Pi_{,m} r_o^m + 
	\meio \langle \Pi_{,mn} p^m p^n \rangle \\
\nonumber
	&-& 
	\Pi_{,m} \langle \Lie p \rangle^m +
	\meio \Pi_{,m}  \langle \Lie^2 b \rangle^m +
	\langle \Pi_{,m} (\Lie p)^m \rangle -
	\meio \langle \Pi_{,m}  (\Lie^2 b)^m \rangle
\eeq
where, as in Eq. (\ref{Pi_r_tr}), we have integrated Lie derivatives
by parts and used the equations of motion for the background and
perturbations.

We assume now that the averaging does not introduce any extra time
dependence, i.e. $d/dt \langle f \rangle = \langle df/dt \rangle$. 
In that case the terms in the second line cancel, because the differential
operator $\Pi_{,m}$ is {\it linear} on its arguments, and therefore the
averaging operation commutes with the action of $\Pi_{,m}$ over $(\Lie
p)^m$ and $(\Lie^2 b)^m$. The fact that the differential operator
$\Pi_{,m}$ has the laplacian built on it is irrelevant, 
since we are interested in the homogeneous modes whose spatial
derivatives are zero, and the inhomogeneous
parts $f_k$ with $k\neq 0$ vanish under average. In other
words, as long as the averaging does not alter the time dependence,
the zero modes of the linear operator $\Pi_{,m}$ evaluated on a function
are the same as the operator evaluated on the zero modes of the function,
$\langle \Pi_{,m} f^m \rangle = \Pi_{,m} \langle f^m \rangle$. 
We have then

\be
\label{Pi_r0_tr}
	\Pi_{,m} \tr_o^m + 
	\meio \langle \Pi_{,mn} \tp^m \tp^n \rangle = 
	\Pi_{,m} r_o^m + 
	\meio \langle \Pi_{,mn} p^m p^n \rangle = 0 \quad .
\ee

It is easy to imagine examples of averaging procedures which
do introduce ``extra" time dependencies. Take, for example, an
average over the domain ${\cal V}(t)$ between the spherical shells at
comoving radii $R_{UV}(t) = e^{-Ht}H^{-1}$ and $R_{IR} = R_i$ constant.
(See Fig. 4.2). The small radius $R_{UV}(t)$ corresponds to
the Hubble horizon of a de Sitter spacetime in comoving coordinates
$R=r/a(t)$, and the
large radius $R_{IR}$ is the comoving radius of some initial bubble,
as defined in the last chapter.
This average corresponds to a
set of perturbations whose spectrum has both ultraviolet (UV) and
infrared (IR) comoving
cut-offs $k_{UV}(t)=R_{UV}^{-1}(t)$ and $k_{IR}= R_{IR}^{-1}$, so we write

\be
\label{Exa_co_ti}
\langle \phi^2 \rangle_{{\cal V}} (t) \, = \, \int_{k_{IR}}^{k_{UV}(t)} \, 
		\frac{dk}{k} \, \left| \delta_k^\phi (t) \right|^2 \quad ,
\ee
where $\delta_k^\phi(t)$ is the power spectrum of the perturbation
$\phi_k(t)$ (see Section 3.3). Clearly, the averaging does not commute
with a time derivative, because the upper limit of the integral is
time dependent. Of course, this time dependence can be very small
in some cases, so that in perturbation theory we can ignore it. This
is indeed the situation that we will encounter when we analyse the
back reaction of scalar perturbations in the chaotic inflation scenario
(see Chapter 7.)

\begin{figure} 
\centerline{\epsfig{file=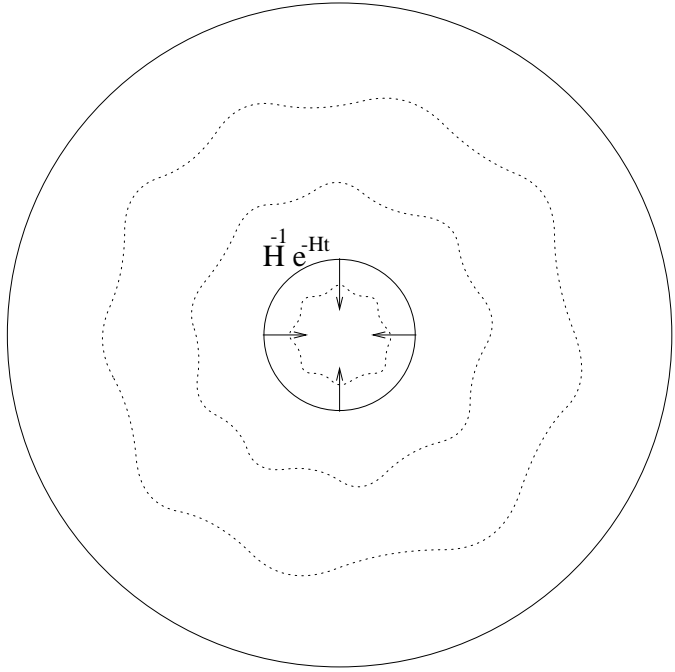,height=3.5in}}
\vspace{10pt}
\caption{\small Perturbations generated during an inflationary epoch exit
the Hubble horizon $H^{-1}/a(t)$ at all times. The number of comoving
momenta included in the spectrum of the ``classical" perturbations are
therefore increasing with time, for example $k_{UV} \propto e^{Ht}$ during
inflation.}
\label{FIG_Cr_Pe}
\end{figure}

However, let us argue that an averaging procedure which introduces a
non-negligible time dependence is not acceptable for reasons other than
the fact that it does not yield a covariant back reaction problem. The
reasoning is best followed in momentum space. Take the expression
(\ref{Exa_co_ti}), in which the UV cut-off is time dependent. The
integration sums up the contributions from all $k$-modes, from the 
longest comoving wavelength $k_{IR}^{-1}$ to the shortest wavelength
$k_{UV}^{-1}$. The fact that the UV cut-off is time dependent means,
though, that the average at a time $t+\Delta t$ will include more
modes than the average at time $t$. In other words, at later times
more (short-wavelength) modes have been ``created" in the calculation.
Where have they come from?

In inflation models, these modes have just exited the horizon, and
from that point on we regard them as classical perturbations, given that
the quantum fluctuations for those scales have a very high occupancy
number. For the purposes of the classical calculations, what that means
is that at every moment $t$ during inflation there are modes being
created, with comoving wavelengths with the size of the horizon scale,
$H^{-1}(t) / a(t)$ (see Fig. 4.2).
Hence, there is effectively a source added to the stress-energy tensor
that creates the small-scale perturbations, and this source is not
conserved in the same sense that the stress-energy tensor is conserved. In
the case of inflation, we will show later that the creation rate is very
small and can be neglected in the conservation equations.

If this source is too strong and effectively destroys the conservation
equations, some more fundamental problems appear. Mainly, if the
stress-energy tensor is not conserved anymore, as a consequence the
Bianchi identities are not satisfied and it is not even clear if
Einstein's equations can be trusted anymore. Clearly,
we would like to avoid such a radical departure from standard physical
practice, so we concentrate on situations where the conservation equations
are obeyed.

Summarizing, we found that the second order equations for the zero modes
(the so called back reaction of perturbations on the background) can be
formulated in a covariant manner. This was done with the help of the
theory of finite gauge transformations, and a crucial ingredient is that
the zero mode variables $r_o^m$ are rescaled by a gauge transformation. We
found that the long-held assumption that the background is unchanged by a
gauge transformation does not hold to second order.

In this context, the paradox that the EEMT is changed under a gauge
transformation could be resolved: if on the one hand the EEMT changes, on
the other hand the zero modes that describe back reaction also change, in
just the right way such that the problem of back reaction is well defined.

\section{Gauge invariant variables to higher orders}

We now want to go beyond this interesting but standard covariant
construction, and try to build variables that can describe back reaction
in a gauge {\it invariant} way. We saw in Section 2.5 that it is
possible to build gauge invariant variables that describe perturbations,

\be
\label{Def_ga_inv_var_1}
P^m \equiv p^m + (\LieX b)^m \quad ,
\ee
where the vector $X$ was suitably defined in terms of its
transformation laws,

\be
\label{Gau_tr_X}
X^\mu \stackrel{\xi}\longrightarrow \tX^\mu = X^\mu +
	\xi^\mu \quad .
\ee
We found earlier that we can express $X$ in terms of the metric
perturbations in a generic gauge, and that it is also a ``small" vector
$\propto \eps$, in the same sense that $\xi$ and the perturbations are
small. Moreover, by fixing
the gauge such that $X=0$ by taking $B=E=0$, the gauge invariant variables
$P^m$ would take the same value as the variables in longitudinal gauge,
$p^m_l$.

This procedure can be extended to second and higher orders, now that
we know the gauge transformation laws to arbitrary orders.
Inspired by the results in first order in the perturbations, consider
the following candidates for gauge invariant variables:

\be
\label{Def_Q_2}
Q^m \, = \, (e^{\LieZ} q)^m \quad ,
\ee
where the Lie differential operator is formally defined as before, but
evaluated with $\xi = -Z$. In a transformed coordinate frame $\tx$, $Q^m$
would be defined by  $\tQ = e^{ {\cal L}_{\tZ} } \tq $.

We want the $Q$'s to be gauge invariant, that is, their expressions must
be equal in both reference frames, $\tQ^m = Q^m$. Using the ansatz above
we have

\be
\label{LieZ_Q}
\tQ = e^{ {\cal L}_{\tZ} } \tq =  
	e^{ {\cal L}_{\tZ} } e^{-\Lie} q =  e^{\LieZ} q = Q  \quad .
\ee
The Lie derivatives cannot be summed in the exponential because,
due to their different arguments ($Z$ and $\xi$), they do not commute.
We can use the Campbell-Baker-Hausdorff formula for the exponential
of an operator and get the conditions on the vector $Z$,

\be
\label{Con_Z}
\tZ^\mu = Z^\mu + \xi^\mu + \meio [ Z,\xi]^\mu + {\cal O} (\eps^3) \quad ,
\ee
where $[Z,\xi]^\mu = Z^\mu_{\,\, ,\nu}\xi^\nu - 
Z^\nu \xi^\mu_{\,\, ,\nu}$ is the commutator of the vectors $Z$ and $\xi$.
To derive this formula, it is useful to note that the Lie derivative of a
contravariant vector is given by the commutator of $\xi$ with that vector,
$(\Lie A)^\mu = [A,\xi]^\mu$.

If we ignore the third term on the RHS of (\ref{Con_Z}) we retrieve the
definition of $X$ used in the construction of the gauge invariant
variables to first order (see Section 2.5).
The commutator gives a correction term of order $\eps^2$, which
must be incorporated into the definition of $Z$ if we want to define gauge
invariant variables to second order. Higher order terms like that must
also be included as we proceed in perturbation theory. To second order
we have, perturbatively, $Z = \eps X + \eps^2 Y + {\cal O}(\eps^3)$
and their transformation laws should be

\beq
\label{Gau_X}
	\tX[\tp] &=& X [p] + \xi \quad \quad \quad , \\
\label{Gau_Y}
	\tY [\tr,\tp^2] &=& Y[r,p^2] + \meio [X,\xi] \quad ,
\eeq
where by $Y[r,p^2]$ we mean that the vector $Y$ is
in general a linear function of $r$ and $p^2$,
by the same token as $X$ is a linear function of the $p$'s.

The gauge invariant variables for perturbations and background to
second order are defined then by the formulas

\beq
\label{GT_pe}
P^m &=& p^m + (\LieX b)^m \quad \quad \quad , \\
\label{GT_ba}
R^m_o &=& r^m_o + \langle (\LieX p)^m \rangle + \langle (\LieY b)^m
\rangle + \meio \langle (\LieX^2 b)^m \rangle \quad .
\eeq

We still should determine what the second-order vector $Y^\mu$ is.
The equations that determine its expression, though, are
very complicated, and there may not even be an answer that is local
on the variables $r^m$ and $p^m$. It would be important to find an
explicit expression for $Y$, but we may be able to set it to zero.

The argument is the following. Consider the relationship between gauge
invariant linear perturbations and the perturbation variables of a fixed
gauge. To linear order, we saw that there is an
infinite number of gauge invariant variables, corresponding to
the infinite number of ways of choosing the vector $X$. Each set
of gauge invariant variables is equal to the perturbation variables in one
specific gauge, for example in longitudinal gauge $\phi_l = \Phi^{(gi)}$
and $\psi_l = \Psi^{(gi)}$, where the variables in the right-hand-sides
are the gauge invariant variables found through the use of the
vector $X_l^\mu = [B_l-E_l',-\nabla^i E_{l}] = 0$. The only requirement
is that the constraints on the metric ($B_l=E_l=0$ for longitudinal gauge)
remove all the gauge freedom and force the shift vector $\xi=0$. 
As long as a gauge is completely fixed, there are gauge invariant
variables corresponding to the perturbations in that gauge.

With respect to the background variables $r_o$ the same rationale is 
applicable: the variables in any well-defined gauge are equal to some set
of gauge invariant variables. If the constraints on the metric
perturbations demand that $\xi$ is zero to first order, all we need are
second order conditions on the $r^m$ that fix $\xi$ to an accuracy of
order $\eps^2$. In other words, we must impose
constraints on the second-order quantities $r^m$ such that no variations
of $\xi$ of order $\eps^2$ are allowed. Let us split the shift vector in
the form $\xi^\mu = \eps \zeta^\mu + \eps^2 \sigma^\mu$, so that the
first-order constraints should fix $\zeta^\mu=0$ and the second-order
constraints should fix $\sigma^\mu=0$.

Let us assume that the first order constraints have been imposed,
so we only have to worry about the second-order ones.
The background variables $r_o^m(t)$ change, under a gauge transformation
with the shift vector $\sigma^\mu$, as

\be
\label{Gau_r0_si}
\tr_o^m (t) = r_o^m(t) - (\Lies b)^m \quad .
\ee
Call the second-order metric perturbations $j_{\mu\nu}$, that
is, $\gmn = \ymn(\eta) + \eps\hmn(\vec{x},\eta) + \eps^2 j_{\mu\nu} (t)$
where $\ymn$ is given in Eq. (\ref{Met_Ba_Co}) and the perturbations
are the usual ones (of the scalar, tensor and vector kinds). The most
general form of $j$ that we can consider is,

\be
\label{Def_j}
j_{\mu\nu} = a^2(\eta) \left( \begin{array}{ll}
                        2z(t)   & v_i(t) \\
                        v_j(t)  & -2w(t)\delta_{ij}
                \end{array}
       \right) \quad ,
\ee
in which case the gauge transformation laws with $\sigma$ give

\beq
\label{Tra_z}
\tilde{z}(t) &=& z(t) - \frac{1}{a} [a\sigma^0]' \quad , \\
\label{Tra_v}
\tilde{v}_i(t) &=& v_i(t) -  \sigma^0_{,i} + {\sigma_i}' \quad , \\
\label{Tra_w}
\tilde{w}(t)\delta_{ij} &=&w(t)\delta_{ij} + 
	\frac{a'}{a} \sigma^0 \delta_{ij} + 
	\meio (\sigma_{i,j}+\sigma_{j,i}) \quad .
\eeq
If we demand that both sides of the equations are homogeneous
functions, from (\ref{Tra_z}) we immediately see that $\sigma^0$ must be
a function only of time. Using this in Eq. (\ref{Tra_v})
we deduce that $\sigma^i$ should also be a homogeneous function,
$\sigma_i(t)$. Inserting this into Eq. (\ref{Tra_w}) we see that the
gradient terms $\sigma_{i,j}$ drop out, and the gauge transformations are
finally

\beq
\label{Tra_z2}
\tilde{z}(t) &=& z(t) - \frac{1}{a} [a\sigma^0]' \quad , \\
\label{Tra_v2}
\tilde{v}_i(t) &=& v_i(t) + {\sigma_i}' \quad , \\
\label{Tra_w2}
\tilde{w}(t) &=&w(t) + 
	\frac{a'}{a} \sigma^0 \quad .
\eeq

We can now fix the gauge to second order, for example by demanding
that $w=v_i=0$. These constraints fix the homogeneous functions
$\sigma^0(t)$ to zero and $\sigma_i(t)$ to an irrelevant
time-independent constant.

The conclusion is, as long as we fix the gauge in first and second orders,
the problem of back reaction will be well defined, since there are
gauge invariant variables corresponding to the perturbations
$p^m(\vec{x},t)$ and $r_0^m(t)$. In addition, we know exactly how
these quantities transform when we go to a different gauge,
hence we can compare the dynamics of back reaction in one gauge
with the dynamics in a different gauge. We have shown that the results
are identical when it is gauge invariant quantities the ones which
are compared.

From now on we only use gauge invariant quantities in the treatment
of the problem of back reaction. We have then,

\be
\label{BR_gi}
\Pi_{,m} R^m_o = -\meio \langle \Pi_{,mn} P^m P^n \rangle \quad ,
\ee
where the variables $R^m$ were defined in Eq. (\ref{GT_ba}) and
$P^m$ in (\ref{GT_pe}). The RHS of this equation is the
gauge invariant effective energy-momentum tensor.


	\chapter{The Back Reaction of Gravitational Waves}
 
%
%

As a first application of the ideas developed in the previous
chapters, let us focus on a simple example: tensor perturbations
in cosmology, or, gravity waves in FRW spacetimes. We will
treat the case of scalar perturbations in Chapter 7.

We can treat gravity waves and their back reaction independently of
the effect from scalar perturbations because the physical processes that
give rise to gravity waves and to scalar perturbations are 
disconnected: while scalar perturbations couple to energy density
and pressure perturbations, gravity waves evolve in a completely
independent way from the matter distribution. Therefore, the two
types of perturbations are not correlated, that is, the 
average $\langle \phi h \rangle=0$, where $\phi$ is the newtonian
potential and $h$ is the strength of the gravitational waves.
Therefore we have two EEMT's, one for gravitational waves and one
for scalar perturbations. The back reaction equations (\ref{BR_gi}) are
linear on the EEMT's, therefore the total back reaction $R_o^m$ is
just the sum of the partial back reactions from the tensor modes,
$R^{m\, (T)}_o$ and from scalar perturbations, $R^{m\, (S)}_o$.

\section{Energy-momentum tensor for gravitational waves}

The metric for tensor perturbations was given by Eq. (\ref{Met_Pe_Te}),

\be
\label{Met_GW}
ds^2 = dt^2 - a^2(t)(\delta _{ik}+h_{ik})dx^idx^k \quad ,
\ee 
where $h_{ij}$ obeys the constraints $h^i_{\, i}=h^i_{\,j,i}=0$.  It was
shown in Section 2.4 that these quantities are unchanged under gauge
transformations, that is, the variables $h_{ij}$ are automatically gauge
invariant.

The starting point to derive the EEMT of gravity waves is
the second variation of the Ricci tensor [see, e.g., \cite{MTW}],

\begin{eqnarray}
\nonumber	
	{\frac 12}R_{\mu \nu ,mn}\,\delta g^m\,\delta g^n 
&=& 	{\frac 12} \left[ {\frac 12}\,\delta g_{|\mu }^{\alpha \beta }\,
	\delta g_{\alpha \beta |\nu } \right.\\
\nonumber
	&+& \delta g^{\alpha \beta }\left( \delta g_{\alpha\beta |\mu \nu}
	+ \delta g_{\mu \nu |\alpha \beta }-\delta g_{\alpha \mu |\nu\beta}
	- \delta g_{\alpha \nu |\mu \beta }\right) \\
\label{Ricci_2} 
	&+& \delta g_\nu ^{\alpha |\beta }\left( \delta g_{\alpha \mu 
	|\beta} 
	-\delta g_{\beta \mu |\alpha }\right) \\
\nonumber
	&-& \left. \left( \delta g_{|\beta }^{\alpha \beta }-{\frac 12} \,
	\delta g^{|\alpha }\right) \left( \delta g_{\alpha \mu ;\nu }
	+ \delta g_{\alpha\nu|\mu }-\delta g_{\mu \nu |\alpha } \right) 
	\right] \quad ,
\nonumber
\end{eqnarray}
where $|$ denotes covariant derivatives with respect to the
background metric.

It is also useful to recall the equations of motion for the
gravity waves in normal time,

\be 
\label{EOM_GW}
\ddot{h}_{ij} + 3 H \dot{h}_{ij} - \frac{1}{a^2} \nabla^2 h_{ij} = 0
\quad,
\ee
which can be used to simplify the expressions for the Ricci and
Einstein's tensor containing gravity waves. After performing the
spatial averaging, we get

\begin{equation}
\label{Tau_GW_00}
	\tau _{00}= {\frac{\stackrel{.}{a}}a}
	\langle \stackrel{.}{h}_{kl}h_{kl } \rangle 
	+ {\frac 18}\left( 
	\langle \stackrel{.}{h}_{kl }\stackrel{.}{h}_{kl} \rangle
	+ \frac {1}{a^2} \langle h_{kl ,m}h_{kl ,m} \rangle \right)  
\end{equation}  
and

\begin{eqnarray}
\label{Tau_GW_ij}
\tau_{ij} &=&\delta _{ij}a^2\left\{ {\frac 3{8a^2}}
	\langle h_{kl,m}h_{kl ,m}\rangle
	-{\frac 38}
	\langle \stackrel{.}{h}_{kl }\stackrel{.}{h}_{kl } \rangle 
	\right\}
\nonumber \\
&+& 
	\frac 12a^2 \langle \stackrel{.}{h}_{ik}\stackrel{.}{h}_{kj} \rangle 
	+ \frac 14 \langle h_{kl,i}h_{kl ,j} \rangle
	- {\frac 12} \langle h_{ik,l }h_{jk,l } \rangle \quad .  
\end{eqnarray}
The first expression can be interpreted as the effective energy density of
gravitational waves,

\begin{equation}
\nonumber
\rho_{gw}=\tau_{\,\, 0}^0 \quad .
\end{equation}
We note that, only in the case of the EMT of gravity waves,
$\tau^\mu_{\,\, \nu} = \gamma^{\mu\sigma} \tau_{\sigma \nu}$ where
$\ymn$ is the background metric (the relationship is more complicated when
the EMT of the matter perturbations is nonzero - see Chapter 7.)
This result also coincides with the result obtained through Landau's
pseudotensor of energy and momentum \cite{Landau} in the gauge
specified in formula (\ref{Met_Pe_Te}).

The relation between $\tau_{ij}$ and the quantity which we could naturally
interpret as an effective pressure is not so straightforward as it looks
at first glance. The problem is that the energy momentum tensor for the
gravity waves $\tau^\mu_{\,\, \nu}$ is not conserved itself, that is,
$\tau^\mu_{\,\, \nu | \mu } \neq 0.$ This is not surprising since the
gravitational perturbations ``interact'' with the background and only the
total EMT must be conserved. Consider the exact Bianchi identities for the
metric perturbations $\gmn=\gamma_{\mu\nu} +P_{\mu\nu}+R_{\mu\nu}$:
\beq
\label{Exp_Bi}
G^\mu_{\, \nu;\mu} &=& \partial_\mu 
	\left( G^\mu_{\, \nu}[\gamma] 
		+ G^\mu_{\, \nu , m} P^m + 
		+ G^\mu_{\, \nu , m} R^m + 
		\meio G^\mu_{\, \nu,mn}P^m P^n  + \ldots \right) \\
\nonumber 
	&-& \left( 
	\Gamma^\alpha_{\mu\nu}[\gamma] +
	\Gamma^\alpha_{\mu\nu,m} P^m +
	\Gamma^\alpha_{\mu\nu,m} R^m +
	\meio \Gamma^\alpha_{\mu\nu,mn} P^m P^n + \ldots \right) \times \\
\nonumber	
	&& \quad\quad \left(
	G^\mu_{\, \alpha}[\gamma] +
	G^\mu_{\, \alpha,m}P^m +
	G^\mu_{\, \alpha,m}R^m +
	\meio G^\mu_{\, \alpha,mn} P^m P^n + \ldots \right) \\
\nonumber
	&-& \left( 
	\Gamma^\mu_{\alpha\nu}[\gamma] +
	\Gamma^\mu_{\alpha\nu,m} P^m +
	\Gamma^\mu_{\alpha\nu,m} R^m +
	\meio \Gamma^\mu_{\alpha\nu,mn} P^m P^n + \ldots \right) \times \\
\nonumber
	&& \quad\quad \left( G^\alpha_{\, \mu}[\gamma]+
	G^\alpha_{\, \mu,m} P^m + 
	G^\alpha_{\, \mu,m} R^m + 
	\meio G^\alpha_{\, \mu,mn} P^m P^n + \ldots \right) \quad ,
\eeq 
where we have expanded the connections $\Gamma$ and the Einstein
tensor $G$ up to second order in the perturbations $P$ and $R$.

These equations are true in each perturbative order, as it should
since the Bianchi identities are geometric identities. We can
greatly simplify these equations by decoupling the several classes
of terms from each other. All zero and first order terms drop out since
they are satisfied independently of the next order equations.
Moreover, all second order terms which are linear in $R^m$ also drop
out, since they are zero by themselves\footnote{$R^m$ are
independent degrees of freedom. Another way of seeing that this is true is
observing that the Bianchi identities for $R^m$ are given by 
$(G^\mu_{\, \nu|\mu}[b])_{,m} R^m$. We could also regard $R^m$ as linear
perturbations in the same sense that $h_{ij}$ are linear perturbations,
in which case the Bianchi identities are automatically satisfied.}.

We are left with the following equations for the EEMT 
$\tau^\mu_{\,\,\nu} = \meio G^\mu_{\,\, \nu,mn} P^m P^n$ :

\beq
\label{Bia_Tau}
&& \meio \langle \left( G^\mu_{\,\, \nu;\mu} \right)_{,mn} P^m P^n \rangle
	= \partial_\mu \tau^\mu_{\,\, \nu} \\
\nonumber
	&-& \meio \langle \Gamma^\alpha_{\mu\nu,mn} P^m P^n \rangle \times 
		G^\mu_{\,\,\alpha}[\gamma] 
	- \langle \Gamma^\alpha_{\mu\nu,m} P^m \times
		G^\mu_{\,\, \alpha,m}P^m \rangle 
	- \Gamma^\alpha_{\mu\nu}[\gamma] \times
		\tau^\mu_{\,\, \alpha} \\
\nonumber
	&+& 
	\meio \langle \Gamma^\mu_{\alpha\nu,mn} P^m P^n \rangle \times
		G^\alpha_{\,\, \mu}[\gamma] +
	\langle \Gamma^\mu_{\alpha\nu,m} P^m \times 
		G^\alpha_{\,\, \mu,m} P^m \rangle +
	\Gamma^\mu_{\alpha\nu}[\gamma] \times 
		\tau^\alpha_{\,\, \mu} \quad .
\eeq 
The terms involving $\tau$ are what we would naively expect from
the conservation equations for the EMT of perturbations, $\tau^\mu_{\,\,
\nu | \mu}$. The remaining
terms in the second and third lines are corrections stemming from
the interactions of the perturbations with themselves and with the
background.

In the case when there are no matter perturbations, the terms on
$G_{,m} P^m$ are proportional to the equations of motion and
therefore vanish by
themselves, so we find that the only correction
comes from the terms involving second variations of the connections.
We have, finally, that in a FRW background metric the
stress-energy tensor of gravity waves obeys the equations

\be
\label{Con_eq_GW}
\dot{\rho}_{gw} + 3 H (\rho_{gw} + \pi_{gw} ) + 
	\meio \langle \Gamma^\alpha_{\alpha 0,mn} P^m P^n \rangle 
		\left( \rho_o + p_o \right) = 0 \quad ,
\ee
where $\pi_{gw}= -\frac{1}{3} \langle \tau^i_{\,\, i} \rangle $, and
$\rho_o$ and $p_o$ are the energy density and pressure of the homogeneous
background matter. Usually, the continuity equations for hydrodynamic 
fluids in FRW spacetimes are
of the form $\dot\rho + 3 H (\rho + p)=0$. Here, instead, due to
interactions between perturbations and background, there are additional
terms corresponding to an effective pressure. We can write then

\be
\label{Eff_Cont_Eq}
\dot{\rho }_{gw}+3H(\rho_{gw}+p_{gw})=0 \quad ,
\end{equation}
where

\begin{equation}
\label{Eff_pressure}
p_{gw}= \pi_{gw} -
	\frac{1}{6H} \langle \dot{h}^{ij} h_{ij} \rangle 
	(\rho_o + p_o)  
\end{equation}
can be interpreted as the pressure of gravitational waves\cite{ABM97}.
The second term in (\ref{Eff_pressure}) does not contribute to the
pressure only in a de Sitter Universe in which $p_o=-\rho_o$. In
this case the term $-\frac 13\tau^i_{\,\,i}$ itself can be interpreted as
a pressure.

\section{Gravitational waves: short wavelength limit}

We will now study the back reaction and the equation of state for
gravity waves in the limit when they have short wavelengths and high
frequencies (HF) in comparison
with the curvature radius $|a/\dot{a}|$. To calculate the EEMT in some
given background we need 1) the mode solutions $h^{ij}_k(t)$ and 2)
the spectrum $|\delta^h_k|^2$ that determines the rms amplitude of each
mode with wavenumber $k$.

The solutions to the equations of motion (\ref{EOM_GW})
in the limit $k^2 \gg (\dot{a}/a)^2$ are simple plane
waves [see, e.g., Eq. (\ref{Sub_mu})], since short-wavelength waves
propagate without being disturbed by the curvature. The 
frequency, of course, depends on the physical wavenumber $k_{ph}=k/a(t)$
so we have

\be
\label{GW_HF}
h_{ij}(\vec{x},t) = \frac{1}{\sqrt{2}} \int \, \frac{d^3k}{(2\pi)^{3/2}} 
	k^{-3/2} e^{i \vec{k} \vec{x} } 
	\left[ \delta^h_k
	\eps_{ij}(k) 
	e^{ik \int \, dt/a(t) } 
	+ {\rm c.c.} \right] \quad .
\ee

If we assume that there is an ensemble of gravitational waves 
which are isotropically distributed, their average value is
given by a sum over all the modes in all directions. If, in
addition, we average over
a time period $T$ such that $k^{-1} \ll T \ll H^{-1}$, the
rapidly oscillating terms vanish and we have

\be
\label{Corr_GW_HF}
\ll h^{ij} h^\ast_{ij} \gg = \int \, \frac{d^3k}{k^3} \,
	\left| \delta_k^h \right|^2 \eps_{ij} \eps^\ast_{ij} \quad 
\ee
where $\ll \cdots \gg$ denotes space and time average.

A single plane gravitational wave propagating in the $z$ direction,
i.e. $\vec{k}_z=(0,0,k_z)$, has a polarization tensor which is traceless 
and transverse to the $z$ direction, hence
the only nonzero elements of the tensor are $\eps_{11} = -\eps_{22}$ and
$\eps_{12}=\eps_{21}$. The strength of the gravity wave is given,
in this case, by 

\be
\label{Exp_kz}
\ll h^{ij}(z) h_{ij}^\ast(z) \gg
	= 2 \left| A \right|^2 (|\eps_{11}|^2 + |\eps_{12}|^2 )
	\quad ,
\ee
where we have used that 
$\delta^h (\vec{k}_z) = A k_z^{3/2} \delta(\vec{k}-\vec{k}_z)$, and $A$ 
is the adimensional constant that sets the amplitude of the wave.
The polarization tensor is usually normalized as
$\eps_{ij}\eps^\ast_{ij} = 1$, 
so that $\eps_{11}=\eps_{12} =1/\sqrt{2}$.
In terms of spherical polarizations, we have 
$\eps_\pm = (1 \pm i)/\sqrt{2}$.

We can now evaluate the effective energy density and pressure of HF
gravitational waves. Notice that the effect of a time derivative acting
on a wave mode $h_{ij}$ is to bring down a factor of $k/a(t)$.
The terms in (\ref{Tau_GW_00}) and (\ref{Tau_GW_ij}) containing 
$\dot{a}/a$ are negligible, since $k \gg H^{-1}$. Likewise, the
correction terms to the pressure in (\ref{Eff_pressure}) can be neglected.
After collecting the diagonal elements we conclude that, for short
wavelength gravity waves, the energy and pressure are given by

\begin{equation}
\label{Rho_P_GW_HF}
p_{gw} = {\frac 13} \rho _{gw} = 
	\frac {1}{12a^2} \ll h_{kl,m}h_{kl,m} \gg \quad .
\end{equation}
In terms of the physical momenta, the integral can be cast in the
form

\be
\label{Rho_P_GW_HF_2}
\rho_{gw}	= \frac{\pi}{a^4} \int \frac{dk_{ph}}{k_{ph}} k_{ph}^2 
		\left| \delta^h_k \right|^2 \quad ,
\ee
where we have used isotropy to transform the integral over all
space into an integral over the norm $k=||\vec{k}||$.

The result above is the usual one\cite{Weinberg}, and expresses the
fact that gravitational waves in the geometrical optics limit (short
wavelength and
high frequency) behave essentially like radiation, independently of the
evolution of the background. Their equation of state is
$p_{gw}=\rho_{gw}/3$, which is consistent with the time dependence
$\rho_{gw} \propto a^{-4}(t)$.  Notice that this time dependence follows
from the continuity equations (\ref{Eff_Cont_Eq}), by neglecting the
correction terms which are small in the context of HF gravity waves.

In the case when the power spectrum is scale invariant,
$|\delta^h_k|={\cal M}$, the integral over physical momenta has a
quadratic divergence on the UV, a familiar feature from quantum gravity
where divergences are usually of the sort $\int dk \, k^{2n-1}$ for $n$
loops. In Einstein's theory there is no consistent way of regularizing
these graphs, so the theory is termed non-renormalizable. That the
integral (\ref{Rho_P_GW_HF_2}) is divergent is a symptom of this fact,
since the space average that we use corresponds to a 2-point correlation
function, or a 1 loop graph.

If the spectrum $|\delta^h_k|$ has time-dependent cut-offs the integral
(\ref{Rho_P_GW_HF_2}) has an additional time dependence coming from the
limits of integration. As we discussed in Chapter 5, though, on the
grounds of the conservation equations we do not expect this additional
time variation to be of much consequence. In the case of HF gravity waves
the physical IR cut-off could be set as the scale $H(t)$ of the
background, and the UV scale could be some fixed microscopic scale that
depends on the cross-sections of the graviton with the other sectors of
matter. Since the Hubble parameter is at most constant, and usually decays
as a power law of time, the energy density of HF gravitational waves is
dominated by the UV, thus a time dependence in the IR is
usually irrelevant.

\section{Gravitational waves: long wavelength limit}

The solutions to the equations of motion for the gravitational waves, Eq. 
(\ref{EOM_GW}), now depends on the background evolution. 
The back reaction effects of long wavelength gravity waves are
then substantially different than the short wavelength ones\cite{ABM97}.
We will treat here the
cases when the background is dominated by radiation
($p=\rho/3$) and vacuum energy ($p=-\rho$): 

\be
\label{ScaFac_GW_SW}
a(t) = 	\left\{
	\begin{array}{ll}
		\left(\frac{t}{t_0}\right)^{1/2} & {\rm Radiation} \\
		e^{Ht} & {\rm De Sitter} . 
	\end{array} \right.
\ee
Since we only need the solutions in the limit where $k/aH \ll 1$ we
can use the trial solutions

\be
\nonumber
h_{ij}(t) =  k^{-3/2} \delta_k^h \eps_{ij} \left[ 
	1 + \alpha \left( \frac{k}{aH} \right)^2 + \ldots 
	\right]
\ee
and solve for the factor $\alpha$
in lowest order in $k$ by inserting into the equations of motion for
$h_{ij}$. For de Sitter
$\alpha=-1/6$ while for radiation $\alpha=1/2$.

In the de Sitter spacetime we obtain, after
substituting the solutions above into Eqs. (\ref{Tau_GW_00}) and
(\ref{Tau_GW_ij}) and taking into account that $\rho_o=-p_o$
we have that, to lowest order in $k$

\be
\label{Rho_GW_LF_dS}
\rho_{gw} \simeq - \frac {7\pi}{2a^2} 
		\int \, \frac{dk}{k} \, |\delta^h_k|^2 \quad ,
\ee
and the pressure is

\be
\label{P_GW_LF_dS}
p_{gw} \simeq {-} \frac 13 \, \rho_{gw}.
\ee

For a scale-invariant spectrum the integral is logarithmic,
therefore in terms of physical momenta the time dependence of the
energy density and pressure is still $a^{-2}$. Notice that this time
dependence is in perfect agreement with the equation of state
(\ref{P_GW_LF_dS}) and the continuity equation.

The fact that the energy density in long wavelength gravitational
waves is negative has been known for some time in connection
with proposals that the cosmological constant should decay by
Hawking radiation\cite{Deser}. However, unless the spectrum
of the waves is severely scale-dependent (not a very persuasive
suggestion in de Sitter spacetime), the influence of LF gravity
waves should decay as $a^{-2}$. The problem with the proposal is that,
once outside the Hubble horizon, the energy density of the gravity
waves decays fastly with time, although their amplitude remains
more or less constant. In Einstein's theory the $0-0$ component
of the Einstein tensor does not contain terms like $H^2 \langle h^{ij}
h_{ij} \rangle$, which would contribute a non-decaying negative
contribution to the energy density. In quantum gravity with a
cosmological constant in the far
infrared, loop corrections yield just that sort of term, and
it has been suggested\cite{Woodard} that in this case the back reaction
from gravitational perturbations will lead to a decay of the cosmological
constant. A central element of this mechanism is the overpopulation
of the low momentum sector of the spectrum, which we described in Fig.
4.2.

In a radiation-dominated universe the situation is a bit more complicated,
since this time the corrections to the pressure (\ref{Eff_pressure})
must be taken into account.
Substituting the solution for the gravity waves into the equation
for the energy density we have that

\be
\label{Rho_GW_LF_Ra}
\rho_{gw} \simeq - \frac{5\pi}{6a^2}
		\int \, \frac{dk}{k} \, |\delta^h_k|^2 \quad ,
\ee
and the {\it effective} pressure is

\be
\label{P_GW_LF_Ra}
p_{gw}=-\frac{\rho_{gw}}{3} \quad .
\ee
The bare pressure is actually $\pi_{gw} = -\frac{21}{15} \rho_{gw}$, but
the second term in (\ref{Eff_pressure}) contributes with $\Delta p =
\frac{16}{15} \rho_{gw}$.

Because of the approximation we have done
(neglecting terms which are of higher order in the momenta) now the
divergence is logarithmic, instead of the quadratic divergence that
happened previously in the case of hard gravitons. 
This integral still necessitates a UV cut-off, but its origin is now
physically sensible: the UV limit is the comoving horizon
scale $H^{-1}a^{-1}$. The nature of the IR cut-off is the same as
we encountered before in the treatment of quantum fluctuations inside
an inflating bubble, and is related to the largest scale over which
we can consider our background to provide accurate information about
our universe. Any time dependence on the cut-offs will enter
in the expression for the energy density of gravity waves only as a
logarithmic factor, therefore deviations from the $a^{-2}$ behavior
can be neglected as small corrections.

The results above show that the energy density and pressure fall with
$a^{-2}$ for a scale-invariant spectrum, as in the previous case. This is
of course consistent with the effective equation of state
$p_{gw}=-\rho_{gw}/3$.
The energy density in long wavelength gravity waves is again negative,
but this time it is decaying less fast than the
background energy density $\rho_m \propto a^{-4}$, which means that
the back reaction from LF modes could become important.


        \chapter{Back Reaction in the Chaotic Inflation Model}

We now want to focus on the case of scalar
perturbations in the early universe. We consider
as a background model the chaotic inflationary scenario\cite{LindeSlava}, 
of Chapter 3, where the inflaton potential is $V=m^2 \vv^2 /2$.
We derive the equations in longitudinal gauge, where
the gauge invariant variables defined in Section 2.5 coincide
with the variables $\phi$ and $\psi$. Furthermore, since the perturbed
energy-momentum tensor of the scalar field is diagonal, $\psi=\phi$.
As we have argued in Chapter 5, the analysis is independent of the
coordinate frame insofar as we consider only gauge invariant variables
in predictions of physical quantities.

The fundamental scale of the chaotic background model is the
self-reproducing scale\footnote{We 
stress once again that our units are such that
$8\pi G=M_{pl}^2=1$, therefore in Planck mass units $\vv_{sr}=
M_{pl} \sqrt{M_{pl}/m} \gg M_{pl}$.},

\be
\label{Self-rep_sca}
\vv_{sr} = m^{-1/2} \quad .
\ee
In regions where the scalar field is above this scale the universe
is foregoing a process of continuous creation of new inflationary
``bubbles" fueled by large quantum fluctuations of the scalar field. 
On the other hand, regions that see the scalar field drop below the
self-reproducing scale behave as an independent bubble that evolves
following the classical equations of motion. 
We will assume that the region over which the FRW description is
approximately valid is this ``classical" bubble of physical radius 
given by $H^{-1}(t_{sr})a(t)/a(t_{sr})$.

We will perform the analysis using covariant components, and it
is a straightforward matter to show that the final results are
equivalent in any possible components one happens to choose.

\section{Effective energy-momentum tensor of scalar perturbations in
Chaotic Inflation}

We use again the expression (\ref{Ricci_2}) for the second
variation of the Ricci tensor, where the metric now is given by

\be
\label{Metric_scalar}
ds^2 = (1+2\phi) dt^2 - a^2(t) [ (1-2\phi) d\vec{x}^2 ] 
	\quad ,
\ee
where $\vec{x}$ are the comoving coordinates.
The energy-momentum tensor of the scalar field matter is 

\be
\label{EMT_Scalar}
T_{\mu\nu} = \varphi_{;\mu} \varphi_{;\nu} - \gmn \left[ \frac{1}{2}
\varphi^{;\alpha} \varphi_{;\alpha} - V(\vv)
\right] \quad .
\ee

With these inputs, the expressions for the EMT of perturbations are

\begin{eqnarray}  
\label{tauzero}
\tau_{0 0} &=&  12 H \langle \phi \dot{\phi}
\rangle
- 3 \langle (\dot{\phi})^2 \rangle + 9 a^{-2} \langle (\nabla \phi)^2
\rangle  \nonumber \\
&+& \meio \langle ({\delta\dot{\varphi}})^2 \rangle + \meio a^{-2} \langle
(\nabla\delta\varphi)^2 \rangle  \nonumber \\
&+& \meio V''(\varphi_o) \langle \delta\varphi^2 \rangle + 2
V'(\varphi_o) \langle \phi \delta\varphi \rangle \quad ,
\end{eqnarray}
and

\begin{eqnarray}  
\label{tauij}
\tau_{i j} &=& a^2 \delta_{ij} \left[ (24 H^2 +
16 \dot{H}) \langle \phi^2 \rangle + 24 H \langle \dot{\phi}\phi \rangle
\right.  \nonumber \\
&+& \langle (\dot{\phi})^2 \rangle + 4 \langle
\phi\ddot{\phi}\rangle
- \frac{4}{3} a^{-2}\langle (\nabla\phi)^2 \rangle +
4 \dot{{\varphi_o}}^2 \langle \phi^2 \rangle  \nonumber \\
&+& \meio \langle ({\delta\dot{\varphi}})^2 \rangle
- {1 \over 6} a^{-2} \langle(\nabla\delta\varphi)^2 \rangle
- 4 \dot{\varphi_o} \langle \delta \dot{\varphi}\phi \rangle  \nonumber \\
&-& \left. \meio \, V''(\varphi_o) \langle \delta\varphi^2
\rangle + 2 V'( \varphi_o ) \langle \phi \delta\varphi \rangle
\right] \quad ,
\end{eqnarray}
where $H$ is the expansion rate.

First we recover some familiar results for short wavelength scalar
perturbations. In the limit $k \gg aH$, both $\phi
$ and $\delta \varphi $ oscillate with a frequency $\propto k$. In this
case [see (\ref{constr})]: 

\begin{equation}
\phi \sim \meio \,{\frac{ia}k}\,\dot \varphi_o\delta \varphi \,.
\end{equation}  
Hence, it follows by inspection that all terms containing $\phi $ in (\ref
{tauzero}) and (\ref{tauij}) are suppressed by powers of $Ha/k$ compared
to the terms without dependence on $\phi $, and therefore

\begin{equation}
\tau _{00}\simeq {\frac 12}<(\delta \dot \varphi )^2>+{\frac 12}
a^{-2}<(\nabla \delta \varphi )^2>+{\frac 12}\,V^{\prime \prime }(\varphi
_o)<\delta \varphi ^2>
\end{equation}  
and

\begin{equation}
\tau _{ij}=a^2\delta _{ij}\,\left\{ {\frac 12}<(\delta \dot \varphi )^2>  
-{\frac 16}a^{-2}<(\nabla \delta \varphi )^2>-{\frac 12}\,V^{\prime \prime 
}(\varphi _o)<\delta \varphi ^2>\right\} \quad .
\end{equation}  
It can be checked that HF scalar perturbations have the equation of state
of an ultra-relativistic gas (radiation), $p=\rho/3$, and the
energy density in a fixed set of modes decays accordingly, $\rho \propto
a^{-4}$.

We now focus on the more interesting long wavelength, low frequency (LF)
perturbations. First, some general remarks about the problem of back
reaction of scalar perturbations in a model with a general potential
$V(\vv)$. We have shown in Section 2.6 [see formula
(\ref{EOM_ph_0i})] that the spectra of $\dv$ and $\phi$ are related by
the
constraint

\begin{equation}
\label{constr}
\dot \phi +H\phi = \meio \dot \varphi _o\,\delta \varphi \quad .
\end{equation}
During inflation, $\dot\phi$ is proportional to a slow-roll parameter and
can be neglected. We have then, using that $\dot\vv_o \simeq
-\frac{V'}{3H}$, that the spectra of LF scalar perturbations are
constrained by

\begin{equation}
\label{constr2} 
\delta \varphi_k =-{\frac{2V}{V^{\prime }}}\,\phi_k = -\vv_o \phi_k 
	\quad .
\end{equation}

When considering the contributions of long wavelength fluctuations to
$\tau_{\mu \nu }$, we can neglect all terms in (\ref{tauzero}) and
(\ref{tauij})
containing gradients and $\dot\phi $ factors. Because of the ``slow
rolling'' condition, the terms proportional to
$\dot \varphi_o^2$ and $\dot H$ are also negligible during inflation (but
they become important at
the end of inflation). Hence, in this approximation
 
\begin{equation}
\tau _{00}\simeq {\frac 12}\,V^{\prime \prime }\,<\delta \varphi
^2>+2V^{\prime }<\phi \delta \varphi >
\end{equation}
and

\begin{equation}  
\tau _{ij}\simeq a^2\delta _{ij}\,\left\{ {\frac 3{\pi G}}H^2<\phi ^2>
-{\frac 12}\,V^{\prime \prime }<\delta \varphi^2 >+2V^{\prime }<\phi
\delta
\varphi >\right\} \,.
\end{equation}  
Making use of (\ref{constr2}), this yields

\begin{equation}
\rho _s\equiv \tau _0^0\cong \left( 2\,{\frac{{V^{\prime \prime }V^2}}
{{V^{\prime }{}^2}}}-4V\right) <\phi ^2>  
\label{tzerolong}
\end{equation}    
and
\begin{equation}
p_s\equiv -\frac 13\tau _i^i\cong -\rho ^{(2)} \quad .  
\label{tijlong}
\end{equation}  
 
Thus, we have shown that the long wavelength perturbations in an
inflationary Universe have the same equation of state $p_s=-\rho _s$ as
the background. When the scalar field $\vv_o$ is slowly rolling down its
potential, the energy density of the EEMT of perturbations is negative,
which means that back reaction lower the effective potential energy
density of the background.

It might be objected that it is not consistent to throw away terms in the
background energy density like $\dot{\vv}^2_o/2$, and still consider
the back reaction terms above. However, the background terms that
have been discarded do not change appreciably during inflation, whereas
we will show below that the contributions from back reaction are
a growing function of time and, as inflation proceeds, their strength
increase by many orders of magnitude.

\section{The back reaction of long wavelength modes}

We will use the spectrum of perturbations created by quantum
fluctuations of the scalar field to calculate the intensity and
the time dependence of the back reaction terms in the model $V=m^2
\vv^2/2$. We will calculate, in addition to what was done in the last
section, next-to-leading order terms in the energy density
and pressure of the EEMT of perturbations, using the slow-roll
approximation $\dot{H}/H^2 \ll 1$. 
From Eq. (\ref{Spe_ph}) we take that, during inflation,
the power spectrum of the field $\phi$ is given by

\be
\label{Spectrum_phi}
\left| \delta^\phi_k(t) \right| = \frac{m}{2\pi\sqrt{6}} 
		\left[
                        1 - \frac{4}{\vv_o^2(t) - \vv_o^2(t_f)}
                        \log{ \left( \frac{k}{H(t)a(t)} \right) }
                \right] \quad ,
\ee
where $\vv_o(t_f)$ is the
time when inflation ends. 
To arrive at this formula we made use of the solution to the Hubble
constant and scale factor during slow-roll given respectively by
Eqs. (\ref{H_vv_m2}) and (\ref{a_vv_m2}).

We can use the spectrum of $\phi$ to derive the spectra of every term
present in Eqs. (\ref{tauzero}) and (\ref{tauij}). For example,
from the definition of the spectrum we have that

\be
\label{Rel_Spectra}
\left| \delta^{\dot\phi}_k \right| 
= \frac{d}{dt} \left| \delta^{\phi}_k \right| \quad .
\ee

The average value of the fields $\langle \phi \dv \rangle$ over
the comoving volume between the Hubble horizon at $H^{-1}(t)/a(t)$ and the
``particle horizon" of the inflating bubble at $H^{-1}(t_i)/a(t_i)$ is
given by the integral

\be
\label{Int_general}
\langle \phi \dv \rangle_t = - \int_{k_i}^{k(t)} \, \frac{dk}{k} 
	\left| \delta_k^\phi \right| \left| \delta_k^{\dv} \right| \quad ,
\ee
where $k_i=H(t_i)a(t_i)$ and $k(t)=H(t)a(t)$ are the comoving
momentum cut-offs. Notice that the negative sign is due to the fact that
the constraint (\ref{constr2}) implies that
the spectra of $\phi$ and $\dv$ are anti-correlated. There is a sensible
physical reason for that: the newtonian potential is lowest where the
mass density is highest, and vice-versa. The interaction term $\langle
\phi \dv \rangle$ thus expresses an interaction energy, which
lower the total energy of the system and therefore has a negative sign.

We could proceed and calculate all the terms of the energy density and
pressure of the perturbations that improve upon the expressions
(\ref{tzerolong}) and (\ref{tijlong}), but
there is a more elegant way of doing this calculation. We can use the
constraint (\ref{constr2}) linking $\phi$ and $\dv$ to express the whole
energy-momentum of perturbations in terms of the function $\phi^2$ times
some time-dependent tensor. After that,
we can make the averaging by simply calculating $\langle \phi^2 \rangle$.

From Eq. (\ref{phi_LF}) we can deduce, after using the expressions for
the scale factor (\ref{a_vv_m2}) and the approximation $H = -\dot\vv_o
\vv_o/2$ (valid for the potential $m^2\vv^2/2$ while inflation lasts), that 

\be
\label{App_phi}
\phi_k \simeq A_k \frac{1}{\vv_o^2(t)} 
\ee
and, due to (\ref{constr2}),
\be
\label{App_dv}
\dv_k \simeq - A_k \frac{1}{\vv_o(t)} = - \vv_o(t) \phi_k \quad .
\ee
From these expressions it is easy to see that, e.g., 

\be
\label{rel_corr}
\frac{\dot\phi_k}{\phi_k} = -2 \frac{\dot\vv_o}{\vv_o} 
	\quad {\rm and} \quad
\frac{\delta\dot\vv_k}{\dv_k} = - \frac{\dot\vv_o}{\vv_o} \quad .
\ee

If we use these constraints into expressions (\ref{tauzero}) and
(\ref{tauij}) for the energy density and pressure, and neglect terms
of order 
$(1/\vv_o^2) \ll 1$ we get after some algebra that

\be
\label{EMT_scalar_2}
\tau^\mu_\nu = 
	\langle \phi^2 \rangle_t \times 
	\left(
	\begin{array}{cc}
		-3\rho_o & 0 \\
		0 & \delta^i_j( 3p_o - 8\dot\vv_o^2)
	\end{array}
	\right) \quad ,
\ee
where $\rho_o$ and $p_o$ are the background energy density and
pressure, respectively. The term $8\dot\vv_o^2$ is a small correction,
but we keep it because we want to discuss the continuity equations
that $\tau^\mu_\nu$ satisfies. 
The fact that the energy density is negative implies that the biggest
contribution to the energy density of long wavelength scalar modes
comes not from their self-energy, but from their interaction energies.
The main contribution to the formula (\ref{tauzero}) for the 
energy density comes
from the term $2V' \langle \phi \dv \rangle$, which has a negative
sign due to the anti-correlation between the two variables [see the
constraint (\ref{constr2}).]
But for the correction term above,
we would have the approximate EEMT of scalar perturbations given by
Eqs. (\ref{tzerolong}) and (\ref{tijlong}), or

\be
\label{Approx_br}
\tau^\mu_\nu = -3\langle\phi^2\rangle T^\mu_\nu [\vv_o] \quad .
\ee
This shows that the tendency of back reaction is to counteract
the effective vacuum energy from the scalar field that drives inflation.
The crucial point is, does this effect grow in time, or does it
amount to a renormalization of the energy of the scalar field?
If it grows, is there a time when back reaction becomes
so strong that it ends the period of inflation?

Clearly, to answer these and other questions we must finally calculate
the correlation $\langle \phi^2 \rangle$. This is straightforward, with
the spectrum $\delta^\phi_k$ given in formula (\ref{Spectrum_phi}) and
the solution for the scale factor, Eq. (\ref{a_vv_m2}). Assuming that we
calculate the average of $\phi^2$ at a time when $\vv_f \ll
\vv_o (t) \ll \vv_i $, where $\vv_{i(f)}=\vv_o(t_{i(f)})$, we have that 

\be
\label{Ave_phi2}
\langle \phi^2 \rangle_t \, = \, \int_{k_i}^{k(t)} \, \frac{dk}{k} \,
	\left| \delta_k^\phi \right|^2 \, = \,
	\frac{m^2}{96\pi^2} \left( \vv_o^2 - \vv_f^2 \right) \,  
	\int_{\frac{\vv_i^2 - \vv_o^2}{\vv_o^2 - \vv_f^2}}^1 
	\, dx (1-x)^2 
\ee
after changing variables to $x=\frac{4}{\vv_o^2-\vv_f^2} \log{ \left(
\frac{k}{Ha} \right) }$, using expression (\ref{a_vv_m2}) for the scale
factor and neglecting small logarithmic corrections of the order of
$\log{ \left[ \frac{H(t_i)}{H(t_f)} \right] }$ in the lower limit of the
integral. 
We have then, approximately,

\be
\langle \phi^2 \rangle_t =
	\frac{m^2}{288\pi^2} 
	\frac{ \left[ \vv_i^2 - \vv_o^2 \right]^3}
	{ \left[ \vv_o^2(t) - \vv_f^2 \right]^2 } 
	\simeq \frac{m^2}{288\pi^2} 
	\frac{ \vv_i^6}{\vv_o^4(t)} 
\quad ,
\ee
that is, $\langle \phi^2 \rangle_t$ is {\it growing} with time
as $\vv_o^{-4}$ and consequently the influence of back reaction becomes
more important the longer inflation takes. The EEMT is 
proportional to $\vv_i^6/\vv_o^2$, where $\vv_i$ is the value of the
scalar field when inflation has started in the bubble that we are
considering. The bigger $\vv_i$ is in a Hubble size patch, 
the larger that patch will be at the end of inflation, and
the more important back reaction will become.

If we consider that inflation ends when $\vv_o \sim \vv_f \sim 1$,
then by demanding that back reaction never becomes important in our
patch of the universe we constrain the initial value of the scalar field
in that patch to be 

\be
\label{bound_1}
\rho_o(t_f) \ll
|\rho_s(t_f)| \quad \Longrightarrow \quad \vv_i \ll m^{-1/3} \quad .
\ee
If inflation started with the scalar field above this scale,
then back reaction would have been important for the dynamics
of the universe. Notice that the scale $m^{-1/3}$ is much smaller
than the self-reproducing scale $\vv_{sr} = m^{-1/2}$.

We can also estimate the scale above which back reaction is 
important at any time: if $\vv > m^{-1}$ the energy in the EEMT of
perturbations is of the same size as the energy density of
the background. Of course, this bound is of limited value
since it is applied to a region where we expect quantum gravity to
be valid ($m^2 \vv^2 > 1 \, [M_{pl}^4]$.)

It is much more interesting to estimate the time when back reaction
would become important, given a certain initial value of the scalar
field $\vv_i$. From (\ref{Approx_br}) and (\ref{Ave_phi2}) we find
that the value $\vv_{br}$ of the scalar field at which the energy density
in the EEMT becomes of the order of the energy density of the background
is

\be
\label{bound2}
\vv_{br} = m^{1/2} \vv_i^{3/2} \quad .
\ee
If we suppose, as suggested by global structure of the Chaotic Inflationary scenario,
that the initial value of the scalar field is $\vv_{sr}=m^{-1/2}$, we get that
back reaction became important when the value of the scalar field reached

\be
\label{bound3}
\vv_{br}^{(CI)} = m^{1/2} (m^{-1/2})^{3/2} = m^{-1/4} \quad .
\ee

It is natural to suppose that when the scalar field reaches
this value inflation should end, since the negative energy density
of perturbations is of the same size of the energy density of the
background. In that case it is valid to say that the effective energy
density of the background has relaxed to nearly zero.

Another estimate can be made for the strength of the EEMT of
perturbations if we consider that the size of the visible universe today corresponds
to an entire inflating bubble. Since this region inflated at least $50$
$e$-folds, we have from the formula for the scale factor (\ref{a_vv_m2})
that $\vv_i^2 - \vv_f^2 \simeq 200$. The relative size of the terms of the
EEMT at the end of inflation is then, with $m \sim 10^{-6}$, given by
$\langle \phi^2 \rangle \simeq 10^{-8}$.

We can thus summarize some of the interesting 
features of the back reaction of LF scalar perturbations:

\begin{quote}

$\bullet$ The energy density is {\it negative};

$\bullet$ $|\rho_s|$ is {\it growing} in time;

$\bullet$ The equation of state is approximately that of a {\it negative}
cosmological constant term, $p_s \simeq -\rho_s$.

\end{quote}

\section{The continuity equation and the 
Klein-Gordon equation}

There are two loose knots that we should secure before resting our case. 
First, the fact that the energy density of the EEMT is growing faster
than the energy density of the background is not consistent with
the equation of state $p_s \simeq -\rho_s$ of Eq. (\ref{Approx_br}).
The corrections that we derived in the last chapter [see Eq.
(\ref{Eff_pressure})] are very small and cannot account for this
discrepancy.
And second, what is the role of the Klein-Gordon equations, and
are they consistent with the picture given by the EEMT that we
derived?

First we compute the continuity equation for the EEMT. Starting from the
expression (\ref{EMT_scalar_2}) and using the relation (\ref{rel_corr})
we find that

\be
\label{rhodot}
\dot\rho_s \simeq 3 \frac{\dot\vv_o}{\vv_o} m^2 \vv_o^2 \langle \phi^2
\rangle \left[ 1 + {\cal O}(\frac{1}{\vv_o^2}) \right] 
\quad ,
\ee
and thus we find that, by using $H \simeq - \dot\vv_o \vv_o/2$, 
the continuity equations are

\be
\label{Cont_eq_sc}
\dot\rho_s + 3H(\rho_s + p_s) = \dot\vv_o \left( -2m^2 \vv_o\right) 
	\langle \phi^2 \rangle \quad ,
\ee
to lowest order in slow-roll parameters\footnote{The terms from Eq.
(\ref{Eff_pressure}) are much smaller than the ones shown also by a
factor of $1/\vv_o^2$.}. This equation would
imply that the EEMT is not conserved separately because of the interaction
with the background. However, let us consider the case
of the Klein-Gordon equation, which can be obtained through the
4-divergence of the total energy-momentum of the scalar field:

\begin{equation}
	\langle T^\mu_{\,\, \nu;\mu} \rangle = \dot\vv_o \langle
	\dalamb_{\gamma+\delta g}(\varphi _o+\delta \varphi) +
	V'(\varphi _o+\delta \varphi) \rangle = 0 \quad .
\end{equation}
Expanding it to second order in the perturbations, and
prior to making any
simplifications we have that

\begin{eqnarray}
(\ddot{\varphi_o} \, + 3 H \dot{\varphi_o} )
        (1 \, + \, 4 \langle \phi^2 \rangle ) \,
+ \, V' \, + \, \frac{1}{2} V''' \langle \delta\varphi^2 \rangle \,
- \, 2 \langle \phi \delta \ddot{\varphi} \rangle \nonumber \\
- 4 \langle \dot\phi \delta \dot\varphi \rangle \,
- \, 6 H \langle \phi \delta \dot\varphi \rangle \,
+ \, 4 \dot{\varphi_o} \langle \dot\phi \phi \rangle \,
- \, \frac{2}{a^2} \langle \phi \nabla^2\delta\varphi \rangle \, =
\, 0 \quad .
\label{breq}
\end{eqnarray}
As usual, the spatial derivatives can be neglected for LF perturbations.

Now, the equation above indicate clearly that the background
scalar field is corrected by some source terms proportional,
roughly speaking, to $\langle \phi^2 \rangle$.
A correction term to the scalar field must be included, if we are not to
hit upon inconsistencies. This correction to the homogeneous value of the
scalar field can be cast in the form

\be
\label{corr_sfield}
\vv_o(t) \, \longrightarrow \, \vv_o(t) + \eps^2 \upsilon(t) \quad ,
\ee
which is precisely the zero mode of the second-order variables $r^m$ that
we mentioned in the $5^{\rm th}$ chapter. Of course, the metric must also
be altered [see Eqs. (\ref{Def_j})], and the simplest ansatz is to change
the zero modes by

\be
\label{corr_sfactor}
a(t) \, \longrightarrow \, a(t) [ 1+ \eps^2 w(t) ] \quad ,
\ee
which implies that

\be
\label{corr_H}
H(t) \, \longrightarrow \, H(t) + \eps^2 h(t) \quad ,
\ee
where $h(t)=\dot{w}(t)$.

These corrections must be included so that
the second-order Einstein's equations can in fact be solved
perturbatively. Doing these
substitutions and after using the relations (\ref{rel_corr}) we have, to
second order,

\be
\label{EOM_sec}
{T^\mu_{\,\, \nu ; \mu}}^{(2)} = \dot\vv_o
\left( \ddot\upsilon + 3 H \dot\upsilon + 3 h \dot\vv_o + m^2 \upsilon - 
	2m^2 \vv_o \langle \phi^2 \rangle \right) = 0 \, .
\ee
Notice that the last term on the RHS of the above
expression is precisely the extra term found in Eq. (\ref{Cont_eq_sc}).
Therefore there is nothing wrong with the continuity equation
calculated in Eq. (\ref{Cont_eq_sc}), except that we did not allow for the
degrees of freedom from back reaction: if we include those, as we should,
then then the continuity equations {\it imply} the equation of
motion above.

In order to close the picture, the Einstein equations must also accomodate
the degrees of freedom from back reaction. We have, to second order in the
perturbations and in the back reaction variables,

\beq
\label{Eins_sec_00}
6Hh &=& 
	\dot\vv_o \dot\upsilon + m^2 \vv_o \upsilon 
	- 3\rho_o \langle \phi^2 \rangle \quad , \\
\label{Eins_sec_ij}
-6Hh - 2 \dot{h} &=& 
	\dot\vv_o \dot\upsilon - m^2 \vv_o \upsilon 
	+ (8 \dot\vv_o^2 - 3p_o) \langle \phi^2 \rangle \quad ,
\eeq
to lowest order in small parameters like 
$\dot{H}/H^2 \ll 1$ and $1/vv_o^2 \ll 1$.

It can be verified that the integrability conditions for the system of 
equations (\ref{Eins_sec_00})-(\ref{Eins_sec_ij}) is given by
the {\it corrected} equation of motion (\ref{EOM_sec}) for
the back reaction $\upsilon(t)$ on the scalar field $\vv_o(t)$.
The LHS of (\ref{Eins_sec_00}) and (\ref{Eins_sec_ij} should really
be interpreted as the effective second-order dynamical corrections to the
energy density and the pressure of the background.
Of course, to answer what these corrections are one should solve the
coupled equations above for the variables $\upsilon(t)$ and $h(t)$ in
terms of $\vv_o(t)$ (the background metric is already in terms of
$\vv_o$.)

Let us conclude the proof that there are well-behaved solutions
$\upsilon(t)$ and $h(t)$ to the
system of equations above. We have already shown that the continuity
equations for the RHS of (\ref{Eins_sec_00}) and (\ref{Eins_sec_ij})
imply the equations of motion for the back
reaction on the scalar field, (\ref{EOM_sec}).
Now, we prove that the Bianchi identities for the
back reaction metric variable $h=\dot{w}$ on the RHS are satisfied
independently of the continuity equations.
Taking the the expression
$G^\mu_{\,\,\nu;\mu}$ for the metric $ds^2 = dt^2 - a^2(t)[1-2w(t)]
d\vec{x}^2$ and expanding it to second order we obtain

\be
\label{Bianchi_sec}
\frac{d}{dt} ( 6Hh ) + 3H (-2\dot{h}) + 3h( -2\dot{H}) = 0 \quad .
\ee
where we remember that the connection term $3H \rightarrow 3H + 3h$ when
we perform the covariant derivatives. The Bianchi identities are
therefore satisfied for the back reaction variable $h=\dot{w}$.

Summarizing, we found that the effective energy density of
the effective energy-momentum tensor of
long wavelength scalar perturbations
is negative, and counteracts the energy density of the background.
It grows with time, and becomes more important as inflation
proceeds. In addition,
the equations of motion obeyed by the variables expressing
the back reaction of the perturbations onto the background are completely
consistent (they obey the Bianchi identities and the conservation
equations.)

Of course, in order to give final answers to the questions about the
dynamics of the back reaction corrections, if they actually change the
background or if this effect eventually vanishes, we must solve the system
equations (\ref{Eins_sec_00})-(\ref{Eins_sec_ij}) complemented by the
equation of motion (\ref{EOM_sec}. This will the subject of a forthcoming 
publication\cite{Us}.

\section{Conclusions and Discussion}

We have analysed the problem of how perturbations can
effect the background in which they propagate. The
object that expresses this nonlinear feedback effect is the
effective energy-momentum tensor (EEMT) of perturbations.
Fundamental problems related to the gauge dependence of
the EEMT have been addressed, and the resolution lies in
keeping track of all second-order terms coming from the gauge
transformations. We found that upon a coordinate transformation
that redefines the perturbations, the background variables themselves
must be redefined due to second-order zero modes coming from
the gauge transformations. Based on these observations we have been
able to set up a gauge-independent back reaction formalism.

We applied that formalism to the problems of back reaction
of gravitational radiation and scalar perturbations in cosmological
spacetimes, and found that
the EEMT of long wavelength fluctuations has a negative energy density,
corresponding to a slowdown of the expansion. In the case of
gravitational waves, after including correction
terms to the pressure that should be accounted for if 
the time dependence of the EEMT is to be consistent with
the continuity equations, the equation of state 
becomes $p_{gw}=-\rho_{gw}/3$.

For scalar perturbations, we deduced a consistent set of equations that
express the back reaction on the background. The equation of
state of the EEMT is that of a negative cosmological constant,
$p_s=-\rho_s$. This corresponds to an instability of de Sitter space to
long wavelength perturbations, an effect which has already been noticed in
Ref. \cite{Deser}.

Our back reaction formalism is purely classical
in contrast
to the quantum effects discussed in Refs. \cite{Ford,Mottola}.
However, the source of the fluctuations which we consider is quantum
mechanical (perturbative quantum gravity), and in this sense back reaction
may be able to probe quantum gravitational effects. 
For the spectrum of perturbations predicted in inflationary cosmologies
our back reaction effect is dominated by the infrared, a result similar to
the one found in the analysis of Ref. \cite{Woodard},
where the effect of tensorial quantum gravitational fluctuations which
overpopulate the infrared was calculated.

In the context of a chaotic inflationary universe scenario, back
reaction effects can become very important to the dynamics of the
background before the end of inflation, if inflation began
in that region with high enough initial values of the scalar field. If
this initial value is considered to be the self-reproducing scale
$\vv_{sr}=m^{-1/2}$, then our conclusion is that back reaction must be
taken into account before the time when the scalar field reaches the value
$\vv_{br} \simeq m^{-1/3}$.

On a more speculative note, back reaction might also provide a dynamical
relaxation mechanism whereby a preexisting cosmological constant is
cancelled. The cosmological constant is hundreds of orders of magnitude
smaller than the value suggested from dimensional analysis or even from
supersymmetry, which indicates that some mechanism, very likely a
cosmological one, is acting to preclude its existence from showing up in
our observations. We have seen how the back reaction from scalar
perturbations have the effect of counteracting the effective
cosmological constant of the scalar field potential. In that picture,
inflation ends when the value of the energy density in the background
scalar field is balanced by the energy density coming of the EEMT
of long wavelength perturbations. Analogously, we speculate
that back reaction will act to relax the cosmological
constant of the universe to an effective value which is comparable
to the energy density in ordinary matter (a conservative bound
given the present observational data.) This would solve one of
the longest-standing problems of theoretical physics.

\vskip 1cm
\noindent {\large \bf Acknowledgments}
\vskip 0.5cm

\noindent I would like to thank my advisor, Prof. Robert Brandenberger,
and Prof. V. Mukhanov for their guidance in this project. Many thanks
are due to Fabio Finelli as well, for his invaluable help in solving some
of the problems tackled here. Jackson Maia, Matthew Parry and Wendy Sigle 
are responsible for the eventual shadows of good style. This
work was supported primarily by CNPq of Brazil, award 20.0727/93.1
and by the U.S. DOE under contract DE-FG0291ER40688, Task A. 
The author also benefited from a NSF Collaborative
Research Award NSF-INT-9312335.

	{
\appendix

	\chapter{De Witt's conventions}

Here we describe the conventions used for functional operations
and functional derivatives of tensor fields, initially described in
Chapter 4 and used throughout in this monograph.

Consider tensor fields $w^{\alpha \cdots}_{\beta \cdots}(x^\sigma)$
($w(x)$ for short) which are defined in the manifold ${\cal M}$ with
respect to the coordinate chart $x^\sigma$. We say that $w$ is a tensor
{\it functional} of other tensor fields $f^{\mu \cdots}_{\nu
\cdots}(x^{\sigma})$ if $w(x)$ can be expressed as a local function of
$f(x)$ and its derivatives $\partial f$. Of course, general covariance
implies that generic tensorial
expressions should also involve the metric $g$ and its derivatives
$\partial g$ in the covariant derivatives of $f$. We collect the
``variables" $f$ and $g$ in the following way: 

\be
\label{define_q}
q^m(x)= \left[
	f^{0 0 \cdots}_{0 0 \cdots}(x),
	f^{1 0 \cdots}_{0 0 \cdots}(x),
	f^{0 1 \cdots}_{0 0 \cdots}(x), \cdots ,
	f^{3 3 \cdots}_{3 3 \cdots}(x), 
	g_{00}(x), g_{01}(x), \cdots , g_{33}(x) \right] \quad ,
\ee
where $m$ is a ``field" index running from $1$ to the total number of
degrees of freedom in $f$ and $g$, provided that degrees of
freedom are not counted twice (for example, we include $g_{01}$ but not
$g_{10}=g_{01}$.)

The functional derivative of $w(x)$ with respect to $q^m(x')$
is defined in terms of the infinitesimal variation $q^m(x') + \delta
q^m(x')$,

\be
\label{functional_derivative}
\frac{\delta w(x)}{\delta q^m(x')} = \lim_{\delta q^m(x') \rightarrow 0}
	\frac{ w[q^n(x) + \delta q^n(x)] - w[q^n(x)] }{\delta q^m(x')} 
	\quad ,
\ee
where it is understood that 

\be
\label{delta_dirac}
\frac{ \delta f^n (x) }{ \delta f^m (x') } = \delta^n_m \delta(x-x')
\ee
for two independent degrees of freedom $f^n$ and $f^m$.

A variation of the functional $w$ under 
a fixed variation $\bar\delta q^m(x)$ of the variable $q^m$ can be
expressed as 

\be
\label{var_w}
\bar\delta w(x) = \int \, \frac{ \delta w(x) }{\delta q^m(x')} 
	\, \cdot \, \bar\delta q^m(x') \, dx' \quad .
\ee
Clearly, from the formula above, $\delta w(x)/\delta q^m(x')$ is
an operator acting on the variation $\bar\delta q^m(x)$. For example,
suppose that $w$ is the Ricci tensor $R_{\mu\nu}$, and the $q^m$ are just
the metric. We have in this case the differential operator

\beq
\frac{\delta R_{\mu\nu}(x)}{\delta g_{\alpha \beta}(x')} 
	&=& \delta(x-x') \left[ \, 
	\delta^\alpha_\mu \delta^\beta_\nu 
	\nabla^\sigma (x') \nabla_\sigma(x')
	+ \nabla_\mu (x') \nabla_\nu(x') g^{\alpha\beta}(x') \right. \\
\nonumber
	&& \left. - \delta^\beta_\mu \nabla_\sigma(x') \nabla_\nu(x')
		g^{\alpha\sigma}(x')
	- \delta^\beta_\nu \nabla_\sigma(x') \nabla_\mu(x')
		g^{\alpha\sigma}(x') \,
	\right]
\eeq
acting on the fixed variation $\bar\delta g_{\alpha\beta}(x')$. After
operating on $\bar\delta g$ and integrating the result, we obtain what
is commonly called the linearization of $R_{\mu\nu}$ with respect
to the metric perturbation $\bar\delta g_{\alpha\beta}(x)$.

To facilitate the discussions, we have adopted De Witt's condensed
notation\cite{DeWitt} which assumes that
the field index $m$ includes the continuous coordinates $x$
as well, so for example $q^m(x') \equiv q^{m'}$. We also use an
extension of Einstein's conventions that whenever field indices appear
twice they should be summed over and the expression should be integrated,
so that for example the contraction and integration of the tensors
$f_{\mu\nu}$ and $g^{\mu\nu}$ in this notation reads

\be
\label{example_conv}
f_{m} g^m = \sum_{\mu\nu} \int \, dx \, f_{\mu\nu}(x) g^{\mu\nu}(x) 
	\quad .
\ee

Finally, for functional derivatives we use the short-hand notation

\be
\label{example_2}
w_{,m'}dq^{m'} 
	\equiv \int \, \frac{ \delta w(x) }{\delta q^m(x') }
	\cdot dq^m(x') \, dx' 
\ee
where $m$ now is a dummy index, $w_{,m'} dq^{m'} = w_{,m} dq^m$.

	\chapter{Transformation Law for the Ricci Tensor}

Here we prove some useful formulas connected with the Lie derivatives of
riemannian objects.

\section{Metric and connections}

Consider a metric that underwent a gauge
transformation $\tx^\mu = x^\mu + \xi^\mu(x)$,

\beq
\label{def_tr_metric}
\tgmn = \gmn - (\Lie g)_{\mu\nu}
	= \gmn - g_{\mu\nu,\alpha} \xi^\alpha -
	g_{\alpha\mu} \xi^\alpha_{,\nu} -
	g_{\alpha\nu} \xi^\alpha_{,\mu} \quad ,
\eeq
and the contravariant components transform as

\be
\label{tr_cometric}
\tilde{g}^{\mu\nu} = g^{\mu\nu} - (\Lie g)^{\mu\nu} 
	= g^{\mu\nu} - g^{\mu\nu}_{,\alpha} \xi^\alpha +
	g^{\alpha\mu} \xi^\nu_{,\alpha} +
	g^{\alpha\nu} \xi^\mu_{,\alpha} \quad .
\ee
It can be easily verified that $\tilde{g}^{\sigma\mu} \tgmn =
\delta^\sigma_\nu + {\cal O}(\xi^2)$, and we remind the reader that
$\xi$ is a ``small" vector in the sense that $\tgmn-\gmn={\cal O}(\eps)$.

Under a gauge transformation the connections $\Gamma$ transform by
the following formula, obtained after a bit of algebra:

\beq
\nonumber
\tilde{\Gamma}^\sigma_{\mu\nu} &=& 
	\Gamma^\sigma_{\mu\nu} -
	\Gamma^\sigma_{\mu\nu,\alpha}\xi^\alpha +
	\Gamma^\alpha_{\mu\nu}\xi^\sigma_{,\alpha} -
	\Gamma^\sigma_{\alpha\nu}\xi^\alpha_{,\mu} -
	\Gamma^\sigma_{\alpha\mu}\xi^\alpha_{,\nu} - 
	\xi^\sigma_{,\mu\nu} \\
\label{transf_connections}
	&=& 	\Gamma^\sigma_{\mu\nu} - (\Lie \Gamma)^\sigma_{\mu\nu} -
	\xi^\sigma_{,\mu\nu} \quad .
\eeq
We have used the definition of the connection, $\Gamma^\sigma_{\mu\nu} =
\meio g^{\sigma\lambda} \left( g_{\lambda\mu,\nu} + g_{\lambda\nu,\mu} -
g_{\mu\nu,\lambda} \right) $ and substituted into it expressions 
(\ref{def_tr_metric}) and (\ref{tr_cometric}) for the metric in the
transformed coordinate frame $\tx$ (we neglect all ${\cal O}(\xi^2)$ terms
here.)

\section{Ricci tensor}

We want to demonstrate now the identity $\Lie R_{\mu\nu} [g]= R_{\mu\nu \,
,m} (\Lie g)^m$. First, consider the Ricci tensor calculated with
metric (\ref{def_tr_metric}),

\beq
\label{Ricci_1}
R_{\mu\nu}[\tilde{g}] &=& 
	\tilde{\Gamma}^\alpha_{\mu\nu,\alpha} -
	\tilde{\Gamma}^\alpha_{\mu\alpha,\nu} +
	\tilde{\Gamma}^\beta_{\mu\nu}
		\tilde{\Gamma}^\alpha_{\beta\alpha} -
	\tilde{\Gamma}^\beta_{\mu\alpha}
		\tilde{\Gamma}^\alpha_{\nu\beta}  \\
\nonumber
	&=& R_{\mu\nu}[g] -
	\left[ (\Lie \Gamma)^\alpha_{\mu\nu} \right ]_{,\alpha} +
	\left[ (\Lie \Gamma)^\alpha_{\mu\alpha} \right ]_{,\nu} \\
\nonumber &-&
	\Gamma^\beta_{\mu\nu} \left[ 
		(\Lie \Gamma)^\alpha_{\alpha\beta} +
		\xi^\alpha_{,\alpha\beta} \right] -
	\Gamma^\beta_{\alpha\beta} \left[ 
		(\Lie \Gamma)^\alpha_{\mu\nu} +
		\xi^\alpha_{,\mu\nu} \right] \\
\nonumber &+&
	\Gamma^\beta_{\mu\alpha} \left[ 
		(\Lie \Gamma)^\alpha_{\nu\beta} +
		\xi^\alpha_{,\nu\beta} \right] -
	\Gamma^\beta_{\nu\alpha} \left[ 
		(\Lie \Gamma)^\alpha_{\mu\beta} +
		\xi^\alpha_{,\mu\beta} \right] \quad .
\eeq

At this point it is useful to derive the following identity for
the Lie derivative of the connections,

\be
\label{useful}
\left[ (\Lie \Gamma)^\alpha_{\mu\nu} \right]_{,\sigma} =
	(\Lie \Gamma)^\alpha_{\mu\nu,\sigma} -
	\Gamma^\beta_{\mu\nu} \xi^\alpha_{,\beta\sigma} +
	\Gamma^\alpha_{\beta\nu} \xi^\beta_{,\mu\sigma} +
	\Gamma^\alpha_{\beta\mu} \xi^\beta_{,\nu\sigma}
\ee
where the first term on the RHS is understood to be the Lie derivative of
$\Gamma^\alpha_{\mu\nu,\sigma}$. 
Using this formula into expression (\ref{Ricci_1}) we get that

\beq
\label{Ricci_1_final}
R_{\mu\nu}[\tg] &=& R_{\mu\nu}[g] - 
	(\Lie\Gamma)^\alpha_{\mu\nu,\alpha} +
	(\Lie\Gamma)^\alpha_{\alpha\mu,\nu} \\
\nonumber &-&
	\Gamma^\beta_{\mu\nu} (\Lie\Gamma)^\alpha_{\beta\alpha} -
	\Gamma^\beta_{\alpha\beta} (\Lie\Gamma)^\alpha_{\mu\nu} +
	\Gamma^\beta_{\mu\alpha} (\Lie\Gamma)^\alpha_{\nu\beta} +
	\Gamma^\beta_{\nu\alpha} (\Lie\Gamma)^\alpha_{\mu\beta} \quad .
\eeq

We must compare now the expression (\ref{Ricci_1_final}) with the
result of the Lie derivative of the Ricci tensor,

\be
\label{Ricci_sec}
(\Lie R)_{\mu\nu} 
	=  R_{\mu\nu,\alpha} \xi^\alpha +
	R_{\alpha\mu} \xi^\alpha_{,\nu} +
	R_{\alpha\nu} \xi^\alpha_{,\mu} \quad ,
\ee
where $R$ is given in the first line of (\ref{Ricci_1}) with $\Gamma$
instead of $\tilde\Gamma$. 
It is easy to check that all the terms on (\ref{Ricci_1_final}) cancel
exactly those in (\ref{Ricci_sec}), therefore

\be
\label{proof_final}
[({1} - \Lie) \cdot R]_{\mu\nu}[g] =
R_{\mu\nu}[({1}-\Lie) \cdot g] \quad ,
\ee
where $1$ is the identity. This completes the proof
that, to first order, the differential operator $\hat{1}-\Lie$
acting on the functional $R[g]$ is equal to the functional $R$
evaluated on the ``operated" variables $({1}-\Lie)g$.

Expanding Eq. (\ref{proof_final}) in powers of $\xi$ we have that,
to first order,

\be
\label{proof_2}
\Lie R[g] = R_{,m} \cdot (\Lie g)^m \quad ,
\ee
which is the desired result.

	}


\begin{thebibliography}{100}

\bibitem[*]{presadd} Present Address: Physics Dept,
University of Florida, PO Box 118440, Gainesville, FL 32611-8440. E-mail:
abramo@phys.ufl.edu.

\bibitem{Isaacson} R. A. Isaacson, {\it Phys. Rev.} {\bf 166}, 5 (1968).


\bibitem{Landau}  L. Landau and E. M. Lifshitz, ``The Classical Theory of
Fields'', 4$^{th}$ Ed. (Pergamon Press, Oxford 1975).


\bibitem{Weinberg}  S. Weinberg, ``Gravitation and Cosmology'' (Wiley, New
York 1972).


\bibitem{MTW}  C. W. Misner, K. S. Thorne and J. A. Wheeler,
``Gravitation''
(W. H. Freeman, New York 1973).



\bibitem{LindeSlava} A. S. Goncharov, A. D. Linde and V. F. Mukhanov, {\it
Int. J. Mod. Phys.} {\bf A2}, 561-591 (1987); see also Ref.
\cite{LindeBook}


\bibitem{COBE} G. F. Smoot {\it et al.}, {\it Ap. J.} {\bf 396}, L1
(1992).

\bibitem{CMBR} For a recent review see, e.g., G. F. Smoot, preprint
{\bf astro-ph/9705135} (1997).

\bibitem{Einstein} A. Einstein, in ``The Principle of Relativity", Notes
by A. Sommerfeld (Dover, London 1953).

\bibitem{Piran} T. Piran, preprint {\bf gr-qc/9706049} (1997).



\bibitem{Grischuk} L. P. Grishchuk, {\it Phys. Rev. }{\bf D50}, 7154
(1994).

\bibitem{SlavaDerruele} N. Deruelle and V. F. Mukhanov, {\it Phys.
Rev.} {\bf D52}, 5549 (1995).


\bibitem{HubbleConst} W. Freedman, preprint 
{\bf astro-ph/9706072} (1996).


\bibitem{RevPaper}  V. Mukhanov, H. Feldman and R. Brandenberger, {\it
Phys. Rep.} {\bf 215}, 203 (1992).

\bibitem{Sachs} R. K. Sachs, in ``Relativity, Groups and Topology'',
ed. C. de Witt and B. de Witt (Gordon \& Breach, New York 1964.)

\bibitem{StewartWalker} J. M. Stewart and M. Walker, {\it Proc. R. Soc.
Lond.} {\bf A341}, 49 (1974).


\bibitem{Wald} R. M. Wald, ``General Relativity" (University of Chicago
Press, Chicago 1984.)

\bibitem{Stewart} J. Stewart, ``Advanced General Relativity'' (Cambridge
University Press, New York 1990.)

\bibitem{Bardeen} J. Bardeen, {\it Phys. Rev.} {\bf D22}, 1882 (1980).


\bibitem{Stewart90} J. M. Stewart, {\it Class. Quantum Grav.} {\bf 7},
1169 (1990).




\bibitem{FirstInflation} A. A. Starobinsky, {\it Phys. Lett.} {\bf B91},
99 (1980); A. H. Guth, {\it Phys. Rev.} {D23}, 347 (1981); K. Sato, 
{\it Mon. Not. R. Astron. Soc.} {\bf 195}, 467 (1981); A. Albrecht and P.
J. Steinhardt, {\it Phys. Rev. Lett.} {\bf 48}, 1220 (1982); A. D. Linde,
{\it Phys. Lett.} {\bf B108}, 389 (1982).

\bibitem{LindeBook} A. D. Linde, ``Particle
Physics and Inflationary Cosmology'' (Harwood, Chur 1990); A. D. Linde,
{\it Phys.  Scr.} {\bf T36}, 30 (1991). 


\bibitem{InflationBook} ``The Early Universe," Eds. M. Turner and
E. Kolb (Addison-Wesley, Redwood City 1988.)

\bibitem{RevRobert} R. H. Brandenberger, {\it Rev. Mod. Phys.} {\bf 57}, 1
(1985).

\bibitem{Reheating} 
J. Traschen and R.
Brandenberger, {\it Phys. Rev.} {\bf D42}, 2491 (1990);
Y. Shtanov, J. Traschen and R.
Brandenberger, {\it Phys. Rev.} {\bf D51}, 172 (1995);
L. A. Kofman, A. D. Linde and A. A. Starobinsky, 
{\it Phys. Rev. Lett.} {\bf 73}, 3195 (1994); 
{\it Phys. Rev. Lett.} {\bf 76}, 1011 (1996); 
{\it Phys. Rev. } {\bf D56} (1997), to appear.

\bibitem{HawkingGibbons} S. Hawking and G. Gibbons, {\it Phys. Rev.} {\bf
D15}, 2738 (1977).

\bibitem{Linde}  A. Linde, {\it Phys. Lett.} {\bf 129B}, 177 (1983).
 
\bibitem{Starob}  A. A. Starobinsky , in ``Current Topics in Field Theory,
Quantum Gravity and Strings'', ed. H. J. de Vega and N. S\'anchez, Lecture
Notes in Physics Vol. 246 (Springer-Verlag, Berlin 1982.)




 
\bibitem{Nucleo} R.L. Davis, {\it Phys. Lett.} {\bf 161B}, 285 (1985);
D. Bennett, {\it Phys. Rev.} {\bf D33}, 872 (1986); R. Brandenberger, A. 
Albrecht and N. Turok, {\it Nucl. Phys.} {\bf B277}, 605 (1986).

\bibitem{Geometrodynamics} J. A. Wheeler, {\it Geometrodynamics} (Academic
Press, 1962.)

\bibitem{BrillHartle} D. R. Brill and J. Hartle, {\it Phys. Rev.} {\bf
135} (1964).

\bibitem{Brill} D. R. Brill, {\it Ann. Phys.} (N.Y.) {\bf 7}, 466
(1959); {\it Nuovo Cim. Suppl.} {\bf II}, 1 (1964).


\bibitem{WheelerGeon} J. A. Wheeler, {\it Phys. Rev.} {\bf 97} (1955);
D. R. Brill and J. A. Wheeler, {\it Rev. Mod. Phys.} {\bf 29} (1957).


\bibitem{BrillDeser} D. R. Brill and S. Deser, {\it Ann. Phys.} {\bf 50},
542 (1968); {\it Phys. Rev. Lett.} {\bf 20}, 8 (1968); See also E. Witten,
\CMP{80}{81}{}.


\bibitem{Fock} V. Fock, {\it Rev. Mod. Phys.} {\bf 29}, 3 (1957).


\bibitem{Choquet} Y. Choquet-Bruhat, \CMP{12}{69}{}.

\bibitem{TaubMacCallum} M. A. H. MacCallum and A. H. Taub, \CMP{30}{73}{}.

\bibitem{Burnett} G. A. Burnett, {\it J. Math. Phys.} {\bf 30}, 1 (1989).

\bibitem{Efroimsky} M. Efroimsky, {\it Class. Quantum Grav.} {\bf 9},
2601 (1992). 

\bibitem{DeWitt}  B. DeWitt, ``Dynamical Theory of Groups and Fields''
(Gordon and Breach, New York 1965.)
 

\bibitem{Ellis} G. F. R. Ellis, in {\it General Relativity and Gravitation:
GR10 Conf. Rep.}, ed. B. Bertotti, Dordrecht, 1984; G. Ellis and W. Stoeger,
{\it Class. Q. Grav.} {\bf 4}, 1697 (1987).

\bibitem{Buchert} T. Buchert, preprint {\bf astro-ph/9512107} (1995).

\bibitem{Ehlers} T. Buchert and J. Ehlers, {\it Astron. Astrophys.} {\bf 320},
1 (1997); {\it Gen. Rel. Grav.} {\bf 29}, 733 (1997).

\bibitem{Zalaletdinov} R. M. Zalaletdinov, {\it Gen. Rel. Grav.} {\bf 24}, 1015
(1992); {\bf 25}, 673 (1993); {\bf 28}, 953 (1996); preprint {\bf
gr-qc/9703016} (1997).

\bibitem{Carfora} M. Carfora and A. Marzuoli, {\bf Class. Q. Grav.} {\bf
5}, 659 (1988); M. Carfora, J. Isenberg and M. Jackson, {\it J. DIff. 
Geom.} {\bf 31}, 249 (1990). 

\bibitem{CarforaPiot} M. Carfora and K. Piotrkowska, {\it Phys. Rev. }{\bf
D52}, 4393 (1995).

\bibitem{Hamilton} R. S. Hamilton, {\it J. Diff. Geom.} {\bf 17}, 255 (1982).


\bibitem{Futamase} T. Futamase, {\it Phys. Rev. Lett.} {\bf 61}, 2175
(1988).



\bibitem{Bildhauer} S. Bildhauer and T. Futamase, {\it Gen. Rel. Grav.}
{\bf 23}, 1251 (1991); T. Futamase, {\it Phys. Rev.} {\bf D53}, 681 (1996).

\bibitem{Shibata} M. Shibata, K. Nakao, T. Nakamura and K. Maeda, {\it
Phys. Rev.} {\bf D50}, 708 (1994).


\bibitem{Uros} U. Seljak and L. Hui, in ``Clusters, Lensing and the Future
of the Universe'', Proceedings, College Park, Maryland (in press).

\bibitem{Turner} X. Shi and M. Turner, preprint {\bf astro-ph/9707101}
(1997).




\bibitem{MAB96}  V. Mukhanov, L. R. Abramo and R. Brandenberger, {\it
Phys. Rev. Lett.} {\bf 78}, 1624 (1997).

\bibitem{ABM97} L. R. Abramo, R. Brandenberger and V. Mukhanov, {\it Phys.
Rev. } {\bf D56}, 6 (1997).

\bibitem{Bruni}  M. Bruni, S. Matarrese, S. Mollerach and S. Sonego,
 preprint {\bf gr-qc/9609040} (1996).


\bibitem{Taub}  A. H. Taub, {\it J. Math. Phys.} {\bf 2}, (1961) 787 .
 
\bibitem{Schouten} J. A. Schouten, ``Ricci Calculus: an introduction to
tensor analysis and its geometrical applications." (Springer, Berlin 
1954.)


\bibitem{Matarrese} S. Mollerach and S. Matarrese, {\it Phys. Rev.  } {\bf
D56} (1997), to be published; S. Matarrese, S. Mollerach and M. Bruni,
{\bf astro-ph/9707278} (1997). 







\bibitem{Deser} L. Abbott and S. Deser, {\it Nucl. Phys.} {\bf B195}, 76
(1982).


\bibitem{Woodard} N. Tsamis and R. Woodard, {\it Phys. Lett.} {\bf 301},
351 (1993); N. Tsamis and R. Woodard, {\it Nucl. Phys.} {\bf B474}, 235
(1996).
 




\bibitem{Ford} L. Ford, {\it Phys. Rev.} {\bf D31}, 710 (1985).
 
\bibitem{Mottola} E. Mottola, {\it Phys. Rev.} {\bf D31}, 754 (1985); {\bf
D33}, 1616 (1986); {\bf D33}, 2136 (1986);
P. Mazur and E. Mottola, {\it Nucl. Phys.} {\bf B278}, 694 (1986); J.
Traschen and C. Hill, {\it Phys. Rev.} {\bf D33}, 3519 (1986).


\bibitem{Us} L. R. Abramo, R. Brandenberger and V. Mukhanov (in
preparation.)



\end{thebibliography}
\end{document}